\documentclass[fleqn,usenatbib]{mnras}

\usepackage[T1]{fontenc}

\DeclareRobustCommand{\VAN}[3]{#2}
\let\VANthebibliography\thebibliography
\def\thebibliography{\DeclareRobustCommand{\VAN}[3]{##3}\VANthebibliography}

\usepackage{graphicx}	
\usepackage{amsmath}	
\usepackage{amssymb}	

\usepackage{mathtools, cuted}

\usepackage{multirow}

\def\gsim{\lower.5ex\hbox{$\; \buildrel > \over \sim \;$}}
\def\lsim{\lower.5ex\hbox{$\; \buildrel < \over \sim \;$}}
\newcommand{\msun}{$M_\odot$}
\newcommand{\angstrom}{\mbox{\normalfont\AA}}
\usepackage{soul}	
\usepackage[dvipsnames]{xcolor}

\usepackage{newtxtext,newtxmath}

\title[The Origin of Galactic Cosmic Rays]{The Origin of Galactic Cosmic Rays as Revealed by their Composition}

\author[V. Tatischeff et al.]{
Vincent Tatischeff,$^{1}$\thanks{E-mail: vincent.tatischeff@csnsm.in2p3.fr}
John C. Raymond,$^{2}$
Jean Duprat$^{3}$
Stefano Gabici$^{4}$
and Sarah Recchia$^{1,5}$
\\
$^{1}$Universit\'e Paris-Saclay, CNRS/IN2P3, IJCLab, F-91405, Orsay, France\\
$^{2}$Center for Astrophysics | Harvard \& Smithsonian, 60 Garden St. Cambridge, MA 02138, USA\\
$^{3}$Institut de Min\'eralogie, de Physique des Mat\'eriaux et de Cosmochimie, CNRS-MNHN-Sorbonne Universit\'e, F-75005, Paris, France\\
$^{4}$Universit\'e de Paris, CNRS, Astroparticule et Cosmologie,  F-75006 Paris, France\\
$^{5}$Dipartimento di Fisica, Universit\'a di Torino \& INFN - Sezione di Torino, via Giuria 1, 10122 Torino, Italy
}

\def\kms{$\rm km~s^{-1}$}

\date{Accepted XXX. Received YYY; in original form ZZZ}

\pubyear{2021}

\begin{document}
\label{firstpage}
\pagerange{\pageref{firstpage}--\pageref{lastpage}}
\maketitle

\begin{abstract}
Galactic cosmic-rays (GCRs) are thought to be accelerated in strong shocks induced by massive star winds and supernova explosions sweeping across the interstellar medium. But the phase of the interstellar medium from which the CRs are extracted has remained elusive until now. Here, we study in detail the GCR source composition deduced from recent measurements by the AMS-02, Voyager 1 and SuperTIGER experiments to obtain information on the composition, ionisation state and dust content of the GCR source reservoirs. We show that the volatile elements of the CR material are mainly accelerated from a plasma of temperature $\gsim 2$~MK, which is typical of the hot medium found in galactic superbubbles energised by the activity of massive star winds and supernova explosions. Another GCR component, which is responsible for the overabundance of $^{22}$Ne, most likely arises from acceleration of massive star winds in their termination shocks. From the CR-related $\gamma$-ray luminosity of the Milky Way, we estimate that the ion acceleration efficiency in both supernova shocks and wind termination shocks is of the order of $10^{-5}$. The GCR source composition also shows evidence for a preferential acceleration of refractory elements contained in interstellar dust. We suggest that the GCR refractories are also produced in superbubbles, from shock acceleration and subsequent sputtering of dust grains continuously incorporated into the hot plasma through thermal evaporation of embedded molecular clouds. Our model explains well the measured abundances of all primary and mostly primary CRs from H to Zr, including the overabundance of $^{22}$Ne. 
\end{abstract}

\begin{keywords}
cosmic rays -- ISM: abundances -- stars: abundances -- supernova remnants
\end{keywords}



\section{Introduction}
\label{sec:intro}

Galactic cosmic rays (GCRs) are believed to be powered by supernova (SN) explosions in the interstellar medium (ISM). The intensity of CRs, measured directly near Earth and inferred indirectly on the Galactic scale from the $\gamma$-ray luminosity of the Milky Way, can be explained if $\sim 10$\% of the mechanical energy released by SN outbursts is converted into kinetic energy of CRs \citep{baa34,str10}. Such a particle acceleration efficiency can be accounted for by the first order Fermi mechanism 
operating at SN blast waves \citep[also called diffusive shock acceleration;][]{kry77,axf78,bel78,bla78}. This is broadly supported by X- and $\gamma$-ray observations of SN remnants \citep[SNRs;][]{hel12,ack13,ace16}, although the SNR paradigm for the origin of GCRs is facing several difficulties from recent measurements \citep{gab19}. 

However, the nature of the interstellar reservoir(s) from which the CRs are extracted is not well known. SNRs are observed in various environments, where the density, temperature and ionisation state of the ambient gas can differ substantially \citep[see, e.g.,][and references therein]{hel12}. For instance, Cassiopeia A (Cas A) is probably still interacting with the stellar wind lost by the progenitor star prior to explosion, Tycho's SNR seems to evolve in a warm and partially ionised ISM, while the Crab SN appears to have exploded in a bubble of hot gas. On the theoretical side, the diffusive shock acceleration theory does not provide any clear answers on the nature of the GCR source reservoir(s). Some calculations favour an origin of these particles in SN shocks propagating into the warm ISM \citep{cap17,eic21}. But there are arguments in favour of GCRs being mainly produced in the hot ionised medium, where most Galactic SNe are expected to explode \citep{par04,lin07} and where diffusive shock acceleration could be more efficient than elsewhere in the Galaxy \citep{axf81,byk14}. 

The full understanding of the origin of GCRs requires to be able to explain in detail the chemical composition of these particles. The most extensive work on this topic was done by \citet{mey97}, who showed that refractory elements are relatively more abundant than volatile ones in the composition of GCRs at their sources, which can be explained if dust grains are injected into the diffusive shock acceleration process more efficiently than ions \citep{ell97}. \citet{mey97} also found that among the GCR volatile elements, the heavier ones are enhanced relative to the lighter ones, which the authors attribute to a dependence of the acceleration efficiency on the magnetic rigidity of ions, and thus on their mass-to-charge ratio. Furthermore, as compared with the elemental composition of the solar system, the GCR source composition was found to be characterised by a general overabundance of heavier elements relative to H and He. As discussed below, this reported overabundance is related to the fact that CR protons and $\alpha$-particles have significantly different source spectra than the heavier elements \citep{tat18,evo19,sch21}, which was not known at the time of \citet{mey97}. 

One of the most noticeable differences between the GCR source composition and the one of the solar system is the $^{22}{\rm Ne}/^{20}{\rm Ne}$ isotopic ratio, which is $\sim 5$ times higher in GCRs than in the Sun \citep{bin08}. The measured overabundance of $^{22}$Ne in the GCR composition suggests that these particles do not come solely from the average ISM, but that they contain a significant contribution from Wolf-Rayet (W.-R.) star winds enriched in helium-burning products \citep{cas82}. But exactly how W.-R. wind material is incorporated into GCRs has been a matter of debate for decades \citep[e.g.][]{mae83,hig03,pra12,bin05,lin19,gup20}. 

In this paper, we aim at a better understanding of the origin of GCRs through a detailed study of their chemical composition. We use recent CR data obtained by the Voyager 1 spacecraft \citep{cum16}, the Alpha Magnetic Spectrometer (AMS-02) on the International Space Station \citep[][and references therein]{agu20,agu21} and the SuperTIGER balloon-borne instrument \citep{mur16}. The paper is organised as follows:
the GCR abundance data are presented and compared to the average composition of the ISM in the solar neighbourhood in Section~\ref{sec:data}. In Section~\ref{sec:modelg}, we develop a model to analyse the GCR abundance data using basic assumptions on the origin of these particles from the work of \citet{mey97}. This allows us to study (i) the fraction of each element that is incorporated in ISM dust, (ii) the ionisation state of each element entering a SNR shock as a function of the phase of the ISM, (iii) the distribution of SNe in the ISM phases, as well as (iv) the composition of winds from massive stars and of SN ejecta. The results of the data analysis are presented in Section~\ref{sec:results}. In Section~\ref{sec:discussion} we discuss the GCR sources in the light of these results and estimate the efficiency of GCR acceleration at their sources from the $\gamma$-ray luminosity of the Milky Way. Conclusions are finally given in Section~\ref{sec:conclusions}. 

\section{GCR abundance data}
\label{sec:data}

The GCR source abundances reported in the literature are usually normalisation factors of the injection spectra providing best fits to measured, propagated spectra \citep[e.g.][]{eng90,duv96,cum16,isr18,bos20}. This is justified as long as the CR data for different elements can be described with the same source spectrum. But data from Voyager~1 and AMS-02 have shown that CR protons and $\alpha$-particles have significantly different source spectra than the heavier elements \citep{tat18,evo19}. On the other hand, data on all CRs heavier than helium can be well described with the same injection spectrum, except maybe for iron (\citealp{sch21}; Tatischeff et al., in preparation).   

To estimate the source spectra and relative abundances of protons and $\alpha$-particles from Voyager~1 and AMS-02 data, we used the CR propagation model of \citet{evo19}, which provides a good fit to local interstellar spectra (LIS) measured by AMS-02 above $\sim 10$~GV \citep[see also ][]{evo20,sch21}. It is a 1D advection-diffusion model \citep{jon01}, where CRs are confined in a low-density infinite slab of half-thickness $H$ representing the Galactic halo, the sources of CRs and the target gas being restricted to an infinitely thin disk of half-thickness $h \ll H$. The particle diffusion is taken to be spatially homogeneous, with a diffusion coefficient of the form:
\begin{equation}
D(R) = \beta^\eta D_0 \frac{(R/{\rm GV})^\delta}{[1+(R/R_b)^{\Delta \delta /s_d}]^{s_d}}~. 
\label{eq:diff}
\end{equation}
Here, $\beta=v/c$ is the CR velocity in units of the speed of light, $D_0$ is the value of the diffusion coefficient at the rigidity $R=1$~GV, and the parameters $R_b$, $\Delta \delta$, and $s_d$ describe a break in the diffusion coefficient required to explain the observed break at $\sim 350$~GV in AMS-02 data \citep[e.g.][]{agu15}. The index $\eta \neq 1$ is not considered in \citet{evo19}, but is introduced here to account for a predicted upturn of the diffusion coefficient at low rigidity due the dissipation of magneto-hydrodynamic waves interacting resonantly with CR particles \citep{ptu06}. \citet{gen19} found that $\eta \approx -0.5$ improves significantly the calculated B/C ratio and proton LIS. The other parameters of the CR propagation model were all taken from \citet{evo19}; they are summarised in Table~\ref{tab:propagation}. 

\begin{table}
\caption{CR propagation model parameters. \label{tab:propagation}}
\begin{center}
\begin{tabular}{|ll|}\hline
    Quantity    & Parameter \\ 
\hline
Diffusion coefficient (Eq.~\ref{eq:diff}) & $D_0=1.1\times10^{28}$~cm$^2$~s$^{-1}$$^a$ \\
                           & $\delta=0.63$$^a$           \\
                           & $\eta=-0.5$$^b$           \\
                           & $\Delta \delta=0.3$$^c$           \\
                           & $s_d=0.1$$^a$           \\
                           & $R_b=312$~GV$^a$           \\
\hline
    CR advection velocity     & $u=7$~km~s$^{-1}$$^a$         \\
\hline
    Surface density of the Galactic disk & $\mu=2.3$~mg~cm$^{-2}$$^a$  \\
\hline
    Size of the halo (half-thickness) & $H=4$~kpc$^a$  \\
\hline
\multicolumn{2}{l}{$^a$ Parameter value from \citet{evo19}.} \\
\multicolumn{2}{l}{$^b$ From \citet{gen19}.} \\
\multicolumn{2}{l}{$^c$ \citet{evo19} give $\Delta \delta=0.2$, but we find better results with $0.3$.} \\
\end{tabular}
\end{center}
\end{table}

We took into account the catastrophic losses of CRs in the Galactic disk with the inelastic spallation cross sections of \citet{mos02} and \citet[][and references therein]{tri99}. The production of secondary $^2$H and $^4$He nuclei was calculated from the cross sections given in \citet{cos12}. For the production of heavier secondary CRs, we used cross sections reported in \citet{tat18}, the \textsc{GalProp} cross section measurement database \citep[][and references therein]{mos13}, as well as the TALYS \citep{kon05} and INCL \citep{bou13} nuclear reaction codes. More details on our cross section database will be given in a forthcoming publication (Tatischeff et al., in preparation). 

\begin{figure}
\begin{center}
 \includegraphics[width=0.9\columnwidth]{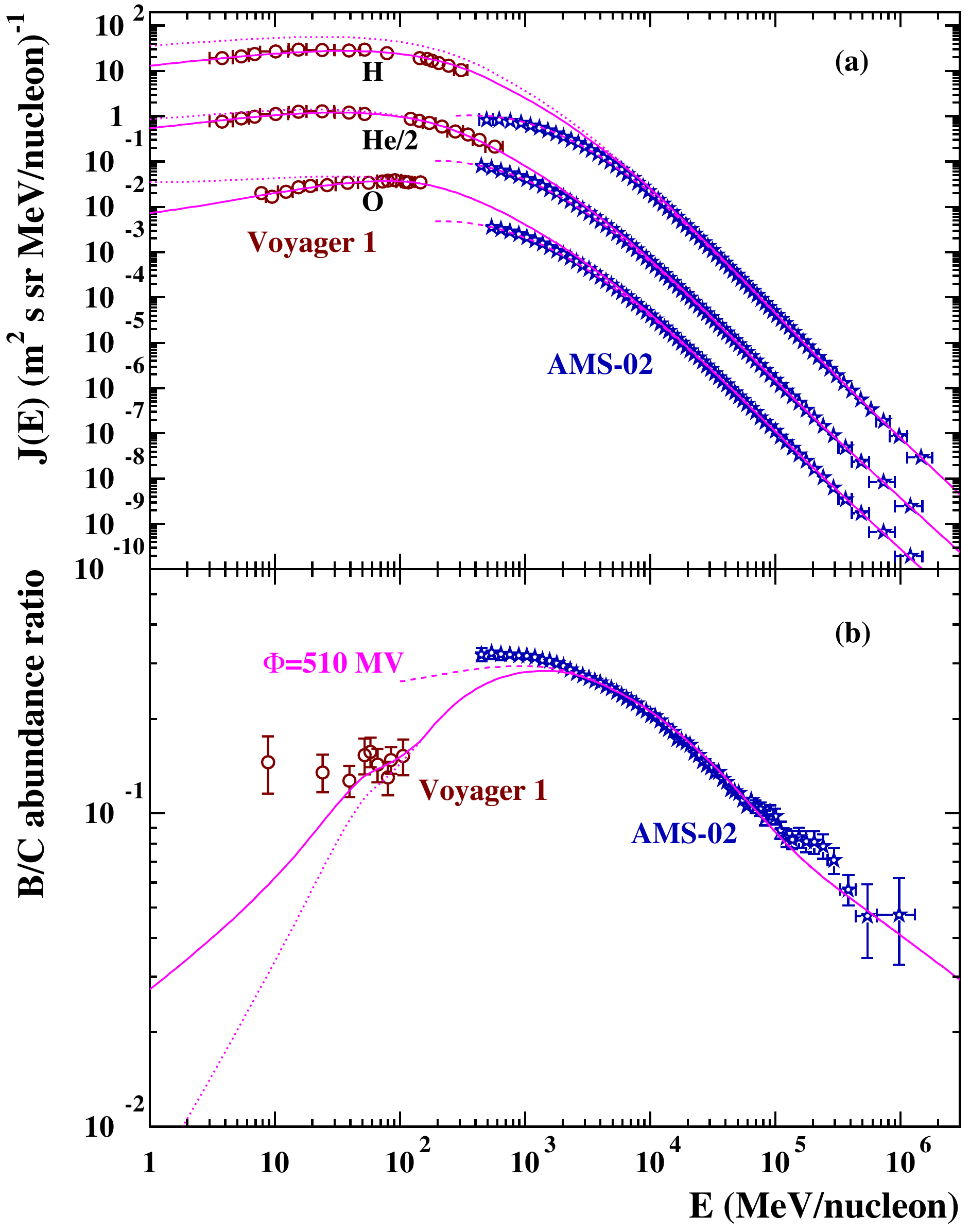}
 \caption{\textit{(a)}: Differential fluxes of GCR protons, $\alpha$-particles and O nuclei measured by Voyager~1 \citep[red circles;][]{cum16} in the local ISM and by AMS-02 \citep[blue stars;][]{agu15,agu17} near Earth. The solid curves show the best-fit LIS assuming a broken power-law source spectrum (Eq.~\ref{eq:sourcespectrum}). The dotted lines show the results for an unbroken power-law source spectrum of slope $\gamma_{\rm h.e.}$ providing a good fit to the AMS-02 data (see Table~\ref{tab:sourcespectrum}). The dashed lines show the solar-modulated spectra with a force-field potential $\Phi=510$~MV corresponding to the epoch of the AMS-02 observations \citep{evo19}. \textit{(b)}: Calculated B/C ratio compared to the AMS-02 \citep{agu16} and Voyager 1 \citep{cum16} data. Solid, dashed, and dotted lines: as in panel~(a).}
 \label{fig:lis}
\end{center}
\end{figure}

The diffusive shock acceleration theory predicts that the CR phase-space distribution function at injection should be a power law in momentum, $f_i(p) \propto p^{-\gamma}$, with $\gamma \approx 4$ \citep[e.g.][]{bla13}. \citet{evo19} found that $\gamma = 4.26$ provides a good fit to AMS-02 data for all nuclei heavier than He, but that the injection of protons (resp. $\alpha$-particles) requires a softer (resp. harder) slope \citep[see also][]{wei20,sch21}. Figure~\ref{fig:lis}a shows propagated CR spectra for protons, $\alpha$-particles, and O nuclei, together with the AMS-02 and Voyager~1 data. We see that the pure power law source spectra adopted by \citet{evo19} do indeed provide a excellent fit to AMS-02 data at high energies, but they overestimate the fluxes measured by Voyager~1 below 1 GeV~nucleon$^{-1}$. In order to fit Voyager~1 data as well, we introduce a break in the injection spectrum, so that the number density of CRs of type $i$ per unit energy interval, $\dot{Q}_i(E) = 4 \pi p^2 f_i(p) dp/dE$, is taken to be of the form:
\begin{eqnarray}
\dot{Q}_i(E) & \propto & \beta^{-1} p^{\gamma_{\rm l.e.}}~{\rm for~} E \leq E_{\rm break}~, \nonumber \\
& \propto & \beta^{-1} p^{\gamma_{\rm h.e.}}~{\rm for~} E > E_{\rm break}~.
\label{eq:sourcespectrum}
\end{eqnarray}
The parameters $E_{\rm break}$ and $\gamma_{\rm l.e.}$ are then fitted to the Voyager~1 data, and the results are given in Table~\ref{tab:sourcespectrum}. We see that the power law break is significant for all three species (i.e. $\gamma_{\rm l.e.} \neq \gamma_{\rm h.e.}$ at more than $\sim$$3\sigma$) and more pronounced for O nuclei than for protons and $\alpha$-particles. Such a low-energy break in the source spectrum can be expected in various scenarios of CR acceleration \citep[see][]{tat18}, but the reasons for the difference in the source spectra between H, He and O are unclear and should be further investigated.

In Figure~\ref{fig:lis}b, we show the ratio of B to C as a function of kinetic energy, which is an important test for GCR propagation models \citep[e.g.][]{gen19}. We see that the calculated B/C ratio compares reasonably well with the data, although the model shows a lower ratio than the Voyager 1 measurements below $\sim 25$~MeV~nucleon$^{-1}$. However, the observed flattening of the B/C ratio at low energies is not predicted by any CR propagation model, and the current result is at least as good as those of the \textsc{GalProp} models shown in \citet{cum16} and \cite{bos20}. We also see in Figure~\ref{fig:lis}b that the calculated B/C ratio assuming an unbroken power-law source spectrum (dotted line in Fig.~\ref{fig:lis}b) is even steeper at low energies. 

\begin{table}
\caption{CR source spectrum parameters (Eq.~\ref{eq:sourcespectrum}).} \label{tab:sourcespectrum}
\begin{center}
\begin{tabular}{|lccc|}\hline
    Parameter & H & He & O \\ \hline
   $E_{\rm break}$ & $10 \pm 2$~GeV/n & $200_{-120}^{+160}$~MeV/n & $160_{-30}^{+40}$~MeV/n \\
   $\gamma_{\rm l.e.}$ & $4.10 \pm 0.03$ & $3.98_{-0.20}^{+0.08}$ & $3.32_{-0.24}^{+0.18}$ \\
   $\gamma_{\rm h.e.}$$^a$ & $4.31$ & $4.21$ & $4.26$ \\
\hline
    $\chi_{\rm min}^2$$^b$ & 16.0 for 13 d.o.f.$^c$ & 7.3 for 14 d.o.f. & 5.9 for 12 d.o.f. \\
\hline
\multicolumn{4}{l}{$^a$ Parameter fixed from \citet{evo19}.} \\
\multicolumn{4}{l}{$^b$ Minimum $\chi^2$ from a fit of the propagated spectrum to Voyager~1 data.} \\
\multicolumn{4}{l}{$^c$ d.o.f.: degrees of freedom.} \\
\end{tabular}
\end{center}
\end{table}

\begin{figure*}
 \includegraphics[width=0.9\textwidth]{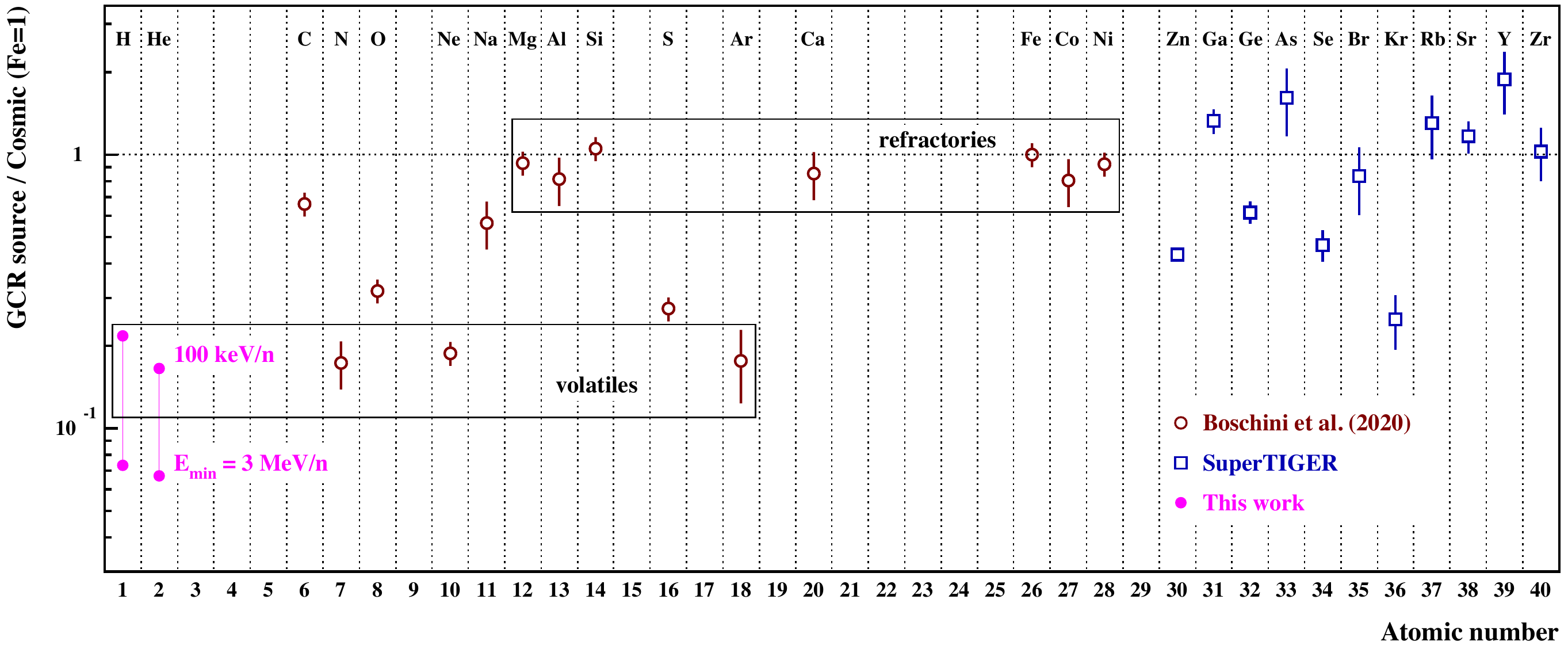}
 \caption{GCR source abundances relative to the SC composition as a function of atomic number (normalised to Fe$\equiv$1). The SC abundances are from \citet[][Table 9]{nie12} for He, C, N, O, Ne, Mg, Si, and Fe (B-type star composition) and \citet[][Table 10]{lod09} for the other elements (solar system composition; see text). The calculated GCR abundances of H and He are shown for two values of the CR minimum energy: $E_{\rm min}=100$~keV~nucleon$^{-1}$ and $3$~MeV~nucleon$^{-1}$. The GCR source abundances of the heavier elements are taken from \citet{bos20,bos21} and \citet[][SuperTIGER data]{mur16}. The two large rectangles group together sets of highly volatile and refractory elements (see text).}
 \label{fig:abund}
\end{figure*}

To estimate the source abundances of protons and $\alpha$-particles relative to that of O nuclei, we finally integrated the injection spectra from a common CR minimum kinetic energy per nucleon, $E_{\rm min}$, assumed to be the same for all species. This assumption is consistent with both the thermal leakage model \citep[e.g.][]{mal95,bla05} and 2D particle-in-cell (PIC) simulations of CR injection and acceleration in collisionless shocks \citep{cap17,han19}, which show that the injection momentum of ions into the diffusive shock acceleration process is proportional to the particle mass. However, the minimum energy of CRs in the ISM should correspond to the lowest energy of nonthermal particles that can effectively escape the CR source --~presumably SN remnants~-- assuming that the diffusive shock acceleration process produces a kappa-like distribution of particles in the source itself. Freshly accelerated, low-energy CRs are thought to escape into the ISM during the radiative phase of SN remnants, when the thermal gas in the shell of the swept-up material gradually recombines. These nonthermal particles are expected to suffer energy losses before being released into the ISM, but how exactly they escape from SN remnants is not well known (and is beyond the scope of the present paper). Here, we adopted a broad range for $E_{\rm min}$ based on phenomenological arguments. As the low-energy CR spectra measured by Voyager~1 \citep[][see Fig.~\ref{fig:lis}a]{cum16} and Voyager~2 \citep{sto19} show no break down to $\sim 3$~MeV~nucleon$^{-1}$, we took this value as an upper limit for $E_{\rm min}$. As a lower limit we assumed that $E_{\rm min} \geq 100$~keV~nucleon$^{-1}$, because lower values of $E_{\rm min}$ would result in too high GCR source abundances of H and He when compared to those of the other highly volatile elements N, Ne and Ar (see Figure~\ref{fig:abund} and Section~\ref{sec:modelg} below on the assumptions of the GCR composition model). 

We took the GCR source abundances of O and other major elements from C to Ni from the recent work of \citet{bos20}, who used the \textsc{GalProp} code to model the propagation of GCRs in the ISM \citep{str98} and the \textsc{HelMod} model to describe the particle transport within the heliosphere \citep{bos19}. The determination of source abundances by these authors is based on the eleven local interstellar spectra (LIS) of CR elements published by the AMS-02 collaboration, from H to O, plus Ne, Mg, and Si \citep[][and references therein]{agu20}, as well as LIS of CRs measured by HEAO-3-C2 \citep{eng90} and Voyager~1 \citep{cum16}. \citet{bos21} have recently provided an updated source spectrum of Fe from the LIS of this element published by AMS-02 \citep{agu21}. 

We have mainly considered primary or mostly primary CRs in our analysis, whose measured LIS and abundance are not very dependent on nuclear spallation reactions that occur during the propagation of CRs in the ISM. Indeed, although the nuclear reaction cross sections used in the \textsc{GalProp} code are accurately selected \cite[e.g.][]{mos13}, the source abundances of several mostly secondary CRs appear to remain uncertain. For example, the Sc/Si abundance ratio reported in \citet{bos20} is about two orders of magnitude higher than in the solar composition (Sc is mostly a secondary CR and Si a primary one), which does not seem to be realistic. However, we use the GCR source abundance of Ar in our analysis, although the LIS of this element includes a significant contribution from spallation of Ca and Fe nuclei. Indeed, noble gas elements are very important to discriminate between the various GCR origin models (as discussed in Section~\ref{sec:discussion} below). By comparing the GCR source abundances reported in \citet{bos20,bos21} to that obtained in previous works \citep{eng90,duv96,cum16,isr18}, we estimated the relative uncertainty on these data to be about 10\% for the mostly primary CRs C, O, Ne, Mg, Si, S, Fe, and Ni, 20\% for the ``half primary" N, Na, Al, Ca and Co, and 30\% for Ar. 

Beyond Ni, we used the GCR source abundances reported by \citet{mur16}, which were obtained from data of the SuperTIGER experiment combined with TIGER abundances \citep{rau09}, weighted by the statistics recorded with each experiment. The reported uncertainties combine the statistical and systematic errors. The abundances given by \citet{mur16} were normalised to the Fe abundance obtained from \citet{bos21}. 

The GCR data used in our analysis are given in Table~\ref{tab:data} and shown in Figure~\ref{fig:abund} relative to the standard cosmic (SC) composition of the ISM in the solar neighbourhood. We used for the SC composition the abundances measured by \citet{nie12} from spectroscopic observations of 29 early B-type stars, which are considered to be ideal indicators for present-day interstellar abundances. Relative to H, the B-type star abundances of C, N, and Si are slightly lower than the protosolar abundances (i.e. those of the solar system 4.56~Gyr ago) assessed by \citet[][see Table~10]{lod09}, and both sets of abundances are consistent for He, O, Ne, Mg and Fe. We used the protosolar abundances of \citet{lod09} for all elements not treated in \citet{nie12}. 


The overabundance of refractory elements over volatiles in the GCR source composition compared to that of the local ISM is striking in Figure~\ref{fig:abund}. As convincingly shown by \citet{mey97} and \citet{ell97}, it can be explained by material locked in dust grains being accelerated to CR energies more efficiently than interstellar gas-phase ions. Another striking feature in Figure~\ref{fig:abund} is that the well-determined GCR abundances of the highly refractory elements Mg, Al, Si, Ca, Fe, Co, and Ni are in the same proportions as those in the SC composition to within $\sim$20\%, which strongly suggests that these elements are accelerated out of the ISM and not from a reservoir with a specific composition different from the SC one. 

\citet{lin98} (see also \citealt{lin19}) suggested that the refractory elements are injected in the GCR population by high-velocity dust grains formed in core-collapse SN ejecta, which are sputtered in the SN reverse shock, when it moves back through the ejecta, and in the forward shock, when fast grains catch up with the slowing blast wave. But the fact that the abundances of the refractory elements in the GCR source composition are in cosmic proportions argues against this scenario. Indeed, if Mg, Al, Si and Ca are mainly produced in core-collapse SNe, $\sim$70\% of Fe, Co and Ni is currently synthesised in thermonuclear SNe \citep{tim95}. The fact that all these elements are found in cosmic proportions in the GCR source composition provides strong evidence that the accelerated particles come from various dust grains of the ISM mix, and not only from core-collapse SN grains. In addition, as already pointed out by \citet{mey99}, the GCR source population contains main s-process elements such as Ba \citep[$Z=56$; see][]{bin89}, which are mainly synthesised by the slow neutron capture in low-mass stars during the asymptotic giant branch phase, and thus are not expected to be present in significant amounts in SN ejecta. 

\begin{table*}
\caption{Summary of all the data used in this analysis: GCR, SC, and massive star wind compositions; element fractions in ISM dust; mass-to-charge ratios in precursors of interstellar shocks and wind termination shocks. \label{tab:data}}
\begin{tabular}{ | l | lccc | l | ccccc | } \hline
Elem. &  & Composition & & & Fraction & & & $(1-X_0)A/Q$$^f$ & &\\
        & GCR Source$^a$ & Cosmic$^b$ & Winds in SB$^c$ & Acc. Winds$^d$ & in ISM dust$^e$ & WNM & WIM & SB/10$^6$~K & SB/10$^7$~K & Winds \\ \hline
 H & (6.58$\pm$0.47)$\times$10$^{6 }$$^g$ & 3.02$\times$10$^{7 }$ & 0.656 & 0.164 & 0\%          &  0.52 &  1.01 &  1.01 &  1.01 &  1.01\\
   & (2.21$\pm$0.16)$\times$10$^{6}$$^h$ & & & & & & & & & \\
He & (4.86$\pm$0.36)$\times$10$^{5 }$$^g$ & 2.93$\times$10$^{6 }$ & 2.03  & 2.89  & 0\%          &  0.92 &  1.60 &  2.00 &  2.00 &  2.00\\
   & (1.97$\pm$0.15)$\times$10$^{5}$$^h$ & & & & & & & & & \\
 C & (4.27$\pm$0.43)$\times$10$^{3 }$	      & 6.46$\times$10$^{3 }$ &  7.03   & 35.7   & (57 $\pm$ 12)\%    &  6.73 &  6.63 &  2.70 &  2.00 &  4.00\\
 N & (3.25$\pm$0.65)$\times$10$^{2 }$	      & 1.87$\times$10$^{3 }$ &  7.46   &  8.97  & (6.5 $\pm$ 6.5)\%  &  3.05 &  7.45 &  2.72 &  2.00 &  4.67\\
 O & (5.52$\pm$0.55)$\times$10$^{3 }$	      & 1.74$\times$10$^{4 }$ &  0.745  &  3.12  & (30 $\pm$ 15)\%    &  8.22 &  8.56 &  2.67 &  2.00 &  5.33\\
Ne & (6.99$\pm$0.70)$\times$10$^{2 }$	      & 3.72$\times$10$^{3 }$ &  1.46   &  2.36  & 0\%  	          &  1.82 & 10.09 &  2.55 &  2.04 &  6.73\\
Na & (3.79$\pm$0.76)$\times$10$^{1 }$	      & 6.74$\times$10$^{1 }$ &  4.44   &  4.66  & (87 $\pm$ 7)\%     & 22.99 & 22.99 &  2.68 &  2.14 &  7.66\\
Mg & (1.02$\pm$0.10)$\times$10$^{3 }$	      & 1.10$\times$10$^{3 }$ &  0.982  &  0.983 & (86 $\pm$ 14)\%    & 23.35 & 22.82 &  2.95 &  2.13 &  8.10\\
Al & (8.03$\pm$1.61)$\times$10$^{1 }$	      & 9.88$\times$10$^{1 }$ &  1.32   &  1.28  & (98.5 $\pm$ 1.5)\% & 23.22 & 22.23 &  3.64 &  2.26 &  8.99\\
Si & (1.01$\pm$0.10)$\times$10$^{3 }$	      & 9.55$\times$10$^{2 }$ &  1.0    &  1.0   & (86 $\pm$ 14)\%    & 28.09 & 20.15 &  3.76 &  2.24 &  9.36\\
 S & (1.35$\pm$0.13)$\times$10$^{2 }$	      & 4.92$\times$10$^{2 }$ &  1.0    &  1.0   & (20 $\pm$ 20)\%    & 20.51 & 19.26 &  4.02 &  2.28 & 10.69\\
Ar & (1.91$\pm$0.57)$\times$10$^{1 }$	      & 1.08$\times$10$^{2 }$ &  1.0    &  1.0   & 0\%		          &  0.55 & 18.06 &  4.37 &  2.28 & 12.10\\
Ca & (6.00$\pm$1.20)$\times$10$^{1 }$	      & 7.04$\times$10$^{1 }$ &  1.0    &  1.0   & (98.5 $\pm$ 1.5)\% & 20.04 & 39.33 &  4.05 &  2.26 & 13.36\\
Fe & (1.00$\pm$0.10)$\times$10$^{3 }$	      & 1.00$\times$10$^{3 }$ &  1.0    &  1.0   & (97 $\pm$ 3)\%     & 55.67 & 20.61 &  6.10 &  2.88 & 18.62\\
Co & (2.20$\pm$0.44)$\times$10$^{0 }$	      & 2.74$\times$10$^{0 }$ &  1.0    &  1.0   & (97 $\pm$ 3)\%     & 58.93 & 21.74 &  6.45 &  2.97 & 19.64\\
Ni & (5.28$\pm$0.53)$\times$10$^{1 }$	      & 5.72$\times$10$^{1 }$ &  1.0    &  1.0   & (97 $\pm$ 3)\%     & 58.20 & 21.70 &  6.25 &  2.93 & 19.56\\
Zn & (6.55$\pm$0.41)$\times$10$^{-1}$	      & 1.52$\times$10$^{0 }$ &  1.0    &  1.0   & (20 $\pm$ 20)\%    & 25.70 & 24.60 &  7.55 &  3.32 & 21.79\\
Ga & (5.67$\pm$0.60)$\times$10$^{-2}$	      & 4.27$\times$10$^{-2}$ &  1.0    &  1.0   & (89 $\pm$ 8)\%     & 29.69 & 28.14 &  8.23 &  3.33 & 23.24\\
Ge & (8.25$\pm$0.78)$\times$10$^{-2}$	      & 1.34$\times$10$^{-1}$ &  1.0    &  1.0   & (66 $\pm$ 14)\%    & 33.73 & 31.59 &  8.70 &  3.34 & 24.20\\
As & (1.15$\pm$0.32)$\times$10$^{-2}$	      & 7.11$\times$10$^{-3}$ &  1.0    &  1.0   & (92 $\pm$ 8)\%     & 35.15 & 33.23 &  9.09 &  3.34 & 24.97\\
Se & (3.68$\pm$0.50)$\times$10$^{-2}$	      & 7.87$\times$10$^{-2}$ &  1.0    &  1.0   & (20 $\pm$ 20)\%    & 37.61 & 35.95 &  9.67 &  3.42 & 26.32\\
Br & (1.04$\pm$0.29)$\times$10$^{-2}$	      & 1.25$\times$10$^{-2}$ &  1.0    &  1.0   & (20 $\pm$ 20)\%    & 40.37 & 38.60 &  9.66 &  3.42 & 26.63\\
Kr & (1.63$\pm$0.37)$\times$10$^{-2}$	      & 6.51$\times$10$^{-2}$ &  1.0    &  1.0   & 0\%		          &  1.14 & 38.80 & 10.37 &  3.51 & 27.93\\
Rb & (1.10$\pm$0.29)$\times$10$^{-2}$	      & 8.43$\times$10$^{-3}$ &  1.0    &  1.0   & (66 $\pm$ 14)\%    & 50.71 & 48.57 &  9.64 &  3.56 & 28.49\\
Sr & (3.17$\pm$0.43)$\times$10$^{-2}$	      & 2.72$\times$10$^{-2}$ &  1.0    &  1.0   & (98.5 $\pm$ 1.5)\% & 44.58 & 44.50 &  8.97 &  3.69 & 29.21\\
 Y & (1.02$\pm$0.27)$\times$10$^{-2}$	      & 5.40$\times$10$^{-3}$ &  1.0    &  1.0   & (98.5 $\pm$ 1.5)\% & 47.88 & 47.05 &  8.63 &  3.76 & 29.64\\
Zr & (1.29$\pm$0.29)$\times$10$^{-2}$	      & 1.26$\times$10$^{-2}$ &  1.0    &  1.0   & (98.5 $\pm$ 1.5)\% & 34.54 & 33.43 &  8.10 &  3.90 & 30.41\\
\hline
\multicolumn{11}{l}{$^a$ GCR source abundances normalized to Fe$\equiv$1000 (Section~\ref{sec:data}).}\\
\multicolumn{11}{l}{$^b$ Present-day standard cosmic (SC) abundances normalized to Fe$\equiv$1000 (Section~\ref{sec:data}).}\\
\multicolumn{11}{l}{$^c$ Abundance enhancement factors of wind material in superbubbles (SBs) relative to the SC composition. $^{22}{\rm Ne}/^{20}{\rm Ne}=0.61$ in the wind material  (Section~\ref{sec:22nesb}).}\\
\multicolumn{11}{l}{$^d$ Abundance enhancement factors of the accelerated-wind composition relative to SC. $^{22}{\rm Ne}/^{20}{\rm Ne}=1.56$ in the accelerated-wind composition (Section~\ref{sec:22newind}).}\\
\multicolumn{11}{l}{$^e$ See Section~\ref{sec:depletion}.}\\
\multicolumn{11}{l}{$^f$ Mass-to-charge ratios of ions in precursors of ISM shocks and wind termination shocks ($X_0$ is the fraction of neutrals, $Q$ the mean ionic charge without the neutrals),}\\
\multicolumn{11}{l}{~~~~see Section~\ref{sec:ioni}.}\\
\multicolumn{11}{l}{$^g$ For the minimum GCR source energy $E_{\rm min}=100$~keV~nucleon$^{-1}$ (see Section~\ref{sec:data}).}\\
\multicolumn{11}{l}{$^h$ For $E_{\rm min}=3$~MeV~nucleon$^{-1}$.}\\
\end{tabular}
\end{table*}

\section{Model assumptions}
\label{sec:modelg}
\subsection{Galactic cosmic ray composition}
\label{sec:model}

Based on the  seminal work of \citet{mey97}, here we attempt to explain the source composition of GCRs from H to Zr with three basic assumptions. First, the overabundance of refractory elements in the GCR composition results from efficient acceleration of ISM dust grains in strong shocks, as studied in detail by \cite{ell97}. Assuming that the measured GCR abundance of any element $i$ comes from two sources, one from a gas reservoir and another associated with ISM dust, i.e. $C_{\rm mes}(i)=C_{\rm gas}(i)+C_{\rm dust}(i)$, the dust contribution can be written as:
\begin{equation}
C_{\rm dust}(i)={\rm SC}(i) f_d(i) \epsilon_{\rm dust}~,
\label{eq:cdust}
\end{equation}
where ${\rm SC}(i)$ is the present-day standard cosmic (SC) abundance of element $i$ (see Table~\ref{tab:data}), $f_d(i)$ is the fraction of its abundance arising from ISM dust (Section~\ref{sec:depletion}), and $\epsilon_{\rm dust}$ the efficiency of injection of refractory elements contained in ISM grains into the GCR population. The latter quantity depends on the evolution of shock waves in the ISM, the ambient medium density, the grain size distribution, but not of the mass and atomic number of the ions sputtered from the grains (i.e. it is independent of $i$; see \cite{ell97}). 

Second, we assume that the reservoir of volatile elements from which GCRs are produced is enriched in W.-R. star wind material in which He-burning products are expelled during the WC and WO phases \citep{cas82,mey97,pra12}. This is required to explain the high $^{22}{\rm Ne}/^{20}{\rm Ne}$ isotopic ratio measured in GCRs, $^{22}{\rm Ne}/^{20}{\rm Ne} = 0.387 \pm 0.007$ (statistical) $\pm 0.022$ (systematic) \citep{bin08}, which is $\sim 5$ times the solar value \citep[see also][Table~3]{bos20}. Assuming that the abundance of volatile elements in the GCR results from two distinct sources, one with a SC composition and the other from W.-R. star winds ($C_{\rm gas}(i)=C_{\rm gas}^{\rm SC}(i)+C_{\rm gas}^{w}(i)$) we can write
\begin{equation}
C_{\rm gas}^{w}(i) \propto f_w(i)  {\rm SC}(i) (1-f_d(i))~,
\end{equation}
with $f_w(i)$ the enhancement of element $i$ in the W.-R. wind reservoir compared to the SC composition. In Section~\ref{sec:22ne}, we study two different scenarios for the overabundance of GCR $^{22}$Ne. The first one assumes that SN shocks in superbubbles (SBs) propagate in a medium enriched by W.-R. winds from the most massive stars of the parent OB associations \citep{hig03,bin05,bin08,lin19}. The second scenario considers that stellar winds in massive star clusters are efficiently accelerated by wind termination shocks \citep[WTSs;][]{gup20}. 

The third and last assumption of our GCR composition model is that the injection efficiency of ions into the diffusive shock acceleration process depends linearly on the magnetic rigidity of the particles and thus on the ion mass-to-charge ratio $A/Q$. This assumption is primarily motivated by the nonlinear shock acceleration theory, which predicts a smoothing of the shock profile caused by the backpressure of CRs on the inflowing plasma \citep[e.g.][]{ell84,ell97}. As a result, ions with a high $A/Q$ ratio feel a larger velocity difference of the background plasma than particles of lower rigidity, as they can diffuse farther back upstream, and consequently are more easily injected into the acceleration process. Such a preferential injection of high $A/Q$ ions was first predicted by \citet{eic79} and later confirmed by \citet{ell81} from Monte Carlo simulations (see also \citealp{ell97}). More recent 2D PIC simulations of CR acceleration in collisionless shocks \citep{cap17,han19} also found that for strong shocks the fraction of ions that enter the acceleration process above the injection energy $E_{\rm inj}$ grows linearly with $(A/Q)$. But the 2D PIC simulation result is not due to nonlinear shock smoothing caused by efficient CR acceleration, as this effect is not studied in these simulations. The obtained enhancement in CR ions with $A/Q \gg 1$ is rather explained by the rate of isotropisation of these particles in the downstream medium, which depends on $A/Q$ and on the self-generated magnetic turbulence \citep{cap17}. But contrary to \citet{cap17}, \citet{han19} found that the injection efficiency starts to saturate at $(A/Q) \sim 8$--$12$ depending on the shock Mach number. This effect is not well understood, and it seems plausible that it is associated with the unavoidable limitations of the simulation box size and run time \citep[see also the discussion in ][]{eic21}. In any case, this saturation effect appears to be contradictory to the preferential acceleration of dust grains, which have very large $A/Q$ ratio, and we did not take it into account in our model.

Combining the three basic assumptions of the model, the GCR abundance arising from the gas reservoir can be written as:
\begin{equation}
C_{\rm gas}(i) = {\rm SC}(i) (1-f_d(i)) \epsilon_{\rm gas} [x_w f_w(i) f_{A/Q}^w(i) + (1-x_w)f_{A/Q}^{\rm SC}(i)]~,
\label{eq:cgas}
\end{equation}
where $\epsilon_{\rm gas}$ is a global efficiency factor for the injection of interstellar ions into the GCR population, $x_w$ the contribution of the W.-R. wind reservoir to the GCR source gas population (in total number of atoms) and $f_{A/Q}^j(i)=(1-X_{0,i}^j)A_i/Q_i^j$, with $X_{0,i}^j$, $A_i$ and $Q_i^j$ the fraction of neutral atoms, the atomic mass and the mean ionic charge (without the neutral fraction; see Section~\ref{sec:ioni}) for ions $i$ in the region immediately upstream of shock waves propagating in the medium of composition $j$ (SC or W.-R. winds). The gas reservoir of SC composition may include several phases of the ISM, such as a mixture of hot gas, warm neutral medium (WNM) and warm ionised medium (WIM). In this case, the mass-to-charge ratio factor in the above equation is obtained from
\begin{equation}
f_{A/Q}^{\rm SC}(i) = \sum_k a_k f_{A/Q}^{{\rm SC},k}(i)~,
\label{eq:aonq}
\end{equation}
where $a_k$ is the relative contribution of the ISM phase $k$ to the GCR volatile production and $f_{A/Q}^{{\rm SC},k}(i)$ the mass-to-charge ratio of element $i$ in the photoionisation precursor of shocks propagating in the $k$-phase. 

Equations~(\ref{eq:cdust}) and (\ref{eq:cgas}) allow us to readily estimate the relative contributions of the ISM dust and gas reservoirs in the measured GCR abundances: 

\begin{align}
& C_{\rm dust}(i) = 
\frac{C_{\rm mes}(i)}{
\left(1 + \frac{1-f_d(i)}{f_d(i)} \cdot \frac{x_w f_w(i) f_{A/Q}^w(i) + (1-x_w)f_{A/Q}^{\rm SC}(i)}{\epsilon} \right)
}~, \label{eq:cdustprim}
\\
& C_{\rm gas}(i)  =
\frac{C_{\rm mes}(i)}{
\left(1 + \frac{f_d(i)}{1-f_d(i)} \cdot \frac{\epsilon}{x_w f_w(i) f_{A/Q}^w(i) + (1-x_w)f_{A/Q}^{\rm SC}(i)} \right)
} ~, \label{eq:cgasprim}
\end{align}
where $\epsilon=\epsilon_{\rm dust}/\epsilon_{\rm gas}$. Similarly, the GCR abundance arising from the gas reservoir of SC composition is obtained from:
\begin{equation}
C_{\rm gas}^{\rm SC}(i) =
\frac{C_{\rm mes}(i)}{
\left(1+\frac{f_d(i)}{1-f_d(i)}\cdot \frac{\epsilon}{(1-x_w)f_{A/Q}^{\rm SC}(i)} + 
\frac{x_w f_w(i) f_{A/Q}^w(i)}{(1-x_w)f_{A/Q}^{\rm SC}(i)} \right)
} ~.
\label{eq:cgassc}
\end{equation}

We present our best estimates of the quantities $f_d$, $x_w$, $f_w$, $f_{A/Q}^{\rm SC}$ and $f_{A/Q}^w$ in Sects.~\ref{sec:depletion} to \ref{sec:ioni}, and then derive $\epsilon$, as well as constraints on the GCR source reservoirs, from a fit of the model to the GCR abundance data (Section~\ref{sec:results}). 

\subsection{Gas-phase element depletions and dust in the interstellar medium}
\label{sec:depletion}

Interstellar dust accounts for $\lsim 1$\% of the total mass of the ISM \citep{dra03,jon17}. Dust grains are condensed in the winds of evolved stars and the ejecta of stellar explosions (novae and supernovae), and also form in dense molecular clouds \citep{zhu08}. The dust population is thought to be composed of a 
heterogeneous mix of silicate and carbonaceous grains, whose composition and structure significantly evolve between diffuse ISM and dense molecular clouds \citep{jon17}. Dust grains are thought to be mainly destroyed by sputtering in slow ($V_s < 200$~km~s$^{-1}$), radiative SN shocks propagating in the warm ISM \citep{boc14}. 

Measurements of gas-phase element depletions are pivotal in determining the mass and composition of dust in the ISM \citep[e.g.][]{jon00}. The fraction of an element that is incorporated into dust can be estimated from measurements of its gas-phase abundance through observations of appropriate absorption lines, and the assessment of its absolute abundance in the ISM based on B-type star observations \citep{nie12} and the solar system composition \citep{lod09}. \citet{car06} studied the gas-phase depletions of Mg, P, Mn, Ni, Co, and Ge as a function of the mean H density along the lines of sight, and identified two abundance plateaus representative of the depletion levels in the warm and cold neutral ISM. In a comprehensive study of abundances reported in more than a hundred papers for 17 elements, \citet{jen09} showed that all (logarithmic) depletions can be represented as varying linearly with a general depletion factor $F_*$ ranging between 0 and 1, where $F_*=1$ is typical of the cold neutral medium (CNM) and $F_*=0$ can be associated with the WIM \citep[see also][]{eic21}. 

The amount and nature of dust contained in the hot ionised medium of the ISM is very uncertain. The lifetime of ISM grains against thermal sputtering in a hot and dilute plasma ($T \sim 10^6$~K, $n \sim 0.01$~cm$^{-3}$) ranges from $\lsim 1$~Myr for small ($\sim$10~\angstrom) carbonaceous grains to $\gsim$20~Myr for large ($\sim$0.1~$\mu$m) silicate grains \citep{tie94}. Moreover, \citet{mck89} showed that the first SN from the most massive star in an OB association exploding in a SB environment should destroy a large amount of dust initially contained in the interior of the growing SB, such that following SNe from the same OB association should process a medium largely cleansed of its dust component. But \citet{och15} argued from multi-wavelength observations of the Orion-Eridanus SB that appreciable amounts of mass are continuously removed from molecular clouds and loaded into the SB interior through thermal evaporation at the edge of the clouds and the so-called ``champagne effect'', that is the leakage of ionised cloud material from an HII region when the ionisation front breaks through the edge of the surrounding cloud \citep{ten79}. Thus each SNR within the SB cavity should in fact interact with a significant amount of newly incorporated matter rich in dust grains. 

In Section~\ref{sec:data} we show that the highly refractory elements Mg, Al, Si, Ca, Fe, Co, and Ni should be in cosmic proportions in the GCR source reservoir. As these elements are expected to be in the same proportions in  all phases of the ISM, we estimated the mean fraction of each element contained in ISM dust considering the same elemental composition for dust in the various phases of the ISM. 
The mean elemental fractions in ISM dust are obtained from the large data set of gas-phase abundances reported in \citet{jen09,jen19} and \citet{rit18}, with the help of the interstellar dust modelling framework THEMIS \citep{jon17} and some general properties of primitive interplanetary dust. We now discuss our results for each element.

\textbf{Hydrogen}. Some H is expected to be present in the dust component of the ISM. In the most primitive interplanetary samples, such as carbonaceous chondrites or interplanetary dust particles that escaped planetary differentiation, H is manly present in their pristine organic matter. The insoluble organic matter, that represents most of this carbonaceous phase, exhibits an average H/C ratio of 0.7--0.8 \citep{ale17}. 
Assuming that the organic component of hydrogenated carbonaceous grains in the ISM have a similar C/H ratio, and considering that, according to the THEMIS dust model, they contain 43--58 ppm of C (relative to the total number of H atoms in the ISM), the H fraction in dust is $f_d=(3$--$5)\times 10^{-5}$. It is thus negligible and we set $f_d=0$ in our calculations. 

\textbf{Helium and Neon}. These noble gas elements are highly volatile species and, although a minor fraction may be trapped in their organic phase or implanted in minerals, the vast majority of these elements is expected to reside in the gas phase of the ISM and we thus consider $f_d=0$. 

\textbf{Carbon}. The weighted mean of the C abundances reported in \citet{jen09} gives ${\rm (C/H)}_{\rm gas}=215$~ppm, which is in agreement with
 the total abundance measured in B-type stars \citep{nie12}: ${\rm (C/H)}_{\rm tot}=(214 \pm 20)$~ppm. However, observations of dust extinction and emission in the diffuse ISM show that the amount of C in solid form is not negligible \citep[e.g.][and references therein]{jon17}. The solution to this problem is provided by \citet{sof11}, who found that C abundances measured from the weak line of ${\rm [C_{II}]}~\lambda$2325\angstrom, as in \citet{jen09}, tend to be overestimated by a factor of $\sim$2.3 compared to those determined from strong line transitions. With the ratio ${\rm (C/H)}_{\rm gas}=(91 \pm 11)$~ppm given by \citet{sof11}, we get ${\rm (C/H)}_{\rm dust}=(123 \pm 23)$~ppm and $f_d=(57 \pm 12)$\%  \citep[see][Table~9]{nie12}. 
 Noteworthy, the THEMIS dust model considers a substantially higher total C abundance, ${\rm (C/H)}_{\rm dust}=206$--$218$~ppm \citep{jon17}, probably because it is based on measured depletions with respect to the C abundance in the early solar system composition \citep[${\rm (C/H)}_{\rm tot}=278$~ppm in][]{lod09}, which is significantly higher than the C abundance in B-type stars assumed to be representative of the present-day ISM.

\textbf{Nitrogen}.
In their work on B-type stars, \citet{nie12} consider a total abundance of N with respect to H of ${\rm (N/H)}_{\rm tot}=(62 \pm 6)$~ppm and an upper limit on the fraction of N locked in dust of ${\rm (N/H)}_{\rm dust}<7$~ppm, which gives $f_d<13$\%. 
The average N/C ratio in organic matter from primitive solar system material is ${\rm N/C} \approx 3.5$\% \citep{dar17}, which, together with the C abundance in ISM dust of ${\rm (C/H)}_{\rm dust}=(123 \pm 23)$~ppm (see above), gives ${\rm (N/H)}_{\rm dust} \approx (4.3 \pm 0.8)$~ppm. Comparing this result with the total N abundance measured in B-type stars, we find $f_d \approx (6.9 \pm 1.5)$\%. However, this has to be considered as an upper limit, because not all C is in organic form in ISM dust. In this first approach, we adopt a conservative value for the N fraction in dust of $f_d = (6.5 \pm 6.5)$\%.


\textbf{Oxygen}.
\citet{nie12} give for the total O abundance ${\rm (O/H)}_{\rm tot}=(575 \pm 66)$~ppm and for the abundance in dust ${\rm (O/H)}_{\rm dust}=(186 \pm 67)$~ppm, such that $f_d = (32 \pm 12)$\%. However, considering that most of O in dust is locked in pyroxene-type silicate grains, the THEMIS model estimates the O abundance in dust to be ${\rm (O/H)}_{\rm dust}=110$~ppm \citep{jon17}, which is slightly lower than the value reported in \citet{nie12}. We have adopted a value consistent with both the study of Nieva \& Przybilla and the THEMIS model: $f_d = (30 \pm 15)$\%. 

\textbf{Sodium}.
The gas-phase depletion of Na is not considered in the studies of \citet{jen09} and subsequent publications. But \citet{sav96} previously reported a depletion measurement towards the star $\zeta$ Ophiuchi of ${\rm [Na/H]}=-0.95 \pm 0.10$, which corresponds to a fraction in dust of $f_d = (89 \pm 3)$\%. The equilibrium condensation temperature of Na, $T_{\rm cond}=958$~K, is close to that of B and Ga, $T_{\rm cond}=908$ and $968$~K, respectively \citep{lod09}, and the latter two elements are included in the depletion analysis of \citet{rit18}. From the weighted mean of the B and Ga gas-phase abundances reported in this work, we obtain $f_d({\rm B}) = (85 \pm 5)$\% and $f_d({\rm Ga}) = (89 \pm 5)$\%. We finally adopt for Na: $f_d = (87 \pm 7)$\%. 

\textbf{Magnesium and Silicon}.
\citet{nie12} give for the total Mg abundance ${\rm (Mg/H)}_{\rm tot}=(36.3 \pm 4.2)$~ppm and for the abundance in dust ${\rm (Mg/H)}_{\rm dust}=(34.8 \pm 4.2)$~ppm, such that $f_d = (96_{-16}^{+4})$\%. This is consistent with the THEMIS model, which considers that interstellar Mg is entirely locked in dust grains. However, the Mg depletions measured by \citet[][and references therein]{jen19} show that in the warm and hot ISM, a significant fraction of Mg is contained in the gas phase, and on average from all the measurements reported in this paper: $f_d = (81 \pm 8)$\%. We have adopted a conservative value consistent with both the depletion measurements and the THEMIS model: $f_d = (86 \pm 14)$\%. The data for Si are very similar to those for Mg and furthermore, both elements have the same condensation temperature, $T_{\rm cond}=1354$~K \citep{lod09}. We thus use $f_d = (86 \pm 14)$\% also for Si. 

\textbf{Aluminium, Calcium, Strontium, Yttrium, Zirconium}. Al, Ca, Sr, Y, and Zr are highly refractory elements with a condensation temperature above $1450$~K. A similar element is included in the study of \citet{jen09}: Ti, with $T_{\rm cond}=1582$~K. The weighted mean of the gas-phase abundances gives $f_d = (98.5 \pm 1.5)$\%, which we have adopted for all the highly refractory elements. 

\textbf{Sulphur, Zinc, Selenium, Bromine}.
S is a volatile element, which can be expected to be mainly contained in ISM gas. However, it can also be chemically bound to refractory elements in the form of sulfides. \cite{jon17} assume that ${\rm (S/H)}_{\rm dust}=3$~ppm in the form of FeS in silicate grains. Assuming that ${\rm (S/H)}_{\rm tot}=16.3$~ppm \citep[][the S abundance is not measured in B-type stars]{lod09}, we get $f_d = 18$\%. An additional amount of S could be contained in the organic matter of the ISM. The insoluble organic matter found in primitive carbonaceous chondrites presents a S/C ratio of 1--4\% \citep{ale17}. Assuming that the organic component of hydrogenated carbonaceous grains in the ISM have a similar S/C ratio, and considering that these grains contain 43--58~ppm of C \citep{jon17}, we obtain between 3\% and 14\% more S in dust. S is discussed in \citep{jen09} as a ``troublesome element'', which presents a significant level of depletion for $F_*=1$, i.e. in the CNM, but zero depletion for $F_*=0$. In light of these different results, we finally adopt for S: $f_d = (20 \pm 20)$\%. Zn, Se, and Br having condensation temperatures close to that of S, in the range $546$--$726$~K ($T_{\rm cond}{\rm (S)}=664$~K; \citet{lod09}), we also use $f_d = (20 \pm 20)$\% for these three elements. 

\textbf{Argon and Krypton}.
The abundances of 
Kr in the cool gas-phase reported in \citet{rit18} and \citet{jen19} exhibit some depletions with respect to solar values, although the origin of these subsolar abundances remains unclear. 
Depletions of Ar were also reported, but \cite{sof98} showed that the measurements of the Ar abundance from its neutral form are not always reliable. 
Ar and Kr are 
noble gases with low, to very low, total abundances with respect to H, ${\rm (Ar/H)}_{\rm tot}=3.6$~ppm and ${\rm (Kr/H)}_{\rm tot}=2.5\times 10^{-3}$~ppm and the depletions of Ar and Kr 
may thus need confirmation. Indeed, only minute amounts of these 
highly volatile elements are expected to be trapped within dust grains, either by implantation in silicates or trapping within the carbonaceous phase \citep[see, e.g.][]{hus96}. 
Their concentration in the carbonaceous or silicate phase of primitive solar system material (i.e. CI chondrites) is very low with ${\rm Ar/Si}<10^{-8}$  and ${\rm Kr/Si}<2\times 10^{-10}$ \citep{lod09}. As a result, the main reservoir of both Ar and Kr is expected to stay in the gas phase, as observed in the solar system where the abundances of Ar and Kr relative to H in the Sun are respectively $9\times10^{6}$ and $3\times10^{5}$ that in early solar system solids \citep[i.e. meteorites; see][]{lod09}.
We thus consider $f_d = 0$ for these two elements. 

\textbf{Iron, Cobalt and Nickel}.
Fe, Co, and Ni are highly refractory elements with similar condensation temperatures, $T_{\rm cond}=1334$, $1352$, and $1353$~K, respectively. The gas-phase depletions of Fe and Ni were studied by \citet{jen09}. From all the measurements reported in this paper, we obtain $f_d({\rm Fe}) = (96.5 \pm 3.5)$\% and $f_d({\rm Ni}) = (98 \pm 2)$\%. We adopt $f_d = (97 \pm 3)$\% for the three elements. 

\textbf{Gallium}.
The gas-phase depletion of Ga was measured by \citet{rit18} along 69 lines of sight. From the weighted mean of the measured abundances we find $f_d = (89 \pm 8)$\%. 

\textbf{Germanium}.
The gas-phase depletion of Ge was recently measured by both \citet{rit18} and \citet{jen19}, and we obtained from these measurements $f_d = (67 \pm 21)$\% and $(65 \pm 20)$\%, respectively. We take the weighted mean value of these results in our calculations: $f_d = (66 \pm 14)$\%.

\textbf{Arsenic}. 
The gas-phase depletion of As was measured by \citet{rit18}, but only 10 sight lines with secure detection of the ${\rm [As_{II}]}~\lambda$1263\angstrom~ absorption line could be identified. As the measured depletion of this element was not sufficiently binding, we used instead Cu as a proxy, given the similarity of their condensation temperature: $T_{\rm cond}=1065$~K for As and $1037$~K for Cu. The gas-phase depletion of Cu is measured in \citet{jen09} and gives $f_d=(92 \pm 8)$\%, which we have also adopted for As. 

\textbf{Rubidium}.
We use for Rb the fraction in dust estimated for Ge, $f_d = (66 \pm 14)$\%, because both elements have similar $T_{\rm cond}$: $800$~K for Rb and $883$~K for Ge. 

In their analysis of SuperTIGER data, \citet{mur16} distinguish the volatile elements from the refractory ones based on their condensation temperatures only: all elements with $T_{\rm cond}<1200$~K are assumed to be volatiles (and thus be found in interstellar gas). Then, \citet{mur16} concluded that several trans-iron ``volatile'' elements are overabundant in the GCR source composition compared to the solar system composition and suggested that the GCR source reservoir is a mixture of $\sim$80\% material with solar-system abundances and $\sim$20\% material from massive star winds and SN ejecta enriched in trans-iron elements \citep[see also][]{lin19}. However, the gas-phase depletion measurements of \citet{jen09,jen19} and \citet{rit18} show that the volatile-versus-refractory selection adopted by \citet{mur16} is not a good approximation. For example, Ga and Ge have $T_{\rm cond}$ of $800$~K and $883$~K, respectively, but the observations of \citet{rit18} and \citet{jen19} show that these two elements are mainly locked in dust grains ($f_d = (89 \pm 8)$\% and $(66 \pm 14)$\%). In the present work, the relatively high abundances of Ga, Ge and other trans-ion elements in the GCR composition is explained by their high fractions in ISM dust and the preferential acceleration of dust grains, and not by an enrichment of the GCR source reservoir in massive star winds and SN ejecta. 

\subsection{Origin of the {22}-Ne-rich component in the Galactic cosmic ray composition}
\label{sec:22ne}

One of the most conspicuous features of the GCR source composition is the high $^{22}{\rm Ne}/^{20}{\rm Ne}$ ratio, which was measured more than 40 years ago \citep[e.g.][]{gar79}, but is still not well understood. All other measured isotopic abundance ratios of the GCR source composition are consistent with the solar composition, except perhaps the $^{58}{\rm Fe}/^{56}{\rm Fe}$ ratio, which is estimated to be $1.69 \pm 0.27$ times the solar value \citep[see][]{bin08}. In the model of \citet{mey97}, the $^{22}$Ne excess is explained by the shock acceleration of W.-R. wind material when the SN blast wave expands in the winds lost the progenitor massive star prior to explosion. This scenario was studied quantitatively by \citet{pra12}, who found that the observed GCR $^{22}{\rm Ne}/^{20}{\rm Ne}$ ratio can be explained if the CRs are accelerated only when the SN blast waves run through the pre-SN winds, and hardly after that, when the shocks propagate in the ISM. However, this scenario seems questionable for several reasons, which are discussed in \citet{tat18}, including the fact that most, if not all W.-R. stars may not end their life in SN explosions but rather collapse to form black holes \citep{sma15}. 

Here we study two other scenarios proposed in the literature for the origin of the $^{22}$Ne excess in GCRs. The first one assumes that GCRs are accelerated from SB material enriched by massive star outflow \citep{hig03,bin05,bin08,lin19}. The second one considers the acceleration of massive star winds in their termination shocks \citep{gup20}. 

\subsubsection{Enrichment of superbubble gas by massive star winds}
\label{sec:22nesb}

\begin{figure*}
 \includegraphics[width=0.9\textwidth]{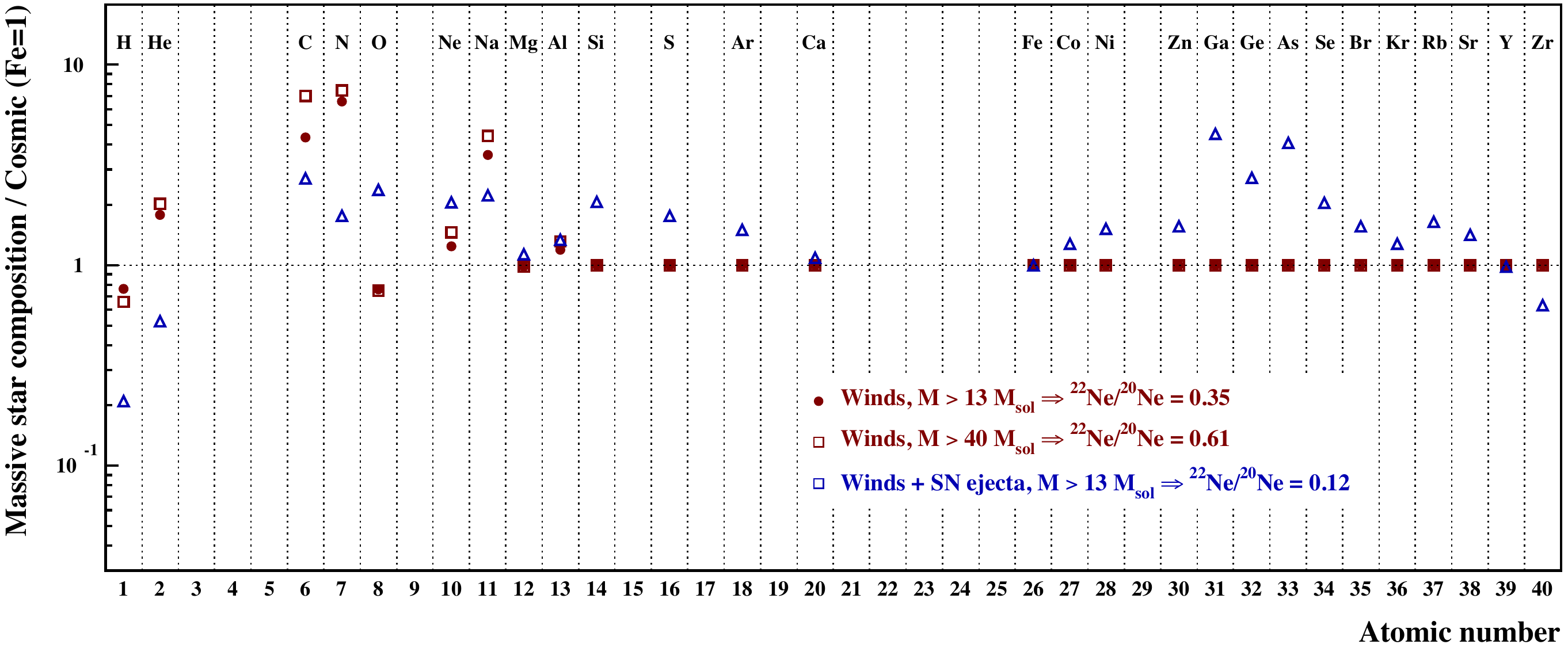}
 \caption{Abundances of elements in massive star winds and SN ejecta relative to the SC composition (normalised to Fe$\equiv$1), based on the stellar yield calculations of \citet{lim18}. The red symbols show the chemical enrichment of the winds obtained by folding the stellar yields over the initial mass function of the stars, assuming a minimum star mass of $13~M_\odot$ (filled circles) and $40~M_\odot$ (empty squares; see text). The blue triangles show the total enrichment provided by massive star winds and SN ejecta. The corresponding $^{22}$Ne/$^{20}$Ne abundance ratio is reported in the Figure.}
 \label{fig:windabund}
\end{figure*}

Most massive stars form in OB associations where the strong stellar wind activity generates large-scale cavities ($\gg 10$~pc) filled with hot ($\gsim 10^6$~K) and diluted ($\lsim 10^{-2}$~cm$^{-3}$) plasma \citep[e.g.][]{mac88}. The interior of these SBs consists of a mixture of massive star wind material, SN ejecta, and material of SC composition that either survived inside the cavity in the form of dense molecular clouds or was evaporated off the SB shell \citep[see][and references therein]{par04}. \citet{hig03} evaluated quantitatively the time evolution of the $^{22}$Ne/$^{20}$Ne ratio within a SB and suggested that the high isotopic ratio of the GCR source composition can be understood as the result of CRs accelerated primarily in SB cores out of a mixture containing a mass fraction of $18\pm5$\% of W.-R. winds and SN ejecta. But \citet{pra12}, using more recent stellar yield calculations \citep{hir05} found that the $^{22}$Ne/$^{20}$Ne ratio in SBs should in fact be close to solar. 

We calculated the SB enrichment in massive star winds and SN ejecta from the models of rotating and non-rotating massive stars (between $13$ and $120~M_\odot$) of \cite{lim18}. We used the ``recommended'' set of models, which assumes that all stars more massive than $25~M_\odot$ fully collapse in the compact-object remnant (failed SNe) and therefore contribute to the chemical enrichment of SBs only through their winds. Using these massive star yields in a consistent chemical evolution model, \citet{pra18} showed that the solar system isotopic composition can be reproduced to better than a factor of two for almost all isotopes up to the Fe-peak. We adopted the observation-motivated recipe of \citet{pra18} for the proportion of rotating massive star at solar metallicity, which assumes that $1/3$ of the stars have an initial surface rotation velocity of $150$~km~s$^{-1}$ and the remaining $2/3$ are not rotating. To obtain the mean composition of the massive star material injected in SBs, we summed the yields of individual stars folded by the standard initial mass function (IMF) of \citet{kro01}.

The SB enrichment provided by the massive star winds and SN ejecta is shown by the blue triangles in Figure~\ref{fig:windabund}. This composition is characterised by its high metallicity: the abundance of heavy elements (Z/H) is much higher than in the SC composition, e.g. by a factor of $\sim$10 for Si and $>10$ for Ga, Ge and As. This composition is similar to the massive star material composition adopted by \citet{mur16} and \citet{lin19} to explain the alleged overabundance of trans-iron “volatile” elements in the GCR source reservoir (see Section~\ref{sec:depletion}). But we find that the $^{22}{\rm Ne}/^{20}{\rm Ne}$ ratio in this composition is significantly less than that in the GCR source composition, $^{22}{\rm Ne}/^{20}{\rm Ne}=0.12$ compared to $0.317$ in GCRs \citep{bos20}. In addition, the massive star material ejected into the SB core are most likely mixed with material of SC composition evaporated off the SB shell, further reducing the $^{22}{\rm Ne}/^{20}{\rm Ne}$ ratio. Thus, we confirm the result of \citet{pra12} that a SB enrichment provided by massive star winds and SN ejecta cannot explain the high $^{22}{\rm Ne}/^{20}{\rm Ne}$ ratio in GCRs. The reason for this is that SN ejecta are rich in $^{20}{\rm Ne}$ synthesised during C burning in stellar cores, and $^{22}{\rm Ne}$ is overabundant with respect to $^{20}{\rm Ne}$ only in massive star winds \citep[see][]{pra12}. 

The red filled circles in Figure~\ref{fig:windabund} show the elemental enrichment obtained by considering only the stellar wind yields from \citet{lim18}, not the SN ejecta. In this case, we obtain a $^{22}{\rm Ne}/^{20}{\rm Ne}$ ratio of 0.35, which is similar to the isotopic ratio found in GCRs. Thus, explaining the GCR source ratio with this wind composition would require that it is hardly diluted with any material of SC composition, which is unlikely. Perhaps a more likely scenario was proposed by \citet{bin08}, who suggested that the bulk of $^{22}{\rm Ne}$-rich material available for acceleration in SB cores is the wind material of very massive stars, of initial mass $M_{\rm ini} \ge 40~M_\odot$. These stars are thought to go through a W.-R. phase where they lose most of their mass during the first $\sim$6~Myr of SB evolution, before collapsing to a black hole without undergoing a SN explosion. Core-collapse SNe resulting from the explosion of less massive stars occur after $\sim$6~Myr. The SN blast waves then propagate in an ambient medium composed of the wind material ejected during the W.-R. epoch of the SB, plus any normal ISM in the SB interior. We calculated the elemental abundances of the $^{22}{\rm Ne}$-rich reservoir in this scenario by averaging the wind yields of \citet{lim18} for stars of mass $M_{\rm ini} \ge 40~M_\odot$ over the \citet{kro01} IMF. The results are shown by the red empty squares in Figure~\ref{fig:windabund} and given in the 4th column of Table~\ref{tab:data}. The $^{22}{\rm Ne}$ abundance is high in this wind material: $^{22}{\rm Ne}/^{20}{\rm Ne}=0.61$. To get $^{22}{\rm Ne}/^{20}{\rm Ne}=0.317$ in the SB cores as in the GCR source composition, the required mixing with a medium of SC composition ($(^{22}{\rm Ne}/^{20}{\rm Ne})_{\rm SC}=7.35\times10^{-2}$; \citep{lod09}) is $x_w=0.54$ and $x_{\rm SC}=0.46$.

\subsubsection{Acceleration of massive star winds in their termination shocks}
\label{sec:22newind}

An alternative scenario to explain the high $^{22}{\rm Ne}/^{20}{\rm Ne}$ ratio of the GCR source composition was recently considered by \citet{gup20}. These authors studied theoretically the relative contributions of WTSs and SN shocks in massive star clusters, and found that WTSs should contribute at least 25\% of the total CR production in these objects and up to $\gsim 50$\% in young ($\lsim 10$~Myr) clusters. They suggested that the acceleration of $^{22}{\rm Ne}$-rich stellar winds in WTSs can explain the GCR isotopic ratio. Independently,  \cite{kal19} calculated that young massive star clusters can account for a significant fraction of $^{22}{\rm Ne}$ in GCRs through efficient particle acceleration in shock waves from interacting winds of massive stars. They used in their calculations massive star yields from the stellar evolution models of \citet{eks12} and \citet{geo12},

\begin{figure}
\begin{center}
 \includegraphics[width=0.7\columnwidth]{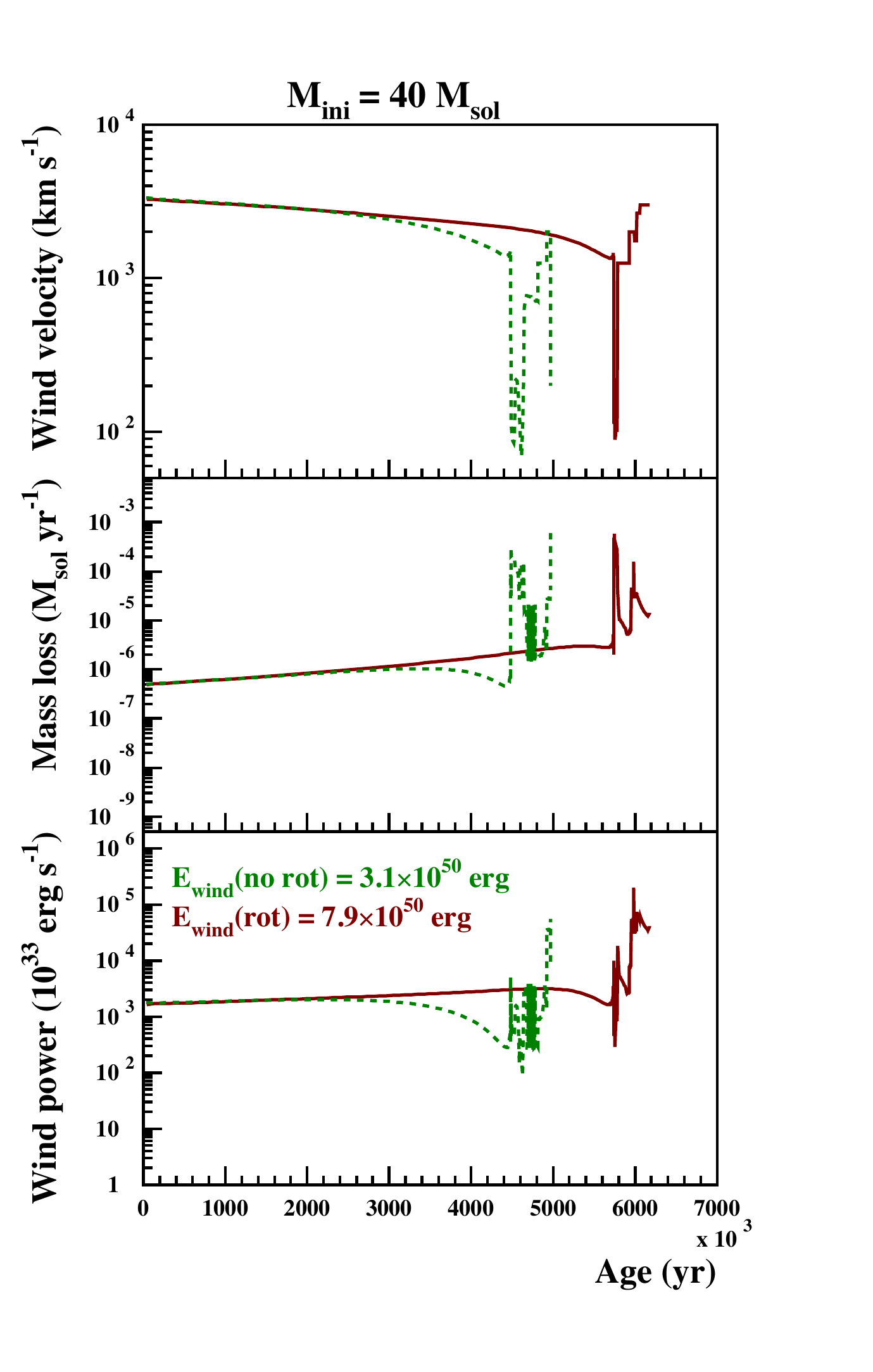}
 \caption{(\textit{Upper panel}) Wind terminal velocity, (\textit{middle panel}) star mass loss, and (\textit{lower panel}) wind mechanical power, as a function of stellar age, for two stars of initial mass $M_{\rm ini}=40~M_\odot$. Red solid curves: rotating star; green dashed curves: non-rotating star. The total kinetic energy of the winds integrated over the star lifetime is reported in the \textit{lower panel} for both the non-rotating and rotating stars.}
 \label{fig:wind40}
\end{center}
\end{figure}

We calculated the composition of particles accelerated by WTSs in star clusters by assuming that at any time $t$, the particle acceleration efficiency in the winds of a star of initial mass $M$\footnote{To simplify the notation, we adopt $M$$\equiv$$M_{\rm ini}$ in the equations of this section.} is proportional to the wind mechanical power
\begin{equation}
P_{M,w}(t) = 0.5 \dot{M}_w(t) v_{M,\infty}^2(t)~,
\label{eq:windpow}
\end{equation}
where $\dot{M}_w$ is the wind mass loss rate of the star and $v_{M,\infty}$ the wind terminal velocity. We did not consider the large-scale WTS that can be created by the collective effect of many stars around a compact cluster of young massive stars \citep[see][]{gup20}, but have treated the WTS of each star individually to represent the case of a loosely bound cluster. Clusters of the latter type are far more numerous in the Galaxy than very massive and compact clusters that can form a collective WTS \citep[see][]{kru19}. 

The mass of an isotope of type $i$ accelerated by the WTSs of all massive stars in a loosely bound cluster is obtained from 
\begin{equation}
m_{\rm acc}(i) \propto \int\displaylimits_{M_{\rm min}}^{M_{\rm max}} \bigg( \int\displaylimits_{0}^{t_M} \dot{M}_w(t) x_{M,i}(t) P_{M,w}(t) dt \bigg) \Phi_{\rm IMF}(M)dM~,
\label{eq:winacc}
\end{equation}
where $t_M$ is the lifetime of the star of initial mass $M$, $x_{M,i}$ the mass fraction of the nuclei of type $i$ in the wind composition of this star, and $\Phi_{\rm IMF}(M)$ the IMF \citep{kro01}. We took $t_M$, $x_{M,i}$, and $\dot{M}_w$ from the stellar evolution models of \citet{eks12} and \citet{geo12}, which are available from the Geneva Observatory database\footnote{ \url{https://www.unige.ch/sciences/astro/evolution/en/database/}}. Based on the massive star models computed by the Geneva group, we set $M_{\rm min}=12~M_\odot$ and $M_{\rm max}=120~M_\odot$ in the above equation. 

The wind terminal velocity $v_{M,\infty}$ for the various types of stars is calculated from the prescriptions of \citet{vos09} for their ``wind08" model. First, stars with $\dot{M}_w > 10^{-3.5}~M_\odot$~yr$^{-1}$ and an effective temperature in the range $3.75 < \log T_{\rm eff} < 4.4$ are classified as luminous blue variables (LBVs) and their wind terminal velocity is set to $200$~km~s$^{-1}$. Stars with $\log T_{\rm eff} > 4.0$ and a fractional abundance of H at the surface below 0.4, are considered to be W.-R. stars, and their wind terminal velocity is taken to be in the range $1250$--$3000$~km~s$^{-1}$ depending on their subclass \citep[see][Table 2]{vos09}. The wind terminal velocity of stars that are not W.-R. or LBV stars is assumed to be proportional to the escape velocity at the stellar surface, which we calculated from \citet[][eq.~10]{how89}: $v_{M,\infty}=1.3 v_{M,{\rm esc}}$ for cool stars ($\log T_{\rm eff} \leq 4.32$) and $v_{M,\infty}=2.6 v_{M,{\rm esc}}$ for hot stars ($\log T_{\rm eff} > 4.32$). 

The evolution of the wind terminal velocity, mass loss and mechanical power are shown in Figure~\ref{fig:wind40} for two stars of $M_{\rm ini}=40~M_\odot$, one rotating and the other not. In the rotating star models of the Geneva group, the initial rotation velocity is set to 0.4 times the critical velocity of the star on the zero-age main-sequence \citep{eks12}. Stellar wind properties calculated for the other massive star models available in the Geneva Observatory database are given in Appendix~A. 

The material expelled by the winds of massive stars is strongly enriched in $^{22}{\rm Ne}$ when the star is a W.-R. in the subclasses WC and WO. At that time, which is quite short in the life of a massive star, both the wind mass loss and terminal velocity are high, such that the corresponding mechanical power is also high. For a rotating star of $M_{\rm ini}=40~M_\odot$ (red solid curves in Fig.~\ref{fig:wind40}), the WC-WO phase begins almost 6~Myr after the birth of the star and lasts about 180,000 years before the star collapses. During this period, the wind power is in the range $(3$--$7)\times 10^{37}$~erg~s$^{-1}$ and the integrated mechanical energy amounts to $2.6\times 10^{50}$~erg, which is $1/3$ of the total kinetic energy released by the winds during the star lifetime. The non-rotating star of $M_{\rm ini}=40~M_\odot$ (green dashed curves in Fig.~\ref{fig:wind40}) becomes a W.-R. star 4.8~Myr after its birth, but never reaches the WC-WO phase. To calculate the mean abundance of the elements in the accelerated wind composition (5th column in Table~\ref{tab:data}), we assumed, based on the chemical evolution model of \citet{pra18}, that $1/3$ of the massive stars formed at the current epoch are rotating. With this assumption, we find that the mean mechanical energy deposited by the winds from a massive star in the Galaxy amounts to $2.6\times10^{50}$~erg, which is about $1/4$ of the kinetic power of a SN explosion. This result is in good agreement with the calculations of \citet{seo18}. 

The Geneva Observatory database contains the calculated mass fractions of the main isotopes of H, He, C, N, O, and Ne at the stellar surface, but does not provide the abundances of Na, Mg, and Al. These three elements are modified by the Ne-Na and Mg-Al nucleosynthesis cycles during the H-burning phase, and thus can be released in non-solar proportions in massive star winds. From the wind yields calculated by \citet{lim18} for 18 stars (9 rotating and 9 non-rotating stars between $13$ and $120~M_\odot$), we checked that the abundances of Na, Mg, and Al are correlated with that of N, which is also produced during H burning, but by the CNO cycle. Based on the stellar models of \citet{lim18}, we derived a simple function to estimate the enhancement factors of these three elements in the accelerated-wind composition from that of N:
\begin{equation}
f_w(i)=k(i) \times (f_w({\rm N})-1)+1~,
\label{eq:namgal}
\end{equation}
with $k({\rm Na})=0.46$, $k({\rm Mg})=-2.08 \times 10^{-3}$, and $k({\rm Al})=3.54 \times 10^{-2}$.

The $^{22}{\rm Ne}/^{20}{\rm Ne}$ ratio in the accelerated wind composition amounts to $1.56$. The Ne isotopic ratio is higher in this composition than in the wind material released in SB cores (Section~\ref{sec:22nesb}), because of the relatively high acceleration efficiency of $^{22}{\rm Ne}$-rich material in WTSs during the W.-R. WC-WO phase, in which the wind mechanical power is higher than during the preceding stellar phases. To obtain the Ne isotopic ratio of the GCR source composition ($^{22}{\rm Ne}/^{20}{\rm Ne}=0.317$), the required mixing with a reservoir of SC composition is $x_w=0.15$ and $x_{\rm SC}=0.85$. However, the required contribution of accelerated wind material depends on the relative particle acceleration efficiencies in SN shocks and WTSs, which in turn depend on the ionisation states of the ions entering the shocks (the full results of the data analysis taking into account this effect are presented in Section~\ref{sec:results}). 

\subsection{Ionisation fractions in shock precursors}
\label{sec:ioni}

The acceleration process depends on the mass-to-charge ratio of the particles entering the shock, so the ionisation state of each element in the preshock region must be specified.  We will discuss the warm and hot phases of the ISM separately, and then the stellar winds accelerated by WTSs.

\subsubsection{WIM and WNM}


The average ionisation states of the WIM and WNM can be obtained from absorption or emission line measurements \citep[e.g.][]{sem00,madsen06}.  Some ionisation will occur in the post-shock gas during the acceleration, but \citet{eic21} find it to be small.  The ionisation state will also be modified in narrow shock precursors produced by cosmic rays and by backstreaming neutral atoms from the postshock region \citep{blasi12, morlino12}, but the gas does not spend enough time there for significant ionisation to occur.  Finally, the ionisation state will be modified by photoionisation due to the SN itself, extreme ultraviolet (EUV) and X-ray emission from the shocked SN ejecta \citep{hamilton88}, and emission from the SNR blast wave. Ionising radiation from the SN and its ejecta will mainly affect the circumstellar gas and the ISM close to the explosion \citep{weil20}, while most of the mass from which the cosmic rays are accelerated is farther from the explosion site, where the SNR is in the Sedov-Taylor phase.  We therefore consider only the photoionisation precursor of the blastwave.  Those precursors are faint, but they have been observed just outside the shocks of Tycho's SNR \citep{ghavamian00}, the Cygnus Loop \citep{medina14}, and N132D \citep{morse96}. 

SNR shock waves slower that about 300 \kms are usually radiative, meaning that nearly all of the post-shock thermal energy is converted into radiation.  These shocks very effectively photoionise the preshock gas \citep{shull79, allen08}, but shocks at those speeds are relatively ineffective in accelerating cosmic rays \citep{raymond20a, raymond20b}.  Therefore, we will consider the faster, nonradiative shocks.  Fast, nonradiative shocks in the WNM are visible in H$\alpha$ as filaments with pure Balmer line emission spectra \citep{chevalier78, heng10}, while similar shocks in the WIM are not visible at optical wavelengths.

X-rays from the SNR interior can ionise the ambient gas, particularly by K-shell photoionisation, but the cross sections are relatively small and the ionisation time scale is of the order of $10^5$ years.  The dominant photoionisation is produced by He I and He II photons from the thin ionisation zone behind the shock.  When a helium atom is suddenly immersed in a very hot plasma, it will be excited and eventually ionised.  It will produce on average a number of photons given by the ratio of excitation to ionisation rates, which is essentially constant at high temperatures \citep{laming96}.  Each neutral He atom will produce about 4.5 He I photons that can ionise hydrogen, mostly in lines close to h$\nu$=21 eV.
It will then produce He II photons before becoming fully ionised, about 1.25 photons at 40.8 eV and 1.14 photons in the 2-photon continuum.  With a He abundance of 0.085 compared to H, there are enough photons to partially ionise the preshock gas.  Because the photoionisation cross section of H declines rapidly with photon energy,
He and heavier elements will absorb a disproportionate fraction of the photons.

For these estimates we consider an SNR near the end of the Sedov phase with a shock speed of 1000 \kms , since that is where most of the ISM mass is swept up and most of the cosmic rays are accelerated.  At that stage, the precursor thickness is around 1/5 the SNR radius (for a density of 1 $\rm cm^{-3}$), so we assume a planar geometry in which the flux of ionising photons balances the flux of particles being ionised.  We assume initial ionisation states of 5\% H I, 40\% He I in the WIM and 90\% H I, 95\% He I in the WNM.  

Photoionisation cross sections are taken from \citet{reilman79} for elements through Zn, from \citet{marr76} for Kr I and from \citet{henke93} for the other atoms heavier than Zn.  Reilman \& Manson give cross sections for all the ions of the elements they consider, and for the others we have scaled the neutral atom cross sections along isoelectronic sequences.  In most cases, the recombination rates are so small that recombination can be neglected.  The exception is that in the WNM, very rapid charge transfer can occur between some specific ions and hydrogen atoms, so that the charge transfer recombination time is shorter than the time to cross the photoionisation precursor.  The $\rm O^{2+}$ ion in particular recombines to $\rm O^+$ in the WNM.  \citet{kingdon96} give rates for the abundant elements, but we have not been able to find rates for the heavier elements.  Therefore, the ionisation states of the heavy elements in the WNM may be overestimated. 

Most elements are 1 or 2 times ionised when they reach the shock.  It is noteworthy that many elements are more highly ionised than hydrogen, because the photoionisation cross section of H at 40.8 eV is small. The predicted values of the factor $f_{A/Q}=(1-X_0)A/Q$ are presented in the 7th and 8th columns of Table~\ref{tab:data} for the WNM and the WIM, respectively ($X_{0}$, $A$ and $Q$ are the fraction of neutral atoms, the atomic mass and the mean ionic charge without the neutral fraction, respectively).

The ionisation state of the preshock gas can also have important effects on the diffusive shock acceleration process. \citet{blasi12} show that collisionless shock can be strongly modified by the presence of neutral atoms through the processes of charge exchange with ambient ions. A substantial neutral fraction can also damp the magneto-hydrodynamic waves needed for diffusive shock acceleration, leading to very steep particle spectra for shock speeds below about $3000$~km~s$^{-1}$. Dust grains interact with waves corresponding to very energetic protons, so the lack of those waves would severely inhibit acceleration of grains and injection of refractory elements into the CR population. These effects are further discussed in Section~\ref{sec:discussion}. 

\subsubsection{Superbubbles and the hot ISM}

Most core-collapse SNe occur in the hot, SB phase of the ISM and not in the warm medium \citep{par04,lin19}.  Most massive stars explode in SBs of less than $\sim 30$~Myr, within a plasma of $\gsim 10^6$~K. The temperature inside SBs can exceed $10^7$~K in young and massive stellar clusters such as the Arches and Quintuplet clusters \citep{wan06}. All atoms are highly ionised in such plasma. Additional photoionisation of the preshock plasma by X-rays from the SNR interior can be safely neglected.  

We obtain the ionisation states as a function of the plasma temperature from \cite{maz98} for C..Ni and \cite{pos77} for Zn, As, Kr, Rb, and Zr. The $A/Q$ values of Ga, Ge, Se, Br, Sr, and Y in the hot SB gas ($X_0=0$) were estimated by interpolation on mass of neighbouring elements. The results are shown in the 9th and 10th columns of Table~\ref{tab:data} for $T_{\rm SB}=10^6$~K and $10^7$~K, respectively. 

\subsubsection{Stellar winds}

The EUV radiation of hot stars ionises their winds, and the lines of C$^{3+}$ and Si$^{3+}$ are generally among the strongest features in their UV spectra \citep[e.g.][]{hillier20}.  Lower ions such as C$^{2+}$ and Si$^{2+}$ are present as well, and more highly ionised species such as N$^{4+}$ and O$^{5+}$ are produced by the EUV and X-ray emission from shocks in the winds \citep[e.g.][]{bouret12}.  For simplicity, we ignore the range of ionisation states and assume that all the elements are triply ionised, except for H and He which are taken to be fully ionised (last column in Table~\ref{tab:data}).  The heavier elements have received less attention, and they quite likely have both higher photoionisation and recombination rates, so they might be somewhat more highly ionised. 

\subsection{Distribution of supernovae in the ISM phases}
\label{sec:snr}

The charge-to-mass ratio of gas entering the shock governs the acceleration efficiency, so we must specify the distribution of SNe among the phases of the ISM.  Some guidance can be obtained from observed Galactic SNe, though the statistics are poor and there are systematic effects such as obscuration by dust.  
We consider only young SNRs of less than a few thousands years old, because it is hard to determine the SN type for older SNRs. 

Type Ia SNe should occur randomly, and they can occur in the halo as well as in the plane of the Galaxy.  SN1006 is 550 pc from the plane, and it is located in a region of 90\% ionised gas \citep{ghavamian02}.  Kepler's SNR is also far from the plane, but it shows Balmer line filaments where the shock sweeps up neutral hydrogen.  However, it is likely that the shock is still within local CSM material.  Tycho's SNR is in the plane.  It shows Balmer filaments over about 1/4 of its circumference, indicating that it encounters about 1/4 WNM and 3/4 WIM.
RCW86 appears to have exploded in hot bubble, and the shock has now reached the shell.  G1.9+0.3 seems to be in a region of low density, about 0.03 $\rm cm^{-3}$, \citep{brose19}.  That suggests that it is part of the hot ionised medium, but the asymmetry might indicate a large range of densities.

Core collapse SNe are expected to occur in the SBs created by their OB associations about 80\% of the time \citep{lin07,lin19}.  Indeed, the Crab, RCW89, HESS1731-347, Vela, Jr. and G330.2-0.1 all seem to have exploded in low density bubbles.  Cas A and G11.2-0.3 seem to still be interacting with their own circumstellar media.

Some observational guidance is available, such as the WHAM measurements of faint, diffuse emission lines that indicate a filling factor of order 30\% for the WIM \citep{hausen02}.  For the models below, we follow \cite{lin07} and assume that 80\% of the core-collapse SNRs occur in SBs. 
We also assume that $1/4$ of the Galactic SNe are of Type Ia, which occur randomly in the warm ISM, with 30\% in the WIM and 70\% in the WNM. The distribution of all Galactic SNe is then: 60\% in SBs, 28\% in the WNM and 12\% in the WIM. This seems to be in keeping with global simulations of the ISM, such as \cite{gur20}. 

\section{Data analysis and results}
\label{sec:results}

\begin{figure*}
\begin{tabular}{c}
\begin{minipage}{0.5\textwidth}
\begin{center}
\includegraphics[width=0.67\textwidth]{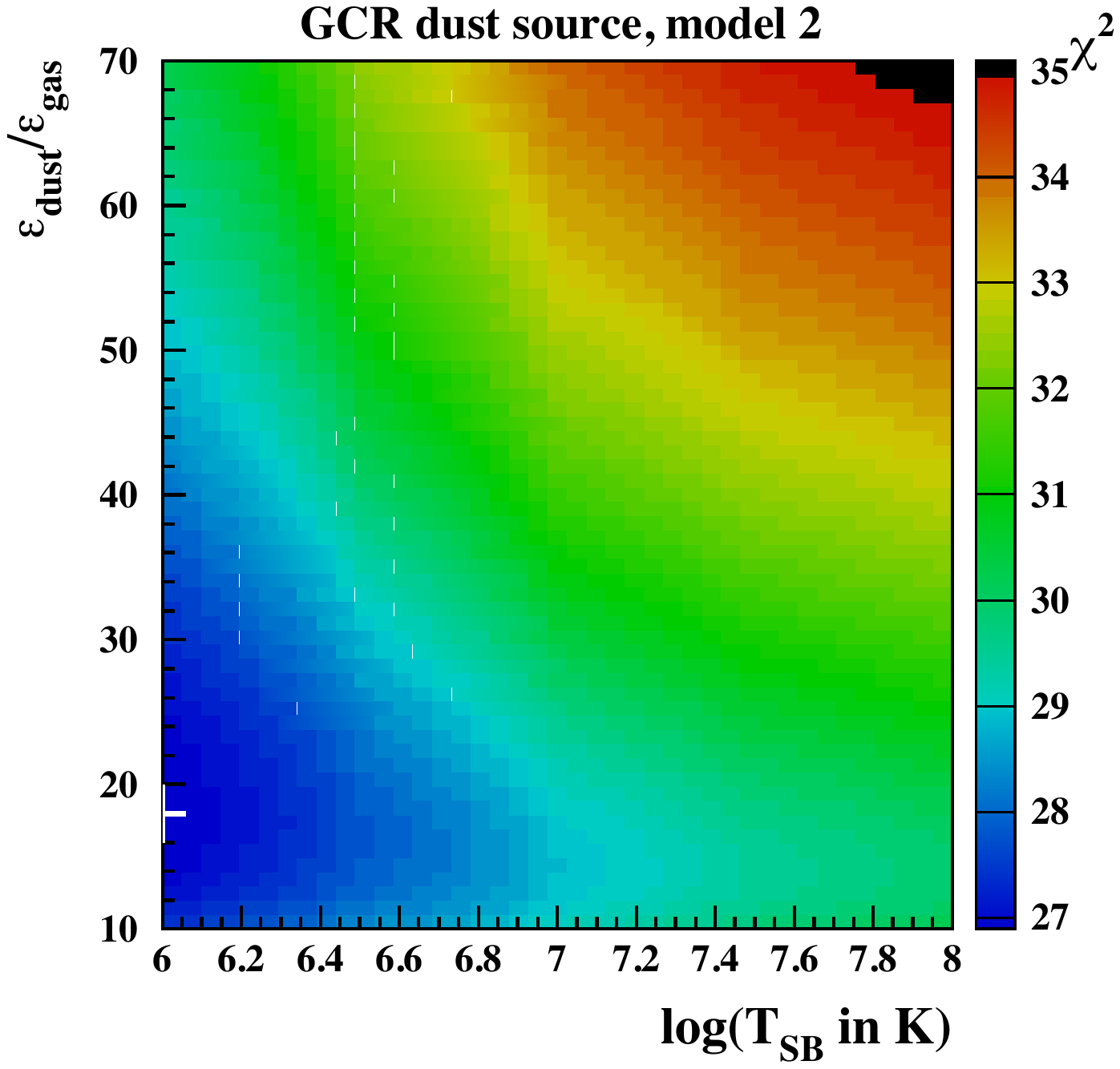}
\includegraphics[width=0.67\textwidth]{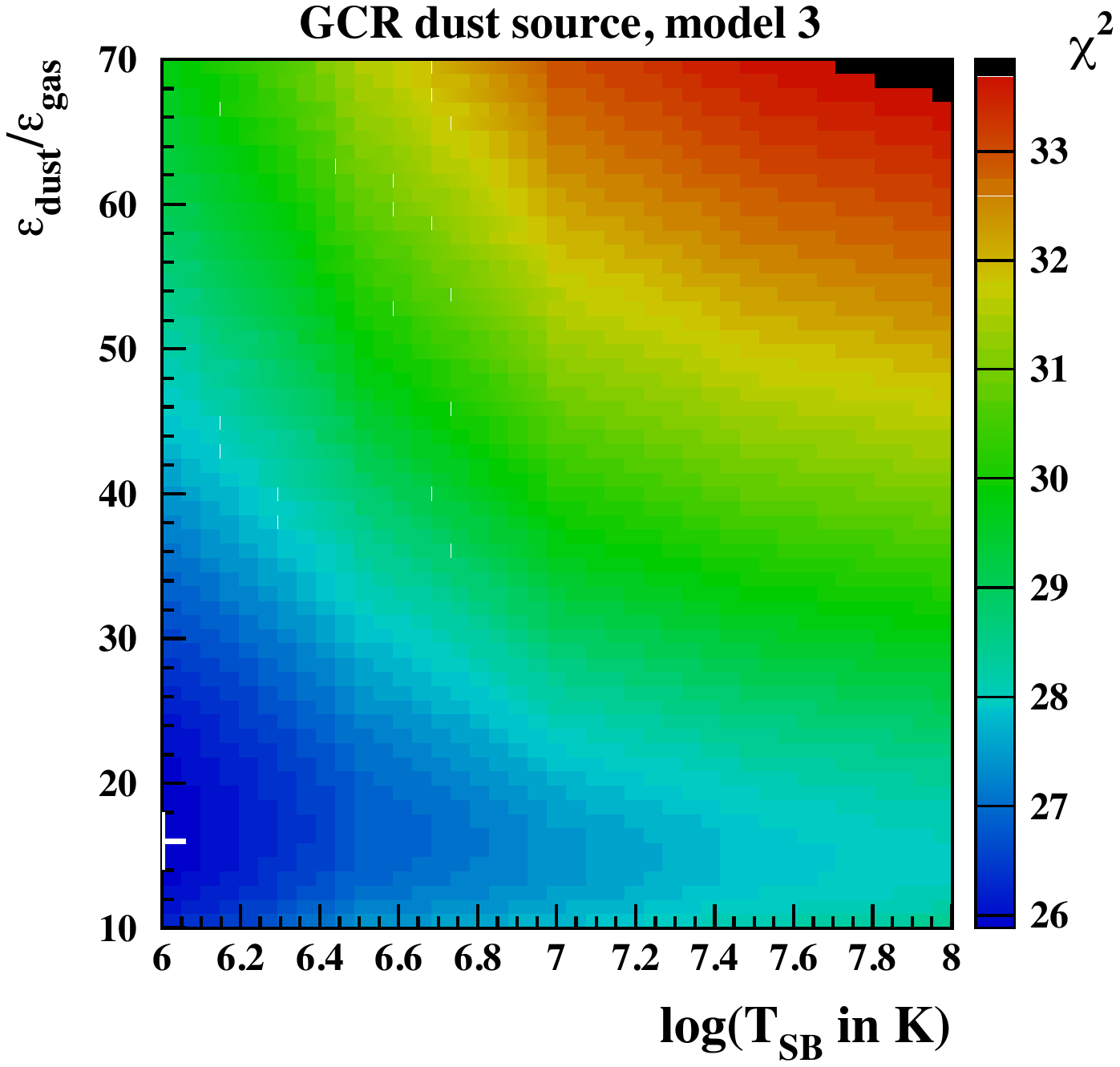}
\includegraphics[width=0.67\textwidth]{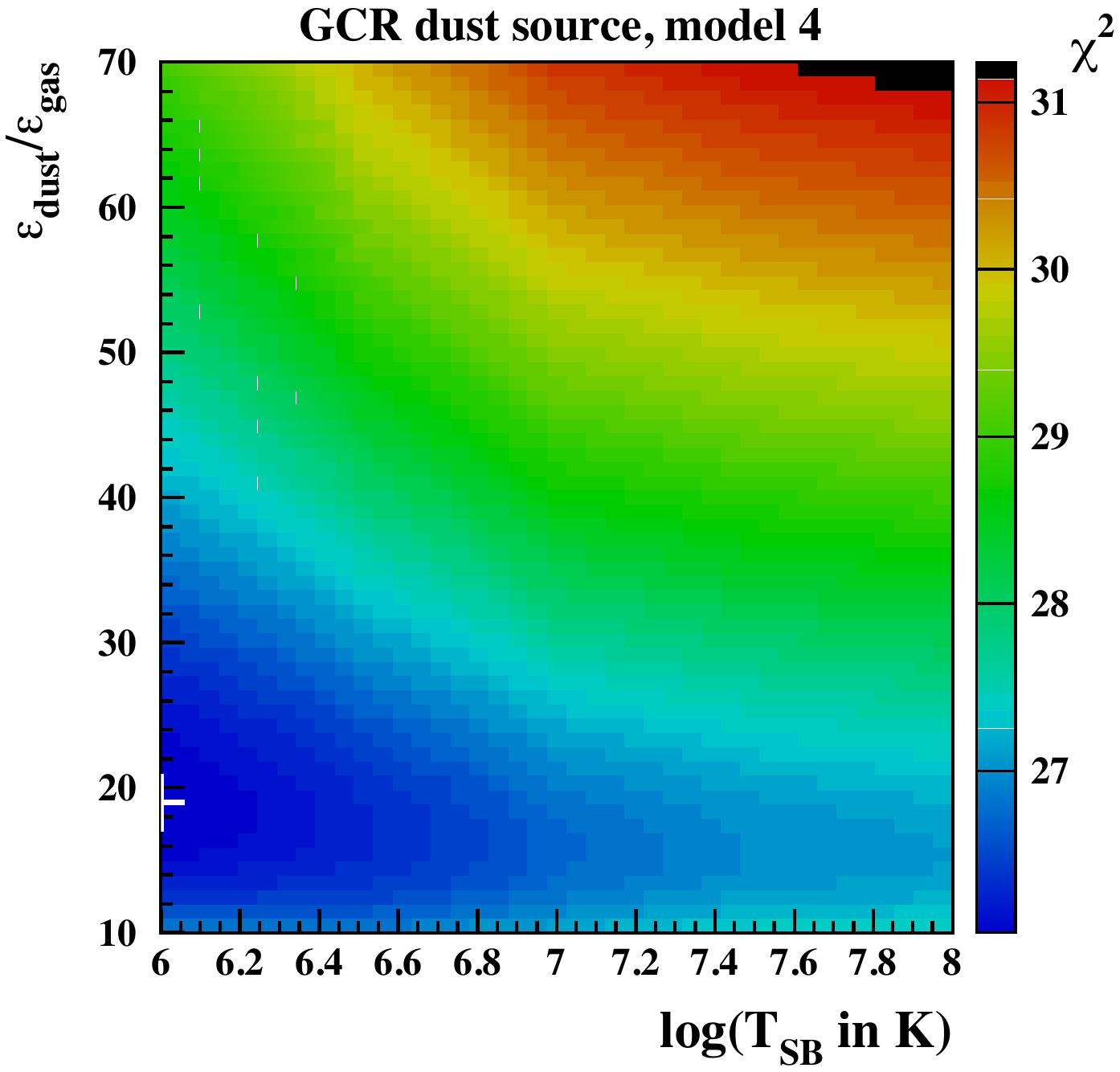}
\includegraphics[width=0.67\textwidth]{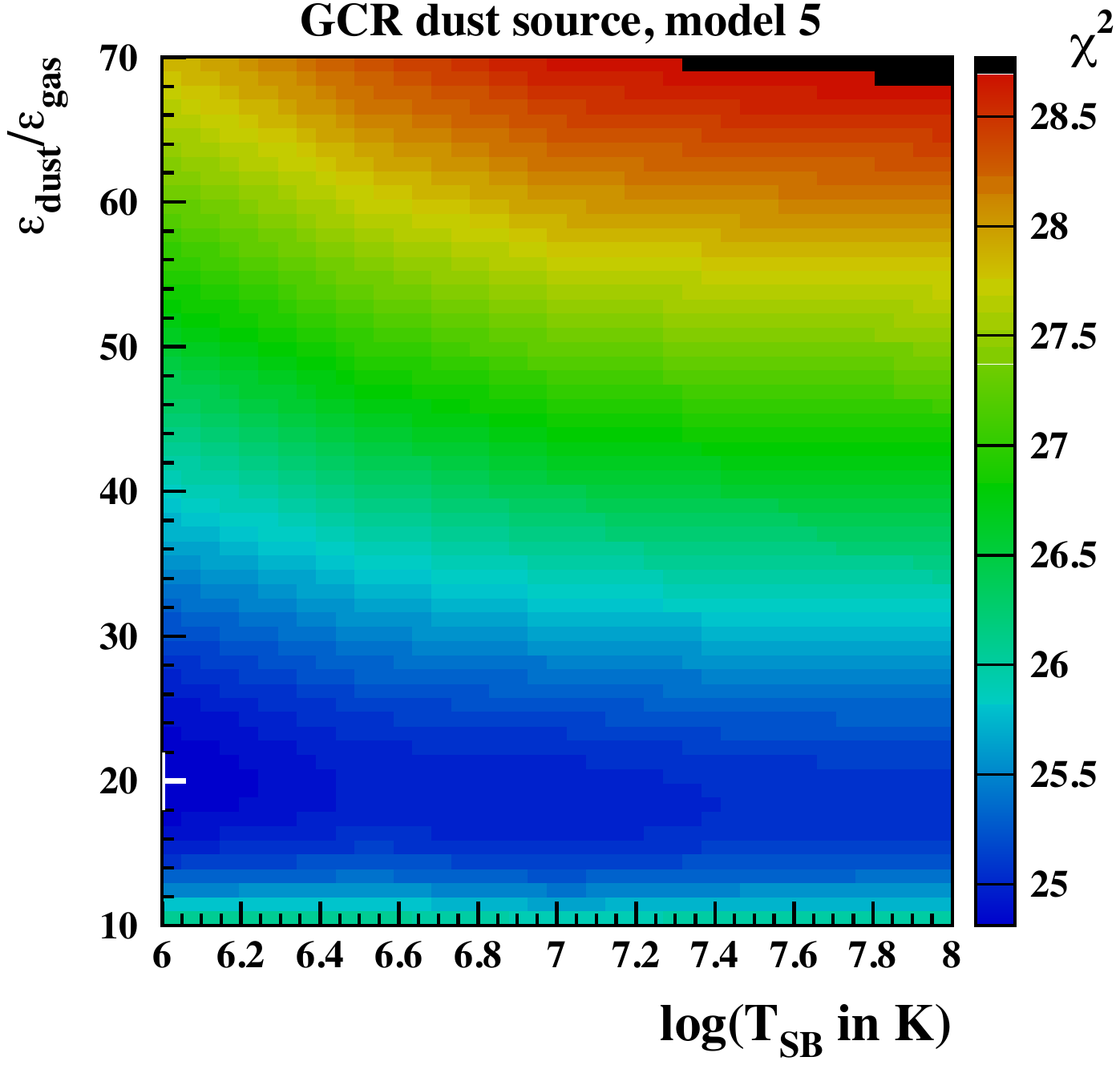}
\end{center}
\end{minipage}
\begin{minipage}{0.5\textwidth}
\begin{center}
\includegraphics[width=0.67\textwidth]{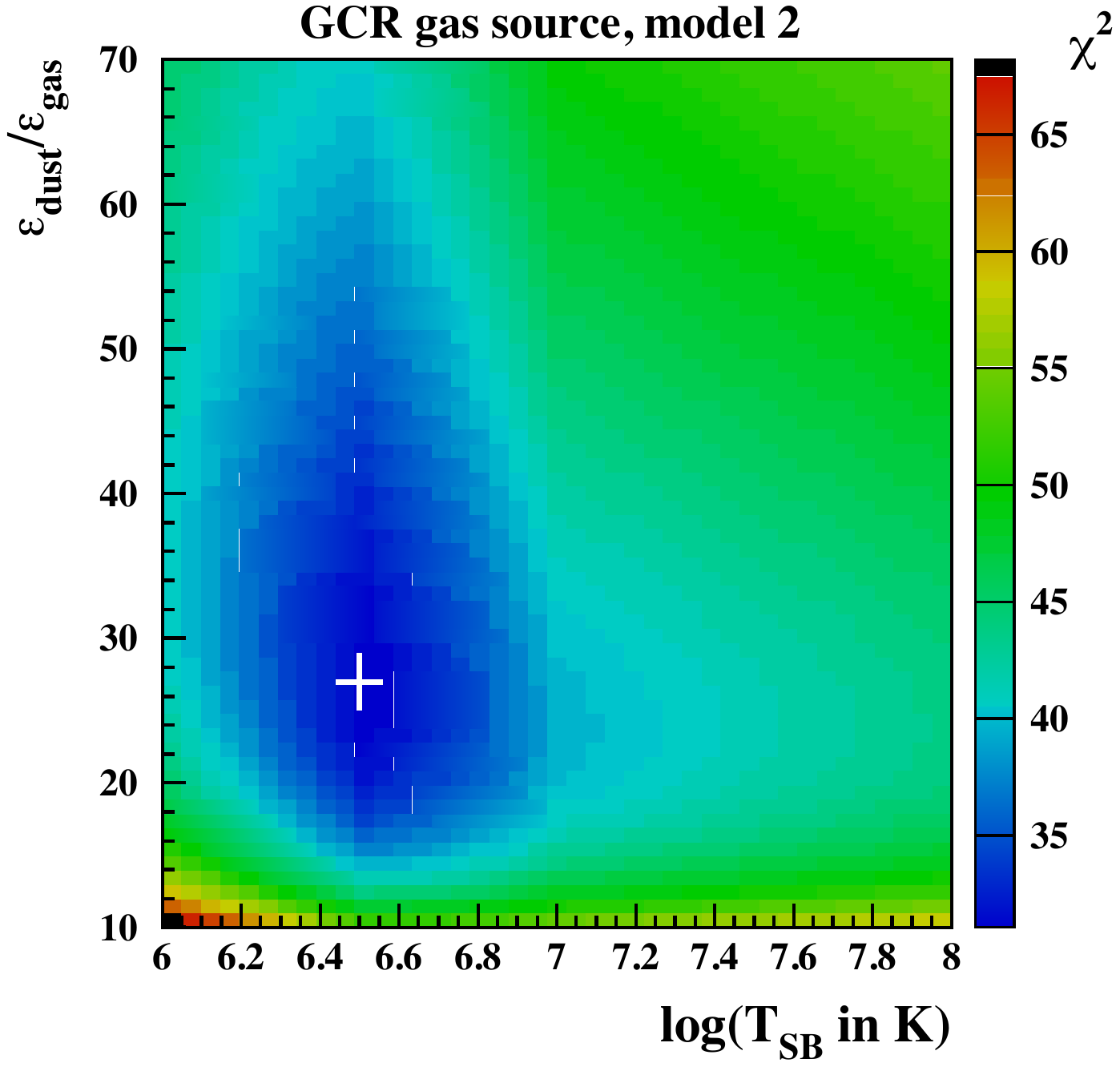}
\includegraphics[width=0.67\textwidth]{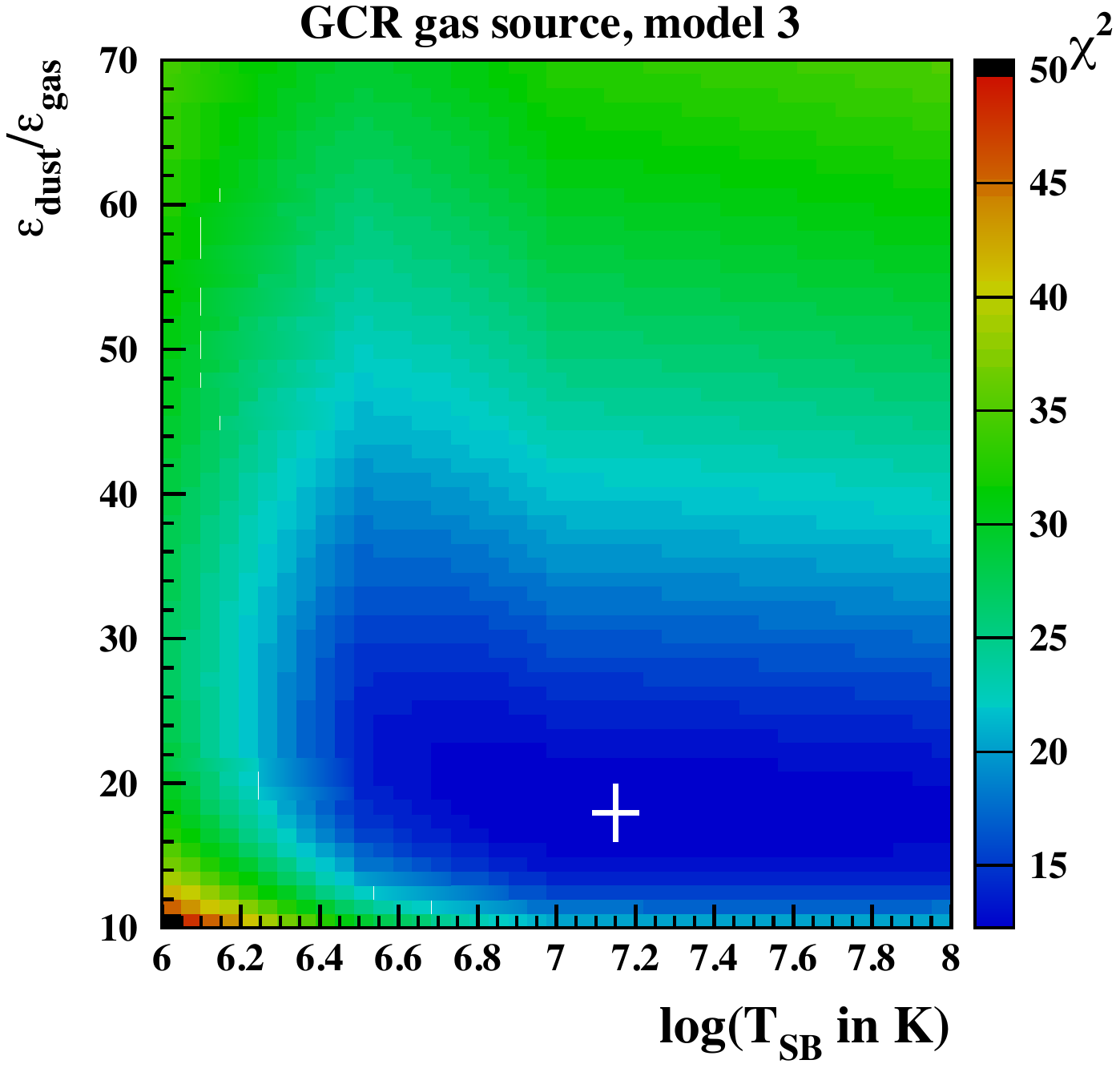}
\includegraphics[width=0.67\textwidth]{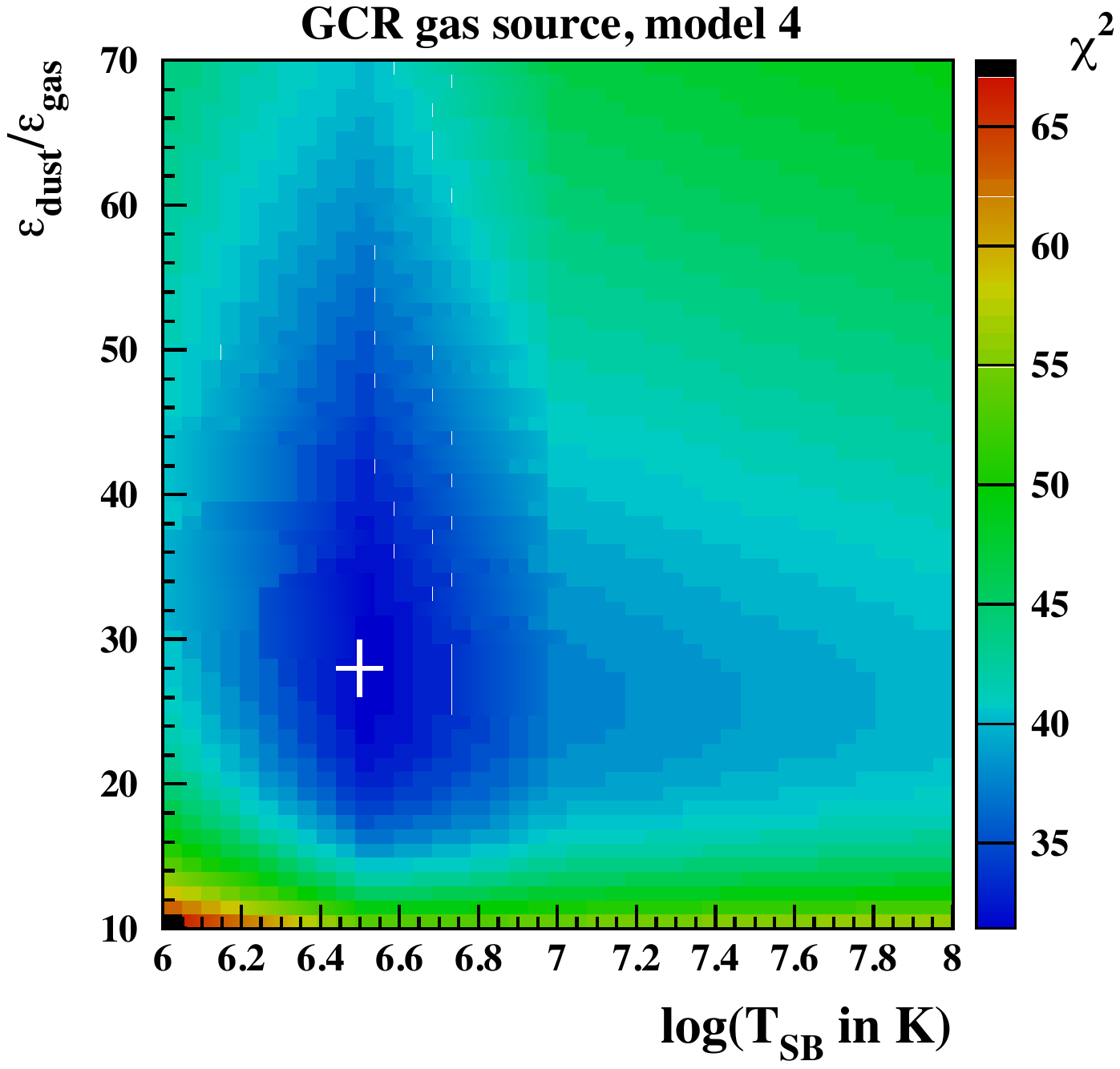}
\includegraphics[width=0.67\textwidth]{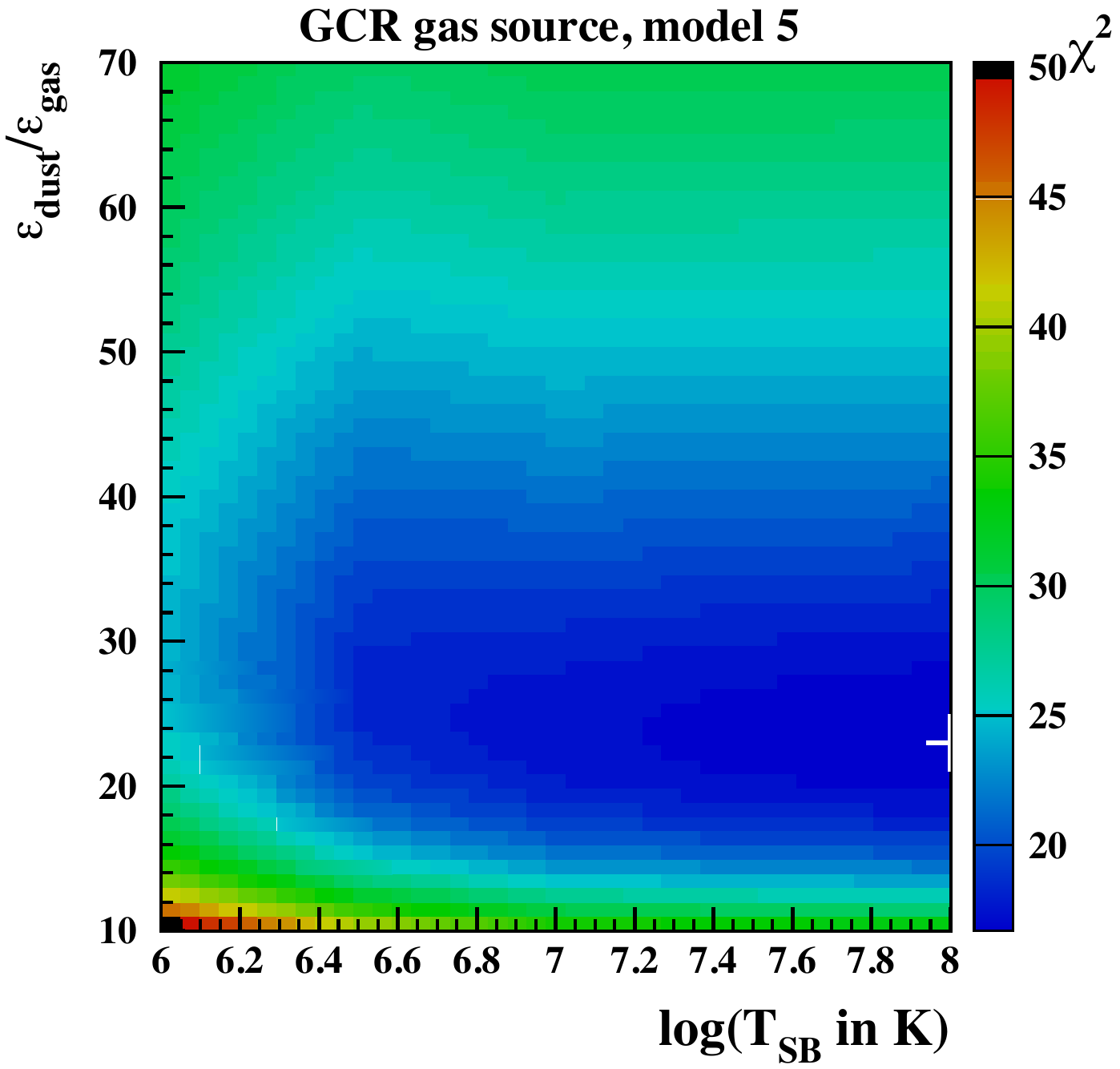}
\end{center}
\end{minipage}
\end{tabular}
\caption{Chi-square maps from fits of the GCR abundances from acceleration of dust grains (\textit{left panels}) and those from acceleration of ions in gas phase (\textit{right panels}) with two free parameters, the SB temperature and the relative efficiency $\epsilon=\epsilon_{\rm dust}/\epsilon_{\rm gas}$, for Models 2 to 5 (from \textit{top} to \textit{bottom}; see Table~\ref{tab:model}). Model 1 does not depend on the SB temperature; $\chi^2$ as a function of $\epsilon$ for this model is shown in Fig.~\ref{fig:chi2}. The white crosses mark the best-fit values.}
\label{fig:chi2bidim}
\end{figure*} 

\begin{table*}
\caption{Models for the origin of the GCR gas source in the ISM and best-fit parameters with associated minimum $\chi^2$. \label{tab:model}}
\begin{tabular}{|l|lllll|}\hline
                          & Model 1            & Model 2      & Model 3           & Model 4            & Model 5            \\\hline
GCR gas source of SC compo. & 70\% WNM, 30\% WIM & SB           & SB                & 60\% SB,           & 60\% SB,           \\
                          &                    &              &                   & 28\% WNM, 12\% WIM & 28\% WNM, 12\% WIM \\
$^{22}$Ne-rich GCR gas source & Accelerated winds  & Winds in SB  & Accelerated winds & Winds in SB        & Accelerated winds  \\\hline
SB temperature $\log(T_{\rm SB})$$^a$ & --               & $6.50\pm0.25$  & $>6.45$     & $6.5_{-0.2}^{+0.3}$      & $>6.35$      \\     
Relative eff. $\epsilon=\epsilon_{\rm dust}/\epsilon_{\rm gas}$$^b$ & $33.8\pm13.4$ & $26.0\pm13.2$ & $17.9\pm9.7$ & $27.0\pm13.2$ & $22.8\pm10.6$ \\
W.-R. wind contribution $x_w$$^c$ & 10.3\%  & 48.9\% & (5.1~--~6.1)\% & $(55.6_{-0.3}^{+1.3})$\%   & (7.3~--~7.9)\%              \\
$\chi^2_{\rm min}$(GCR dust source)$^d$ & 24.6     & 26.9         & 25.9              & 26.0               & 24.8               \\
$\chi^2_{\rm min}$(GCR gas source)$^e$  & 24.7     & 31.1         & 12.2              & 31.4               & 16.7               \\\hline
SB temperature $\log(T_{\rm SB})$ & --                 & $6.6$ (fixed)& $6.6$ (fixed)     & $6.6$ (fixed)      & $6.6$ (fixed)      \\     
Relative eff. $\epsilon=\epsilon_{\rm dust}/\epsilon_{\rm gas}$$^b$ & $33.8\pm13.4$ & $23.2\pm9.4$ & $20.2\pm7.2$ & $24.6\pm10.2$ & $24.4\pm9.2$ \\
W.-R. wind contribution $x_w$$^c$          & 10.3\%   & 48.9\%       & 5.9\%             & 56.0\%             & 7.7\%              \\
$\chi^2_{\rm min}$(GCR dust source)$^d$ & 24.6     & 28.0         & 26.9             & 26.4               & 25.0               \\
$\chi^2_{\rm min}$(GCR gas source)$^e$  & 24.7     & 32.3         & 13.2             & 32.4               & 18.3               \\
\hline
\multicolumn{6}{l}{$^a$ Best-fit values with $1\sigma$ errors of the SB temperature ($\log(T_{\rm SB})$ with $T_{\rm SB}$ in K) as obtained from the GCR gas source data.} \\
\multicolumn{6}{l}{$^b$ Weighted mean of the best-fit values obtained from the GCR gas and dust source data.} \\
\multicolumn{6}{l}{$^c$ Required mixing of W.-R. wind material with the reservoir of SC composition to get $^{22}{\rm Ne}/^{20}{\rm Ne}=0.317$ in GCRs \citep{bos20}.} \\
\multicolumn{6}{l}{$^d$ Minimum $\chi^2$ from a least-squares fit of the $A_{\rm dust}(f_d)$ data (Eq.~\ref{eq:a}) to a straight line.} \\
\multicolumn{6}{l}{$^e$ Minimum $\chi^2$ from a least-squares fit of the $B_{\rm gas}(f_g)$ data (Eq.~\ref{eq:b}) to a straight line.} \\
\end{tabular}
\end{table*}

Once we have estimated (i) the elemental fractions in ISM dust, (ii) the composition of the $^{22}$Ne-rich component arising from W.-R. winds, and (iii) the ionisation state of each element in the preshock region of SNR blast waves and stellar WTSs, the GCR composition only depends on the acceleration efficiency of dust grains compared to that of ions, and on the ISM phase(s) from which the GCR volatiles are extracted. To constrain the relative acceleration efficiency $\epsilon$ and the nature of the GCR source gas reservoir from the abundance data, we exploit the fact that the quantities 
\begin{eqnarray}
A_{\rm dust}(i) & = & C_{\rm dust}(i)/{\rm SC}(i)~{\rm ,~and} \label{eq:a}
\\
B_{\rm gas}(i) & = & C_{\rm gas}(i) /({\rm SC}(i) [x_w f_w(i) f_{A/Q}^w(i) + (1-x_w)f_{A/Q}^{\rm SC}(i)]) \nonumber \\
& = & C_{\rm gas}(i) /[{\rm SC}(i) F_{\rm gas}(i)]~, \label{eq:b}
\end{eqnarray}
are expected to be proportional to $f_d(i)$ and $f_g(i)=1-f_d(i)$, respectively (see Eqs.~\ref{eq:cdust} and \ref{eq:cgas}). In practice, we first adopt a model for the origin of the GCR volatiles, both for the reservoir of SC composition and the component enriched in $^{22}$Ne, then, for each value of $\epsilon$ within a reasonable range, we calculate the abundance values $C_{\rm dust}(i)$ and $C_{\rm gas}(i)$ from Equations~\ref{eq:cdustprim} and \ref{eq:cgasprim}, and finally compute the goodness of linear fits to the data sets $A_{\rm dust}(f_d)$ and $B_{\rm gas}(f_g)$. 

We consider five models covering the main assumptions for the origin of the GCR gas reservoir, as summarised in Table~\ref{tab:model}. Noteworthy, the ISM phase(s) from which the GCR refractories are accelerated does not need to be specified at this point, since we assumed in first approximation that dust has the same composition in all phases and that the acceleration of dust grains does not depend on ambient conditions (temperature, density etc..). We will come back to the nature of the accelerated dust reservoir in the discussion, once we have established the origin of the GCR volatiles  (Section~\ref{sec:grain}). 

In Model 1, we assume that the GCR volatiles of SC composition are extracted from SNRs randomly distributed in the warm ISM, with 70\% in the WNM and 30\% in the WIM. We further assume that the high $^{22}{\rm Ne}/^{20}{\rm Ne}$ ratio of the GCR composition is due to the acceleration of $^{22}{\rm Ne}$-rich stellar winds in WTSs. In Model 2, the GCR volatiles are accelerated in SBs. In this model, the high $^{22}{\rm Ne}$ abundance in GCRs is explained by the enrichment of SB cores by the wind material of very massive stars before the first SN explosions. Model 3 also assumes that the GCR volatiles are produced in SBs, but here the ambient medium swept up by SN blast waves is supposed to be not significantly enriched in $^{22}{\rm Ne}$. Instead, the Ne isotopic ratio at the GCR source is explained by the acceleration of massive star winds in WTSs. In Model 4, we assume that all SNRs in the ISM contribute equally to the production of the GCR volatiles, whose source reservoir is thus a mixture of 60\% SB material, 28\% WNM and 12\% WIM (Section~\ref{sec:snr}). The $^{22}{\rm Ne}$-rich component in this model is taken to be accelerated in SB cores enriched by massive star wind material. Finally, Model 5 is similar to Model 4, except that the high abundance of GCR $^{22}{\rm Ne}$ is due to a contribution from particle acceleration in WTSs. 

In the first step of the analysis, we consider the SB temperature as a free parameter. Indeed, this quantity can vary between $\sim$ $10^6$~K and more than $10^7$~K from one SB to another, depending on the size of the parent OB association and the age of the SB \citep[see][]{kav20}, and the ionisation states of heavy elements are significantly different for $T_{\rm SB}=10^6$ and $10^7$~K (see Table~\ref{tab:data}). Figure~\ref{fig:chi2bidim} shows chi-square maps from linear fits of the data sets $A_{\rm dust}(f_d)$ and $B_{\rm gas}(f_g)$, as a function of the parameters $T_{\rm SB}$ and $\epsilon$, for Models 2 to 5 (in Model 1, the GCR volatiles are not produced in SBs). The GCR dust source composition slightly depends on the SB temperature through elements that are partially in solid form in the ISM, such as carbon and oxygen (see Eq.~\ref{eq:cdustprim}). However, the left panels of Figure~\ref{fig:chi2bidim} show that this dependence is weak. The best-fit values of the SB temperature reported in Table~\ref{tab:model} were obtained only from the GCR gas source data. We see that the results are consistent with $\log(T_{\rm SB}) \sim 6.6$ ($T_{\rm SB} \sim 4\times 10^6$~K) for all models. 

The best-fit values of $\epsilon$ are calculated from the weighted mean of the results obtained from the GCR gas and dust source data. They are in the range $\sim$ $10$--$50$ depending on the model, which confirm that refractory elements contained in dust grains are injected into the diffusive shock acceleration process much more efficiently that ions from ISM gas. The acceleration efficiencies of the various GCR components are further discussed in Section~\ref{sec:acceff}. 

Also given in Table~\ref{tab:model} is the relative contribution of the W.-R. wind reservoir to the GCR source gas population ($x_w$) for each model. In Models 3, 4, and 5, the required mixing depends on the fitted SB temperature, which affects the relative acceleration efficiency of ions within SBs. This is not the case in Model 2, because both the volatile material of SC composition and the W.-R. wind material are taken from the same reservoir. We see that the required mixing is of the order of $x_w \sim 50$\% in the SB model for the origin of the $^{22}{\rm Ne}$-rich GCR component, compared to $x_w \lsim 10$\% in the accelerated wind scenario. In addition, the minimum $\chi^2$ obtained from the GCR gas source data clearly favours a contribution of WTSs to the GCR production: $\chi^2_\nu = \chi^2_{\rm min}/\nu \lsim 1$ in Models 1, 3 and 5, compared to $\chi^2_\nu \cong 1.3$ in Models 2 and 4 ($\nu=24$ is the number of degrees of freedom). Unsurprisingly, the GCR dust source data do not allow to constrain the origin of the GCR volatiles: $\chi^2_\nu \approx 1$ for all models. 

\begin{figure}
\begin{center}
 \includegraphics[width=0.8\columnwidth]{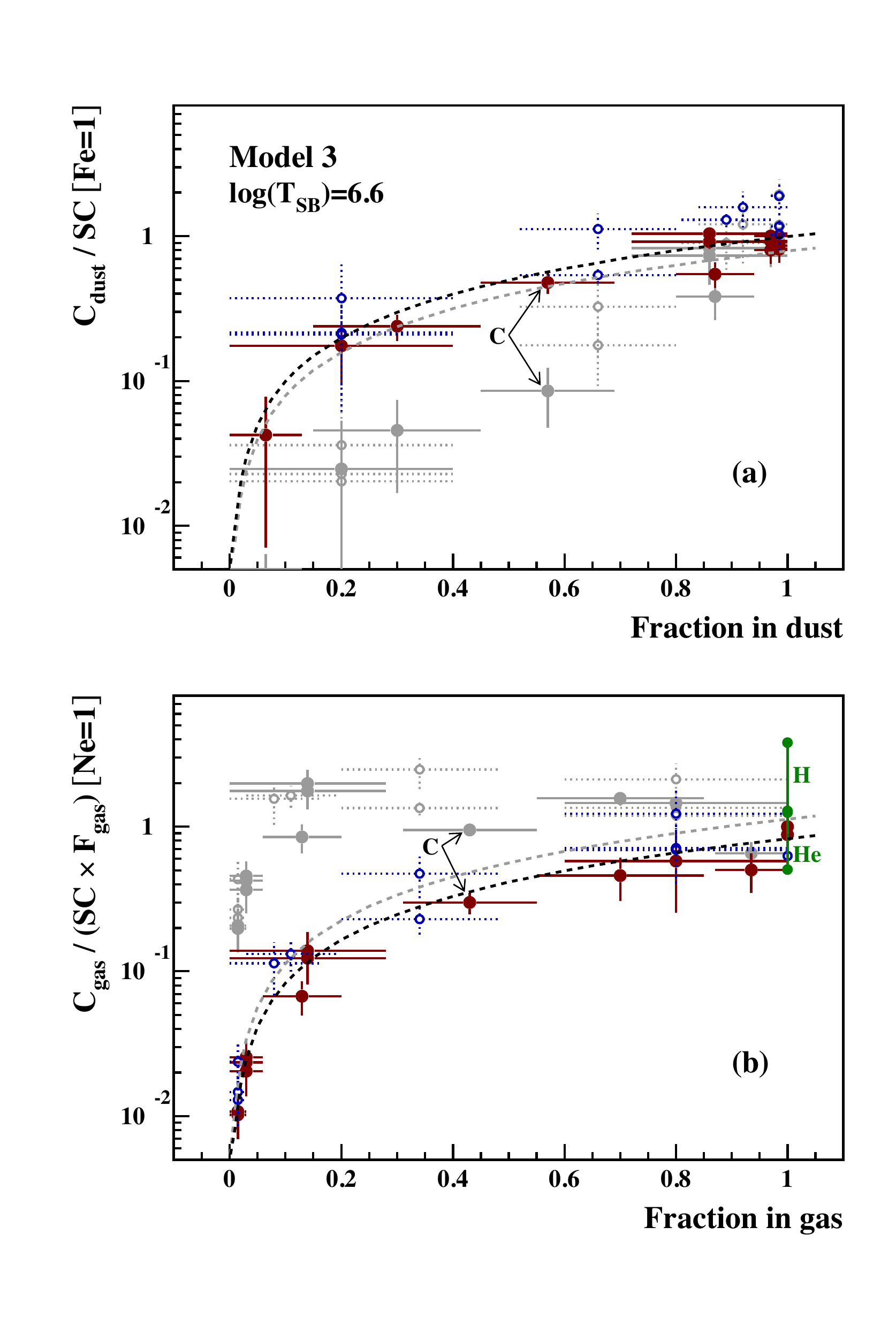}
 \caption{Normalised GCR abundance data for the dust (\textit{a}) and gas (\textit{b}) sources, as a function of the element fractions in ISM dust and gas, respectively. In the best-fit solution, the plotted quantities $A_{\rm dust}$ and $B_{\rm gas}$ (Eqs.~\ref{eq:a} and \ref{eq:b}) are expected to be proportional to $f_d$ and $f_g$, respectively (see Eqs.~\ref{eq:cdust} and \ref{eq:cgas}). Each data point corresponds to one element (e.g. C is at $f_d=1-f_g=0.57$; see Table~\ref{tab:data}) and the coloured symbols show the data obtained from the best linear fit (black dashed line) in Model 3 with $\log(T_{\rm SB})=6.6$ (fixed) and $\epsilon=20.2$ (see Table~\ref{tab:model}) -- red filled circles: GCR data from \citet{bos20,bos21}, blue open circles: GCR data from \citet{mur16}. The grey points and dashed curve show the data and best linear fit obtained with $\epsilon=1$ ($\chi^2=40.0$ and $235$ for the GCR dust and gas sources, respectively).  The H and He data in \textit{panel} (\textit{b}) are represented by two connected points obtained for the minimum GCR source energy $E_{\rm min}=100$~keV~nucleon$^{-1}$ (upper point) and $3$~MeV~nucleon$^{-1}$ (lower point). }
 \label{fig:depgcrmod3}
\end{center}
\end{figure}

\begin{figure*}
\begin{center}
 \includegraphics[width=0.65\textwidth]{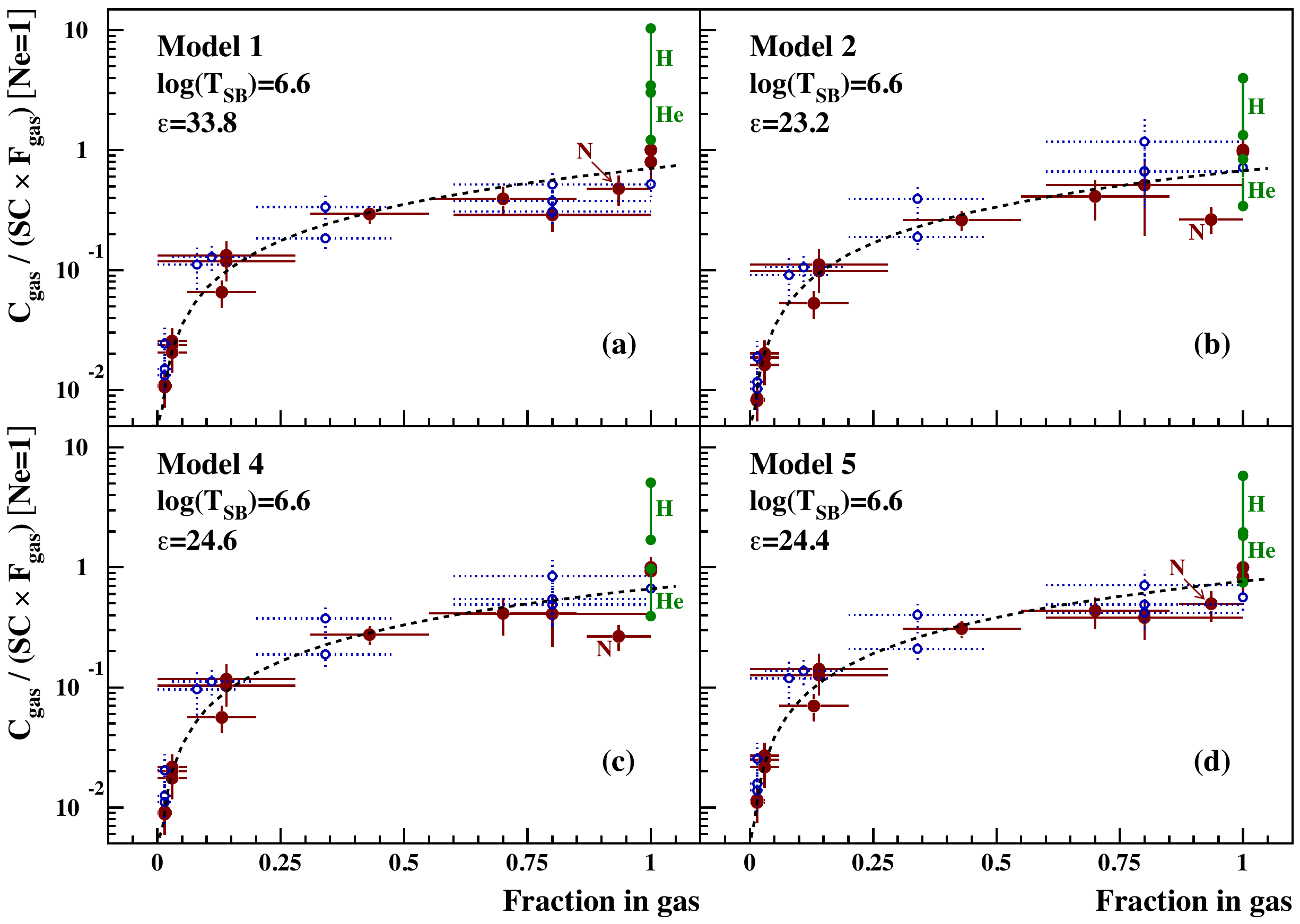}
 \caption{Normalised GCR abundance data for the gas source, as in Fig.~\ref{fig:depgcrmod3}b, for Models 1, 2, 4, and 5. The SB temperature is fixed to $\log(T_{\rm SB})=6.6$ and the best-fit values of $\epsilon$ are given in the Figure.}
 \label{fig:depgcrx4}
\end{center}
\end{figure*}

Next, we fix $\log(T_{\rm SB})=6.6$ in Models 2 to 5, such that there are now two free parameters: $\epsilon$ and a normalisation constant. Figure~\ref{fig:depgcrmod3} shows the quantities $A_{\rm dust}$ and $B_{\rm gas}$ calculated in Model 3 (the best model from a statistical point of view, see Table~\ref{tab:model}) as a function of $f_d$ and $f_g$, respectively, for two values of $\epsilon$: the best-fit value $\epsilon=20.2$ (Table~\ref{tab:model}) and $\epsilon=1$. With the latter value, the normalised GCR abundance data for the dust and gas sources are clearly not proportional to $f_d$ and $f_g$, respectively. For example, we find $A_{\rm dust}({\rm C})/A_{\rm dust}({\rm Fe})\approx0.09$ with $\epsilon=1$ (see Fig.~\ref{fig:depgcrmod3}), whereas $f_d({\rm C})/f_d({\rm Fe})=0.59\pm0.13$. For $\epsilon=1$, least-squares linear fits to the $A_{\rm dust}(f_d)$ and $B_{\rm gas}(f_g)$ data sets return $\chi^2_\nu = 1.5$ and $9.0$, respectively, compared to $\chi^2_\nu = 1.0$ and $0.51$ for $\epsilon=20.2$ (note that $\nu=26$ when both $T_{\rm SB}$ and $\epsilon$ are fixed). 

In Figure~\ref{fig:depgcrx4} we show the best-fit normalised abundances at the GCR gas source for the four other models. These abundances are roughly proportional to the elemental fraction in gas, but there are nevertheless significant differences for some elements from one model to another, which explains the different values of $\chi^2_{\rm min}$ (Table~\ref{tab:model}). In particular, we see that the N abundance is much lower in Models 2 and 4 than in Models 1 and 5 (and than in Model 3; see Fig.~\ref{fig:depgcrmod3}). This is because in the SB model for the origin of GCR $^{22}{\rm Ne}$ (Models 2 and 4) the SB cores are strongly enriched in $^{14}$N expelled by massive stars before their W.-R. WC and WO phases, and the higher value of $f_w({\rm N})$ has the effect of decreasing $B_{\rm gas}({\rm N})$ (see Eq.~\ref{eq:b}). The $^{14}{\rm N}/^{22}{\rm Ne}$ ratio in the SB wind composition amounts to 10.8, compared to 3.1 in the accelerated wind composition (Models 1, 3 and 5). The $^{14}$N abundance is lower in the latter composition, because the wind mechanical power and thus the wind acceleration efficiency are lower during the stellar phases preceding the W.-R. WC and WO phases (Section~\ref{sec:22newind}).

We also see in Figures~\ref{fig:depgcrmod3} and \ref{fig:depgcrx4} that the H and He normalised abundances vary significantly from one model to another. These two elements were included in the fitting procedure by using the mean of the GCR source abundances calculated with $E_{\rm min}=100$~keV~nucleon$^{-1}$ and $3$~MeV~nucleon$^{-1}$ (see Table~\ref{tab:data}) as the central value, and the difference of the abundances obtained with these two values of $E_{\rm min}$ as the error. Whereas the He normalised abundance appears to be in fairly good agreement with expectation, $B_{\rm gas}({\rm He})/B_{\rm gas}({\rm Ne})\approx1$ within a factor of two, the H normalised abundance is systematically too high, especially in Model 1: $B_{\rm gas}({\rm H})/B_{\rm gas}({\rm Ne})=6.9$. The reason for this is that in the warm ISM, the acceleration efficiency of H is generally lower than that of heavier elements, which have higher mass-to-charge ratios (see Table~\ref{tab:data}). Thus, we have $f_{A/Q}^{\rm SC}({\rm H}) / f_{A/Q}^{\rm SC}({\rm Ne}) =0.15$ in Model 1, compared to $f_{A/Q}^{\rm SC}({\rm H}) / f_{A/Q}^{\rm SC}({\rm Ne}) =0.43$ in Models 2 and 3, and the normalised abundance $B_{\rm gas}$ increases as $f_{A/Q}^{\rm SC}$ decreases (Eq.~\ref{eq:b}). This result on the H abundance is one of the reasons why Model 3 (where the GCR volatiles are accelerated in SBs) is favoured. But other elements also support this conclusion. Thus, excluding the H and He GCR data from the analysis, we obtain $\chi^2_{\rm min}{\rm (GCR~gas~source)}=18.9$, $29.9$, $11.4$, $29.7$, and $14.7$ for Models 1 to 5, respectively. 

\begin{figure}
\begin{center}
 \includegraphics[width=0.83\columnwidth]{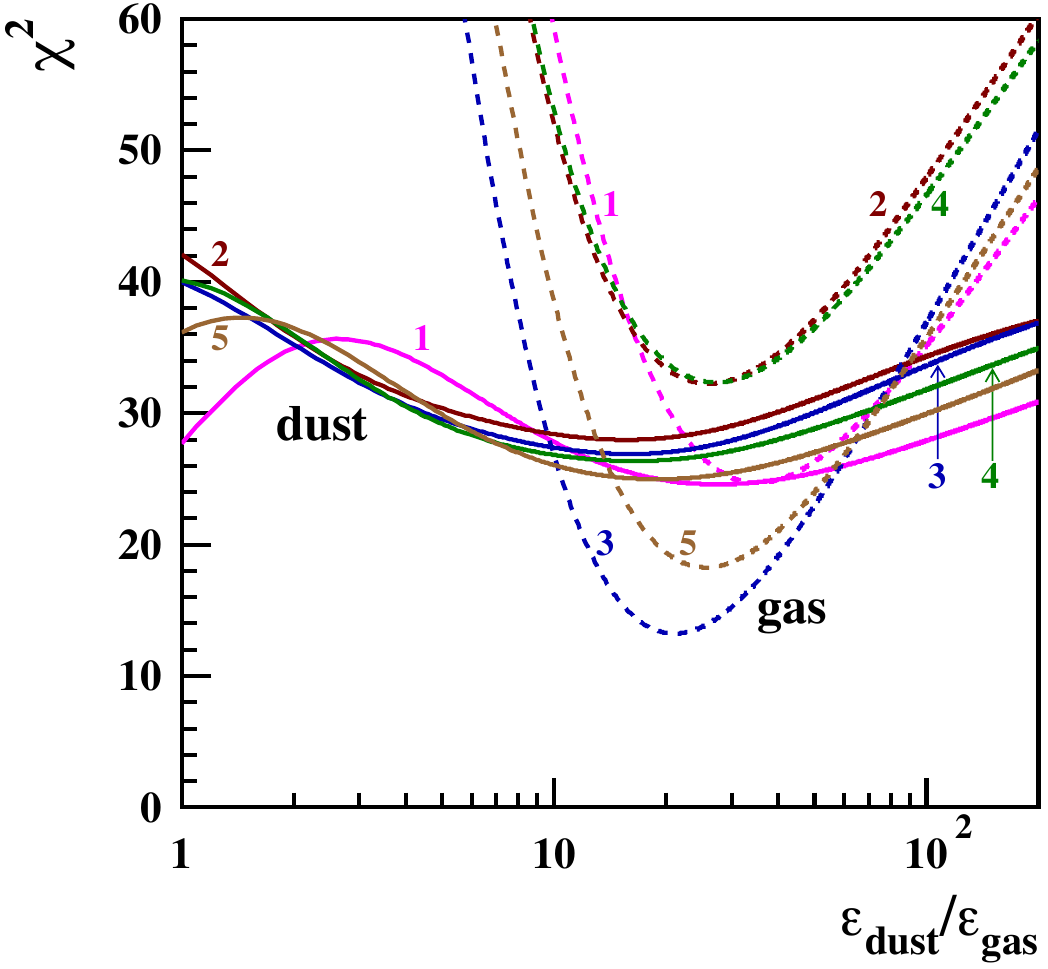}
 \caption{Chi-square values as a function of the relative efficiency $\epsilon=\epsilon_{\rm dust}/\epsilon_{\rm gas}$ when the SB temperature is fixed to $\log(T_{\rm SB})=6.6$. Solid lines: GCR dust source; dashed lines: GCR gas source. Numbers between $1$ and $5$ refer to the models summarised in Table~\ref{tab:model}.}
 \label{fig:chi2}
\end{center}
\end{figure}

Figure~\ref{fig:chi2} shows calculated chi-square values as a function of $\epsilon$. We see that the GCR dust and gas source data provide consistent values for the best-fit relative efficiency. Thus, in Model~3, $\epsilon=15.9_{-10.4}^{+25.6}$ for the dust source and $21.0_{-5.9}^{+9.7}$ for the gas source. The weighted mean of the best-fit values obtained from the GCR dust and gas source data are reported in Table~\ref{tab:model}.   

\begin{figure*}
\begin{center}
 \includegraphics[width=0.75\textwidth]{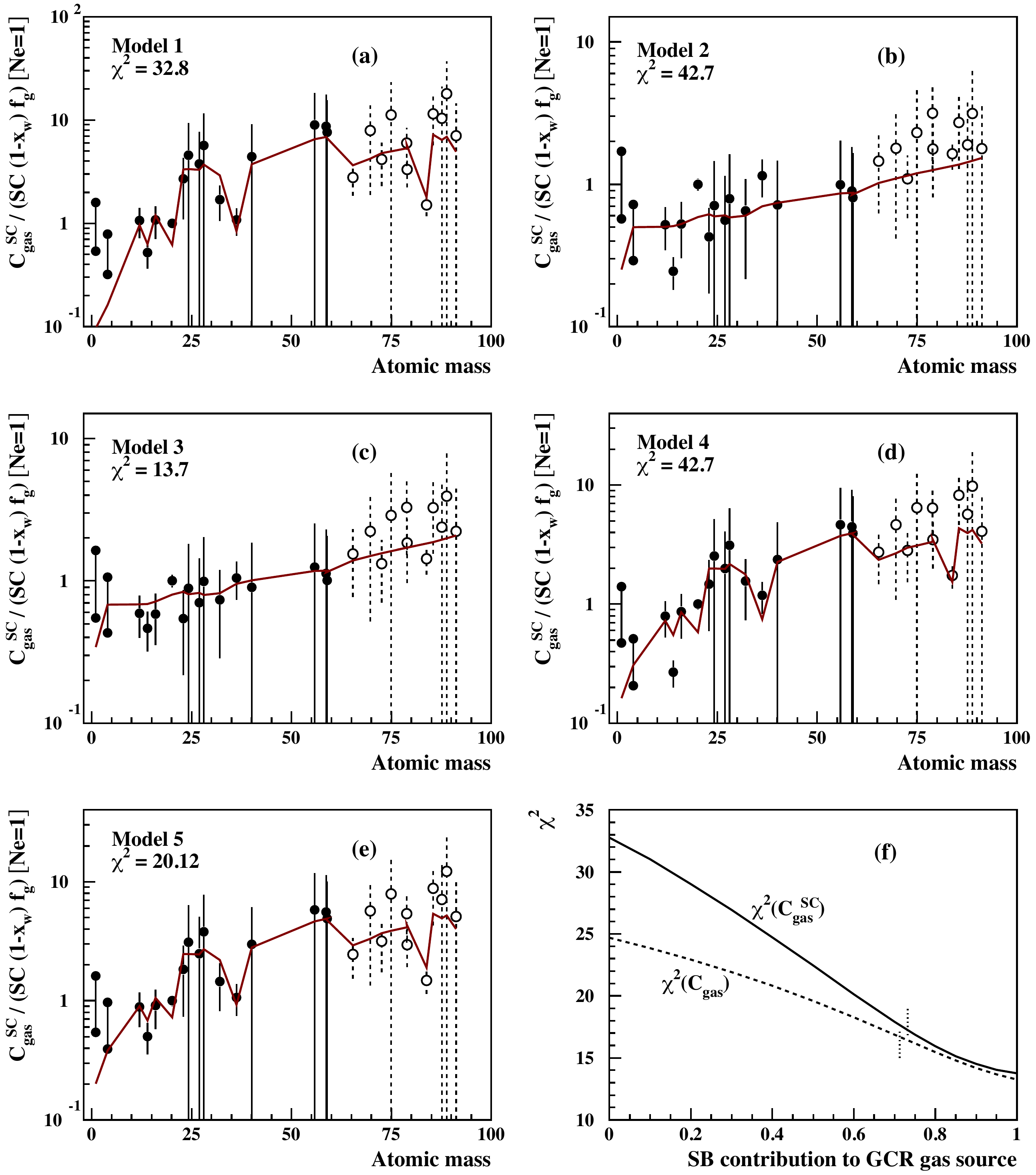}
 \caption{\textit{Panels (a)~--~(e)}: Normalised GCR data for the gas source of SC composition as a function of atomic mass. The plotted data are $B_{\rm gas}^{\rm SC}(i)$ calculated from Equation~(\ref{eq:bsc}), with $\log(T_{\rm SB})=6.6$ (fixed) and the best-fit value of $\epsilon$ given in Table~\ref{tab:model} -- filled circles: GCR data from \citet{bos20,bos21}, open circles: GCR data from \citet{mur16}. The H and He data are represented by two connected points obtained for the minimum GCR source energy $E_{\rm min}=100$~keV~nucleon$^{-1}$ (upper point) and $3$~MeV~nucleon$^{-1}$ (lower point). The red solid line shows the normalised mass-to-charge ratio factor, $p f_{A/Q}^{\rm SC}$, where $p$ is the best-fit normalisation of $f_{A/Q}^{\rm SC}$ to the data; the corresponding chi-square value is given in each panel. \textit{Panel (f)}: Chi-square values as a function of the SB contribution to the GCR gas source ($a_{\rm SB}$; see Eq.~\ref{eq:aonq}), in a model where the reservoir of SC composition is a mixture of WNM, WIM and SB gas, and the $^{22}$Ne-rich component is produced in WTSs. Solid line: $\chi^2$ calculated from the comparison of $B_{\rm gas}^{\rm SC}$ and $f_{A/Q}^{\rm SC}$, as in panels (a)~--~(e); dashed line: $\chi^2$ calculated from the comparison of $B_{\rm gas}$ and $f_g$, as in Figs.~\ref{fig:depgcrmod3}b and \ref{fig:depgcrx4} (see also Table~\ref{tab:model}). The two vertical, dotted line segments at $a_{\rm SB} \sim 0.72$ mark the $1\sigma$ limits for a fit with three free parameters ($a_{\rm SB}$, $\epsilon$ and $p$) at $\chi^2_{\rm min}+3.5$.}
 \label{fig:gcrsc}
\end{center}
\end{figure*}

In panels (a)~--~(e) of Figure~\ref{fig:gcrsc}, we show normalised GCR data for the gas source of SC composition:
\begin{equation}
B_{\rm gas}^{\rm SC}(i) = C_{\rm gas}^{\rm SC}(i) / [{\rm SC}(i) (1-x_w) f_g(i)]~,
\label{eq:bsc}
\end{equation}
where the abundances $C_{\rm gas}^{\rm SC}(i)$ are calculated from Equation~(\ref{eq:cgassc}). By construction, $B_{\rm gas}^{\rm SC}$ is expected to be proportional to the mass-to-charge ratio factor $f_{A/Q}^{\rm SC}$ (see Eq.~\ref{eq:cgas}), which is shown by red solid lines in the Figure. The pattern of $f_{A/Q}^{\rm SC}$ is more irregular in Models 1, 4, and 5 than in Models 2 and 3, because the effects of atomic shell closure on the ionisation states are more pronounced in the warm ISM than in the hot SB medium. This is particularly obvious for the noble gases Ne ($A=20.2$), Ar ($A=36.3$), and Kr ($A=83.8$). This data representation again highlights the goodness of Model 3, in which the GCR volatiles are produced in SBs and the $^{22}$Ne-rich GCR component comes from particle acceleration in WTSs (note the difference in N abundance between Models 2 and 3). The second best model is Model 5, which is similar to Model 3 except that SNRs in the warm ISM also contribute to the production of GCR volatiles.  

Panel~(f) of Figure~\ref{fig:gcrsc} shows $\chi^2$ values calculated by two methods, as a function of the SB contribution to the GCR gas source, $a_{\rm SB}$ (see Eq.~\ref{eq:aonq}). Models 1, 3, and 5 correspond to $a_{\rm SB}=0$, $1$, and $0.6$, respectively. The dashed line in Figure~\ref{fig:gcrsc}f shows $\chi^2$ calculated from the comparison of $B_{\rm gas}$ and $f_g$, and the solid line shows $\chi^2$ calculated from the comparison of $B_{\rm gas}^{\rm SC}$ and $f_{A/Q}^{\rm SC}$. We can conclude from this Figure that, at the $1\sigma$ level, SNRs in the warm ISM do not contribute to the GCR volatile composition for more than 30\%. 

\section{Discussion}
\label{sec:discussion}

\begin{figure*}
 \includegraphics[width=0.9\textwidth]{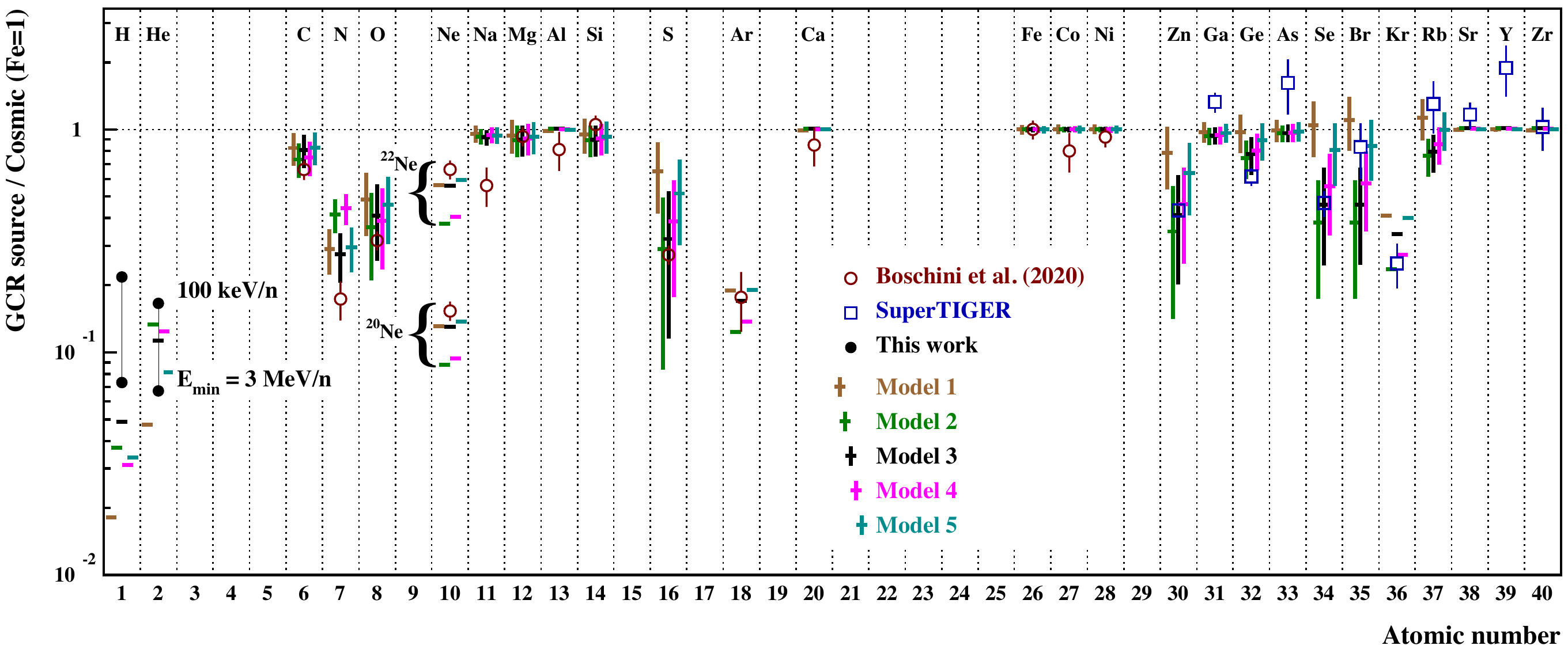}
 \caption{GCR source abundances relative to the SC composition, as in Fig.~\ref{fig:abund}, except that the abundances of $^{20}$Ne and $^{22}$Ne are shown separately and that the measured abundances are compared to those predicted in the five models. Error bars on the model points take into account the uncertainties on the element fraction in ISM dust (see Table~\ref{tab:data}).}
 \label{fig:abund_model}
\end{figure*}

The recent GCR abundance data obtained from AMS-02, Voyager~1 and SuperTIGER measurements shed a new light on the origin of CRs in the ISM. We have developed a new method of analysis of the GCR data that allows us to specify the nature of the source reservoirs of these particles. Our model explains well the measured abundances of all primary and mostly primary CRs from H to Zr (see Figure~\ref{fig:abund_model}), including the overabundance of $^{22}$Ne, which has remained problematic until now. We showed that the GCR source abundances of H and He relative to the SC composition are similar to those of the other volatile elements N, Ne and Ar, provided that the minimum CR source energy is of the order of a few hundred keV~nucleon$^{-1}$ (Section~\ref{sec:data}). We found that the GCR volatiles are mostly accelerated from a plasma $\gsim 2$~MK, which is typical of the hot medium found in SBs (see Section~\ref{sec:superbubble} below). We also found that the overabundance of $^{22}$Ne in GCRs is most likely due to a small contribution of particle acceleration in WTSs of massive stars (Section~\ref{sec:wind}). Finally, we confirmed that the GCR refractories are significantly overproduced compared to the volatile elements (by a factor of $\epsilon \sim 20$), which suggests than dust grains are injected into the diffusive shock acceleration process with a higher efficiency than ions (Section~\ref{sec:grain}). As already discussed in Section~\ref{sec:data}, the accelerated refractories most likely come from various dust grains of the ISM mix, and not only from core-collapse SN grains as suggested by \citet{lin19}.

Recently, \citet{eic21} independently proposed a similar model to explain the GCR composition, taking into account the ionisation states of elements contained in ISM gas swept-up by SN shocks, and the fraction of each element locked in dust grains. But contrary to the main result of the present work, they found that GCR acceleration in the WIM, including HII regions, provides a very good description of the GCR composition data. 

A first difference between the two approaches is due to the GCR abundance data themselves. For elements heavier than He, \citet{eic21} mainly used the observations of the Cosmic Ray Isotope Spectrometer (CRIS) on board the Advanced Composition Explorer (ACE) satellite \citep{isr18}. For H and He, they used the \textsc{GalProp} results presented in the Voyager 1 paper of \citet{cum16}, which give, for the diffusive reacceleration model of GCR propagation: ${\rm H/Fe}=221 \pm 3$ and ${\rm He/Fe}=77.6 \pm 1.0$ (the results obtained with the plain diffusion models are similar). In comparison, by integration of the source spectra obtained in Section~\ref{sec:data}, we find: $1940 < {\rm H/Fe} < 7390$ and $172 < {\rm He/Fe} < 546$, depending on the minimum CR source energy (see Table~\ref{tab:data}). Higher abundances of GCR protons and $\alpha$ particles relative to heavy elements favour an origin of these particles in the hot ISM, where the $A/Q$ selection effect is less important (Table~\ref{tab:data}). 

We also note that \citet{eic21} did not consider in their model the strong overabundance of $^{22}$Ne in the GCR composition, which provides evidence for a significant contribution of W.-R. wind material in the GCR source gas population. As discussed in Section~\ref{sec:22ne}, W.-R. stars are mainly found in massive star clusters and SBs \citep[see also, e.g.,][]{hig03}, so that GCR acceleration only in the WIM cannot explain the measured $^{22}{\rm Ne}/^{20}{\rm Ne}$ ratio. 

Another significant difference between the present model and that of \citet{eic21} is that these authors did not consider the photoionisation precursor of SNR blastwaves, which can modify the ionisation state of the particles entering the shock (Section~\ref{sec:ioni}). This additional photoionisation of the preshock plasma is not important for SNR shocks propagating in the hot ISM, but it should be taken into account for shocks in warm ISM environments. 

\subsection{Origin of the GCR volatiles in Galactic superbubbles}
\label{sec:superbubble}

The temperature of the GCR volatile reservoir can be compared with the plasma temperature inside a SB according to the standard wind bubble theory \citep{wea77,mac88}: 
\begin{equation}
T_{\rm SB} \simeq (4.3 \times 10^6~{\rm K}) t_{\rm Myr}^{-6/35} N_{*,30}^{8/35} n_{\rm H}^{2/35} \times (1-x)^{2/5}~,
\label{eq:tsb}
\end{equation}
where $t_{\rm Myr}$ is the time in units of Myr after onset of massive star formation (assumed to be coeval for all stars), $N_{*,30}=N_*/30$ where $N_*$ is the number of massive stars in the 8--120~\msun~mass range, $n_{\rm H}$ is the mean H number density in the external ISM in units of cm$^{-3}$, and $x=r/R_{\rm SB}$ is the relative distance from the SB centre ($R_{\rm SB}$ being the SB radius). The SB temperature is often given as a function of the stellar wind and SN mechanical power, $L_w$, instead of the number of massive stars \citep[e.g.][]{mac88,par04}. But Equation~(\ref{eq:tsb}) uses the  result of \citet{vos09} that the mean power per star from a coeval population of massive stars is nearly constant with time and amounts to $\approx 1.5 \times 10^{36}$~erg~s$^{-1}$. 

When compared to Equation~(\ref{eq:tsb}), the fitted value of $T_{\rm SB}$ and associated $1\sigma$ error (i.e. for model~3 $\log(T_{\rm SB}) > 6.45$ corresponding to $T_{\rm SB} > 2.8 \times 10^6$~K; see Table~\ref{tab:model}) do not provide meaningful information on the characteristic size of the parent massive star clusters. Thus, the detected GCRs could be accelerated in SBs from relatively modest OB associations, such as the nearby Scorpius-Centaurus association, which is thought to have formed the Local Bubble surrounding the solar system from the explosion of 14-20 SNe during the past $\sim 13$~Myr \citep[][]{bre16}. But the fitted value of $T_{\rm SB}$ is also consistent with a GCR production in prominent clusters of young massive stars, such as the Cygnus~OB2 and Westerlund~1 clusters, which contain hundreds of massive stars with ages between 3 and 6~Myr \citep[see][and references therein]{aha19}. 

However, a significant result of the present analysis is that SNRs in the warm ISM contribute to the GCR volatile composition for less than 28\% ($1\sigma$ limit), whereas about 40\% of Galactic SNe occur in this phase and not in SBs  (Section~\ref{sec:snr}). This could be explained by the abundance of neutral atoms in the warm ISM, which can have profound effects on the diffusive shock acceleration process \citep{blasi12,morlino12,ohi12,mor13}. Estimating the ionisation fractions in the photoionisation precursors of SNR blastwaves with a shock speed of 1000~km~s$^{-1}$ (Section~\ref{sec:ioni}), we estimated that in the WNM nearly 50\% of the H atoms cross the shock front as neutral particles (see Table~\ref{tab:data}). In the WIM, H is fully ionised in the preshock region, but 60\% of He is neutral. 

The most important effect induced by neutral particles in the shock acceleration region is the so-called neutral return flux \citep[NRF;][]{blasi12}. It occurs when a neutral atom crossing the collisionless shock undergoes a charge exchange with a hot H$^+$ ion downstream of the shock, thus producing a hot neutral hydrogen that may cross the shock front again and deposit some of its kinetic energy in the upstream medium. This has the effect of heating the upstream plasma, thus reducing the shock Mach number and the particle acceleration efficiency. For a neutral fraction of $\sim 50$\% as expected for a shock propagating in the WNM, the effects of the NRF are most significant for shock velocities $V_s \lsim 3000$~km~s$^{-1}$ \citep[][and references therein]{mor13}. The effects are less important for higher shock speeds, because the relative velocity between neutrals and ions is high enough then that the cross section for H-H$^+$ charge exchange becomes lower than that for H ionisation. The effects are also expected to be less important for a shock propagating in the WIM, because the overall neutral fraction in photoionisation precursors is lower ($f_N\approx6$\%) and  the cross section for charge exchange between He and H$^+$ is relatively small. 

To provide a first quantitative estimate of the NRF effect on the GCR composition, we consider a simple model where the diffusive shock acceleration efficiency is independent of shock speed as long as $V_s$ is higher than a minimum value $V_{s, {\rm min}}^k$, which depends on the ISM phase $k$. CR volatiles are assumed to be accelerated with the same efficiency in all ISM phases as long as $V_s \geq V_{s, {\rm min}}^k$. We take $V_{s, {\rm min}}^{\rm WNM}=3000$~km~s$^{-1}$ in the WNM to account for the NRF effect and $V_{s, {\rm min}}^{\rm WIM}=300$~km~s$^{-1}$ in the WIM assuming that the NRF effect is not important in this phase and that SNR shocks slower than $300$~km~s$^{-1}$ should be radiative and relatively ineffective in accelerating CRs \citep{raymond20a, raymond20b}. For SNRs in SBs, one has to take into account that the blast waves are expected to become subsonic in the hot plasma before reaching the radiative phase \citep{par04}. We take $V_{s, {\rm min}}^{\rm SB}=\mathcal{M}_{S, {\rm min}}c_S\approx600$~km~s$^{-1}$, where $\mathcal{M}_{S, {\rm min}}\approx 3$ is the minimum sonic Mach number for efficient diffusive shock acceleration \citep[see, e.g.,][]{ptu03} and $c_S \cong 200$~km~s$^{-1}$ is the sound velocity in a plasma of temperature $T_{\rm SB}=4 \times 10^6~{\rm K}$. 

The relative contribution of the ISM phase $k$ to the GCR volatile production can readily be estimated from:
\begin{equation}
a_k=\frac{f_k M^k_{\rm s.-u.}}{\sum_k' f_k' M^{k'}_{\rm s.-u.}}~, 
\label{eq:ak}
\end{equation}
where $f_k$ is the fraction of Galactic SNe exploding in $k$ ($f_{\rm SB}=0.6$, $f_{\rm WNM}=0.28$, and $f_{\rm WIM}=0.12$; see Section~\ref{sec:snr}) and $M_{\rm s.-u.}^k$ is the mass of ISM gas swept up by a SNR shock until its velocity slows down to $V_{s, {\rm min}}^k$. The swept-up masses can be estimated by exploiting the constancy of $M^{\rm s.-u.}_k V_s^2$ during the adiabatic, Sedov-Taylor stage of the shock evolution \citep{mck89}: 
\begin{equation}
M_{\rm s.-u.}^{k} \approx 68~M_\odot \bigg(\frac{E_{\rm SN} }{10^{51} {\rm~erg}}\bigg) 
\bigg(\frac{V_{s, {\rm min}}^k}{1000 {\rm~km~s^{-1}}}\bigg)^{-2}~, 
\label{eq:sumass}
\end{equation}
where $E_{\rm SN} \approx 10^{51} {\rm~erg}$ is the total kinetic energy of a SN outburst. We then obtain from Equation~(\ref{eq:ak}) and the values of $V_{s, {\rm min}}^k$ discussed above: $a_{\rm SB}=0.55$, $a_{\rm WNM}=0.01$, and $a_{\rm WIM}=0.44$. Using these values to fix the relative contributions of the ISM phases to the GCR volatile production and assuming that the high abundance of GCR $^{22}{\rm Ne}$ is due to a contribution from particle acceleration in WTSs, the model returns a minimum $\chi^2$ from the comparison of $B_{\rm gas}$ and $f_g$ of $24.0$ ($\log(T_{\rm SB})$ being fixed to $6.6$), which is significantly worse than the result of Model 3: $\chi^2_{\rm min}=13.2$ (Table~\ref{tab:model}). This suggests that the limitation of GCR production in the WNM due to the NRF effect is not enough to explain why the bulk of the GCR volatiles come from SBs. However, the relative contributions of SNRs in SBs and in the WIM depend directly on the values of $V_{s, {\rm min}}^{\rm SB}$ and $V_{s, {\rm min}}^{\rm WIM}$, which are uncertain. 

\subsection{Origin of the {22}-Ne-rich GCR source}
\label{sec:wind}

The high GCR $^{22}$Ne/$^{20}$Ne ratio suggests that a moderate fraction of CRs originate from acceleration in massive star WTSs. The popular SB model for the origin of GCR $^{22}$Ne, which assumes that SNR shocks in SBs propagate in a medium enriched by W.-R. winds from the most massive stars of the parent OB association \citep{hig03,bin05,bin08,lin19} appears to be unlikely in the light of our results. First, massive stars loose large amounts of $^{14}$N during the late main sequence, red supergiant and W.-R. WN phases, but the GCR composition is not remarkably enriched in nitrogen. The N/Ne ratio in the GCR source composition is $0.92\pm0.21$ times that in the SC composition, whereas this ratio in the mean composition of massive star winds is about five times that in the local ISM (Table~\ref{tab:data}). This overabundance of $^{14}$N in the winds of massive stars compared to the GCR source composition has already been noticed by \citet{bin05}, who suggested that it could be explained by a mass dependence of the injection efficiency of ions into the diffusive shock acceleration process (N would then be less efficiently accelerated than Ne). But the acceleration efficiency is in fact proportional to the particle rigidity (\citealp{ell97}; see also \citealp{cap17} and \citealp{han19}), that is on its mass-to-charge ratio, which is similar for N and Ne in a hot SB plasma (see Table~\ref{tab:data}). 

Another problem faced by the SB model for the origin of GCR $^{22}$Ne is that it requires that the SB gas is strongly enriched in W.-R. wind material, at the level of $x_w \approx 50$\% (Table~\ref{tab:model}), which is not supported either by theory or by observations. In the classical wind bubble theory \citep{mac88}, the bulk of the material in the SB interior is provided by conductive evaporation from the cold outer shell of the swept-up ISM, not by stellar winds. Moreover, such a level of mixing would imply that SB gas has a highly non-solar composition, which is not supported by X-ray observations. The latter show on the contrary that SB plasmas have a metallicty close to the ISM average \citep[see][and references therein]{kav20}. 

Models that assume the acceleration of $^{22}$Ne-rich stellar winds in WTSs provide a much better description of the GCR data (compare, e.g., $\chi^2_{\rm min}$(GCR gas source) for Models 2 and 3 in Table~\ref{tab:model}). The $^{14}{\rm N}/^{22}{\rm Ne}$ abundance ratio is lower in the accelerated wind composition ($3.1$ compared to $10.8$ in the SB wind composition), because $^{14}{\rm N}$ is expelled during stellar phases where the wind mechanical power and thus the wind acceleration efficiency are lower than during the W.-R. WC and WO phases. In the best-fit model (Model~3), where the volatile elements of the GCR are mainly coming from SBs, the relative contributions of WTSs and SN shocks to the GCR source gas population is $x_w=5.9$\% and $x_{\rm SC}=94.1$\%. The required contribution of accelerated wind material is that low, because (i) the  $^{22}{\rm Ne}/^{20}{\rm Ne}$ ratio is high in this composition ($1.56$ compared to $0.317$ in the GCR source composition) and (ii) heavy ions being only partially ionised in stellar winds, they are injected with a higher magnetic rigidity in WTSs than in SN shocks propagating in SBs (see the mass-to-charge ratios in Table~\ref{tab:model}). From the global energy budget of strong shocks in massive star clusters, \citet{gup20} estimated that WTSs should contribute at least 25\% of the total GCR production in these objects. Our analysis suggests that this estimate is somewhat too high. 

The GCR composition data do not allow us to distinguish if the $^{22}$Ne-rich material is accelerated in winds of individual massive stars born in loosely-bound clusters or in collective WTSs formed by the overlap of stellar wind bubbles in massive and compact clusters \citep[see][]{gup20}. However, the spectrum of these particles at high energies could differentiate the two scenarios, as powerful compact clusters could produce CRs of PeV energies  \citep{mor21}, whereas WTSs from individual stars could not. 

\subsection{Origin of the GCR refractories from dust grains}
\label{sec:grain}

Having assumed in first approximation that dust has the same composition in all phases of the ISM, our GCR data analysis does not provide any quantitative constraints on the ISM phase(s) from which the refractory elements are accelerated. However, having found that the GCR volatiles are primarily accelerated in SBs, it is worth studying if the GCR refractories could also come from SBs. As already discussed in Section~\ref{sec:depletion}, the amount and nature of dust contained in SBs is very uncertain. Multiwavelength observations of the prototypical Orion-Eradinus SB have shown that it is a complex structure composed of a series of nested shells associated with successive bursts of stellar activity \citep{och15}. Dust is continuously replenished in the SB interior through thermal evaporation of embedded molecular clouds swept up by SN shocks. The typical timescale between two successive SN explosions from an OB association is $\lsim 1$~Myr \citep{par04,bre16}. During this time, carbonaceous and silicate grains mixed into the hot plasma suffer thermal sputtering at a rate of $\sim 10^{-5}$~--~$10^{-4}~\angstrom$~yr$^{-1}$ \citep{tie94}. Thus, large silicate grains of typical size $a_{\rm peak} \approx 140$~nm \citep{jon17}, bearing Mg, Al, Si and other refractory elements from the iron group, should not be strongly altered before being swept-up by a following SN shock. But nano carbonaceous grains and polycyclic aromatic hydrocarbon (PAH) molecules can be destroyed by thermal sputtering, thus returning C atoms to the gas phase. This suggests that a careful study of the C abundance at the GCR source might be key to distinguishing whether the fast refractory elements are produced in SBs or in the warm ISM. 

UV measurements of elemental abundances in a high-velocity ionised gas along a line of sight through the Orion-Eradinus SB found  Al, Si, and Fe to be significantly depleted relative to C  \citep{wel02}, which appears to be consistent with the scenario sketched above. Depletions of highly refractory elements in SB plasmas might also be detectable with the forthcoming X-Ray Imaging and Spectroscopy Mission \citep{xrism}.

The grain acceleration model of \citet{ell97,ell98} was implicitly worked out for the case of a SNR in the warm ISM (with $n_{\rm H} \sim 1$~cm$^{-3}$ and $T \sim 10^4$~K as standard parameters). In this model, the refractory element injection rate into the diffusive shock acceleration process is estimated from the ratio of the shock acceleration timescale for a nonrelativistic grain, $t_{\rm acc}$, to the grain destruction timescale for (nonthermal) collisional sputtering, $t_{\rm sput}$. The latter is expected to be proportional to the momentum-loss timescale resulting from direct collisions of the grains with the ambient gas,  $t_{\rm sput} \sim 5~$~--~$10~t_{\rm loss}$, depending on the grain composition. The refractory element injection rate can then be written as \citep{ell97}:
\begin{equation}
q_{\rm refra} \sim O(10^{-4}) n_G A_G V_s \frac{t_{\rm acc}}{t_{\rm loss}}~,
\label{eq:grain1}
\end{equation}
where $n_G$ is the number density of dust grains in the shock upstream medium, $A_G$ the grain mean atomic weight, and $V_s$ the shock speed, such that the quantity $n_G A_G V_s$ represents the flux of nucleons contained in grain material coming from far upstream. In this equation, both $t_{\rm acc}$ and $t_{\rm loss}$ should be taken at the time when the grain velocity has reached its maximum value $\beta_{G,\rm max}c$ ($c$ is the speed of light). 

In the warm ISM, the maximum grain velocity is set by the coincidence of the acceleration timescale and the momentum-loss timescale, i.e. $t_{\rm acc} \sim t_{\rm loss}$, such that the refractory element injection efficiency is of the order of $10^{-4}$ \citep{ell97}. But in the hot and diluted SB plasma, the grain collisional loss timescale can be longer than the acceleration timescale, and even longer than the duration of the acceleration phase at the SNR shock,  $t_{\rm SNR} \sim 10^5$~yr \citep[see][]{par04}: 
\begin{equation}
t_{\rm loss} \approx 6 \times 10^5 \bigg(\frac{a}{0.1~{\rm \mu m}}\bigg)  \bigg(\frac{\mu_G}{25}\bigg) \bigg(\frac{n_{\rm SB}}{0.01 \rm~cm^{-3}}\bigg)^{-1} \bigg(\frac{\beta_{G,\rm max}}{0.034}\bigg)^{-1}~{\rm yr}~, 
\label{eq:grain2}
\end{equation}
where $a$ is the characteristic size of the grains and $\mu_G$ their mean atomic weight. Here, the maximum grain velocity, $\beta_{G,\rm max} \sim 0.034$, was estimated by equating the grain acceleration timescale \citep[][eq.~13]{ell97} to the age of the SNR at the end of the  acceleration phase, $t_{\rm SNR}$. Thus, the refractory element injection rate is expected to be less efficient in SBs than in the warm and denser ISM by a factor $t_{\rm loss} / t_{\rm SNR}$ of the order of $6$ (depending on the grain properties and the ambient medium density; see Eq.~\ref{eq:grain2}). This reduce efficiency is due to the longer timescale for grain sputtering in SBs, which is partly compensated by the longer lifetime of SNRs and the higher velocity reached by the accelerated grains. 

GCR refractories may also be significantly produced in the warm ISM. As we found that SNe exploding outside SBs can contribute up to 30\% to the GCR volatile composition  (Section~\ref{sec:results}), these objects may in fact be the main source of the fast refractory elements, depending on the relative efficiencies of dust acceleration in the different ISM phases. However, shock acceleration of dust grains in the WNM should be limited by the NRF effect (see Section~\ref{sec:superbubble}). In the model of \citet{ell97,ell98}, dust grains interact with the same magneto-hydrodynamic waves as very energetic protons, so the lack of those waves due to ion-neutral damping should inhibit acceleration of grains and injection of refractory elements into the CR population. More work is needed to refine the grain acceleration model in the light of current knowledge about both interstellar dust and diffusive shock acceleration, and apply it to the different phases of the ISM. 

\subsection{Acceleration efficiency}
\label{sec:acceff}

The efficiency of GCR acceleration at their sources can be estimated from the $\gamma$-ray luminosity of the Milky Way related to CR propagation and interaction in the ISM. Using \textit{Fermi}-LAT $\gamma$-ray data and the \textsc{GalProp} code, \citet{str10} found that the total injected kinetic power in CR protons between 0.1 and 100~GeV is $\approx 7 \times 10^{40}$~erg~s$^{-1}$. With the proton source spectrum derived in Section~\ref{sec:data}, the corresponding proton injection rate in this energy range is $\approx 3.2 \times 10^{43}$~s$^{-1}$, and the total rate for proton energies $E_p > E_{\rm min}$ is 
\begin{equation}
\dot{N}_{\rm GCR}(p) \approx (0.2 {\rm~-~} 1.5)\times 10^{45}~{\rm protons~s^{-1}}~,
\label{eq:eff1}
\end{equation}
with $E_{\rm min}$ in the range $0.1$~--~$3$~MeV. This result can be compared to the theoretical rate for fast proton production by SN shocks in SBs: 
\begin{equation}
\dot{\rm GCR}^{\rm SB}_{\rm gas}(p)=\dot{\rm SN} f_{\rm SB} \frac{M_{\rm s.-u.}^{\rm SB} X_{\rm H}}{m_p} \eta^{\rm SB}_{\rm gas}~,
\label{eq:eff2}
\end{equation}
where $\dot{\rm SN} \approx 1/{48~\rm yr}$ is the SN frequency in our Galaxy \citep{fer01}, $f_{\rm SB}=0.6$ is the fraction of Galactic SNe exploding in SBs (Section~\ref{sec:snr}), $M_{\rm s.-u.}^{\rm SB}$ is the mass of gas swept up by a SN shock in a SB during the CR acceleration phase, $X_{\rm H}=0.71$ is the mass fraction of H in the SB gas, assumed to be of SC composition, $m_p$ is the proton mass, and $\eta^{\rm SB}_{\rm gas}$ the acceleration efficiency, i.e. the fraction of ions in the shocked media that end up as CR particles. The swept-up mass can be estimated from Equation~(\ref{eq:sumass}) with $V_{s, {\rm min}}^{\rm SB}=600$~km~s$^{-1}$(see Section~\ref{sec:superbubble}): $M_{\rm s.-u.}^{\rm SB}=190$~\msun. By equating $\dot{\rm GCR}^{\rm SB}_{\rm gas}$ to $\dot{N}_{\rm GCR}$, we get 
\begin{equation}
\eta^{\rm SB}_{\rm gas} \approx (0.4 {\rm~-~} 2.3)\times 10^{-5}~,
\label{eq:eff3}
\end{equation}
which is on the low side of the range of efficiency predicted by the diffusive shock acceleration theory for strong shocks \citep[e.g.][]{cap10,bla13}: $\eta \sim 10^{-5} .. 10^{-3}$. 

The rate of wind material becoming GCRs in the WTSs of massive stars can be estimated from the required mixing of this material with a reservoir a SC composition to explain the $^{22}{\rm Ne}/^{20}{\rm Ne}$ ratio of the GCR source composition ($x_w \approx 6$\% in Model 3, see Table~\ref{tab:model}):
\begin{equation}
\eta^{\rm wind}_{\rm gas} \approx \eta^{\rm SB}_{\rm gas} \frac{x_w M_{\rm s.-u.}^{\rm SB}}{(1-x_w) M_w} \approx 0.8 \eta^{\rm SB}_{\rm gas}~,
\label{eq:eff4}
\end{equation}
where $M_w \approx 16~M_\odot$ is the mean mass of wind material (averaged over the IMF) processed by the termination shocks. As discussed in Section~\ref{sec:22newind}, the acceleration efficiency is expected to be lower than the mean value during the main sequence and supergiant phases of the massive stars and higher during their W.-R. phase, where the wind power gets greater. 

Finally, the rate of GCR refractories produced from acceleration of dust grains can be estimated in the two scenarios outlined in Section~\ref{sec:grain}. If the grain acceleration also occurs in SBs, the injection efficiency of sputtered, refractory elements is simply: 
\begin{equation}
\eta^{\rm SB}_{\rm dust} \approx \eta^{\rm SB}_{\rm gas} \epsilon / {F^{\rm SB}_{\rm dust}} \approx (0.8 {\rm~-~} 4.6)\times 10^{-4} / {F^{\rm SB}_{\rm dust}}~,
\label{eq:eff5}
\end{equation}
where $\epsilon=\epsilon_{\rm dust}/\epsilon_{\rm gas}\approx 20$ (see Table~\ref{tab:model}) and $F^{\rm SB}_{\rm dust} \leq 1$ is the fraction of dust in the material swept up by SN shocks in SBs compared to the dust fraction in the warm ISM. \citet{och15} assume that $F^{\rm SB}_{\rm dust} = 1$ (this parameter is noted $q_{\rm II,OB}$ in this paper), considering that mass loading replenishes the dust content inside a SB after each SNR from the parent OB association. But a decrease in the depletion of gas-phase elements in the hot ionised medium would suggest that $F^{\rm SB}_{\rm dust}$ is lower than one. 

If the GCR refractories are predominantly produced in the WIM, their acceleration efficiency can be estimated as
\begin{equation}
\eta^{\rm WIM}_{\rm dust} \approx \eta^{\rm SB}_{\rm gas} \epsilon \frac{f_{\rm SB}}{f_{\rm WIM}} \frac{M_{\rm s.-u.}^{\rm SB}}{M_{\rm s.-u.}^{\rm WIM}} \approx (1.0 {\rm~-~} 5.8)\times 10^{-4}~,
\label{eq:eff6}
\end{equation}
where $f_{\rm SB}=0.6$, $f_{\rm WIM}=0.12$, and $M_{\rm s.-u.}^{\rm WIM}=760$~\msun~(as calculated from Eq.~(\ref{eq:sumass}) with $V_{s, {\rm min}}^{\rm SB}=300$~km~s$^{-1}$; Section~\ref{sec:superbubble}). This result is consistent with the prediction of the grain acceleration model \citep{ell97,ell98}. 

\section{Conclusions}
\label{sec:conclusions}

Inspired by the seminal work of \citet{mey97}, we have developed a new method of  analysis of the GCR abundance data that allows us to specify the origin of these particles in the ISM. The measured abundances of all primary and mostly primary CRs from H to Zr are well explained in our model, including the overabundance of $^{22}$Ne, which has remained problematic until now. We showed that the reported overabundance of C and heavier elements relative to H and He \citep[e.g.,][]{mey97} can be explained by the fact that CR protons and $\alpha$-particles have significantly different source spectra than the other elements. 
With a minimum CR source energy of the order of a few hundred keV~nucleon$^{-1}$, the GCR source abundances of H and He relative to the SC composition are similar to those of the other volatile elements N, Ne and Ar. We found that the CR volatiles are mostly accelerated in Galactic SBs, from SNR shocks sweeping up a plasma of temperature $\gsim 2\times 10^6$~K. SNRs in the warm ISM contribute to the GCR volatile composition for less than 28\%, whereas about 40\% of Galactic SNe occur in this phase and not in SBs. We also found that the overabundance of $^{22}$Ne in GCRs is most likely due to a  contribution of particle acceleration in WTSs of massive stars. From the CR-related $\gamma$-ray luminosity of the Milky Way, we estimated the particle acceleration efficiency in both SN shocks and WTSs to be of the order of $10^{-5}$, which is consistent with the prediction of the diffusive shock acceleration theory for strong shocks.  

The high $^{22}$Ne/$^{20}$Ne ratio in the GCR composition has been used as an argument that GCRs originate in SBs, based on the assumption that SNR shocks within SBs should accelerate a medium enriched by W.-R. winds from the most massive stars of the parent OB association \citep[][and references therein]{hig03,lin19}. But, although the GCR volatiles do seem to come mainly from SBs, this model for the origin of the excess $^{22}$Ne is unlikely for two main reasons. First, the N/Ne ratio is too high in the mean composition of massive star winds, as calculated from the stellar yields of \cite{lim18},  to explain the N abundance of the GCR source composition. Also, the SB gas would need to be enriched in W.-R. wind material at the level of $x_w \approx 50$\%, which is not supported by X-ray observations. 

The GCR refractory elements most likely originate from the acceleration and sputtering of dust grains in SNR shocks. The fact that the highly refractory elements Mg, Al, Si, Ca, Fe, Co, and Ni are in the same proportions in the GCR composition as in the SC composition (to within 20\%) shows that the accelerated refractory elements come from various dust grains of the ISM mix, and not only from grains formed in core-collapse SN ejecta. We suggest that the GCR refractories could be mainly produced in SBs, if dust is continuously replenished in the SB interior through thermal evaporation of embedded molecular clouds swept up by SN shocks, as argued by \citet{och15} for the Orion-Eradinus SB. However, the quantitative assessment of this hypothesis requires further work to specify the amount and nature of dust contained in the hot ionised medium of the ISM and apply the grain acceleration model of \citet{ell97,ell98} to a SB environment. Alternatively, the GCR refractories could be predominantly produced in SN shocks propagating in the WIM. In this case, their injection efficiency into the shock acceleration process would be $\gsim 10^{-4}$. The GCR C abundance might be key to distinguishing whether the fast refractory elements are produced in SBs or in the warm ISM, because the fraction of C in dust could be markedly different in these two environments. 

A possible hint for the production of GCR refractories in SBs is provided by the recent detection of radioactive $^{60}$Fe in CRs with the ACE-CRIS instrument \citep{bin16}. $^{60}$Fe is a mostly primary CR \citep[among the 15 $^{60}$Fe events detected by CRIS with 16.8 years of data, $\sim 1$ nucleus could be the result of interstellar fragmentation of heavier nuclei;][]{bin16} with half-life of $2.62 \times 10^6$~years. Its detection in CRs implies that no more than a few million years elapsed between its nucleosynthesis in core-collapse SN(e) and transport to Earth and that the $^{60}$Fe source(s) is(are) located at less than about 1~kpc, which corresponds to the distance CRs can diffuse over this time. A very likely source of the detected $^{60}$Fe nuclei is the large and nearby ($< 150$~pc) Scorpius-Centaurus OB association \citep{bin16}, which is thought to have produced several SNe in the last few million years and also to have formed the Local Bubble of hot gas surrounding the solar system \citep{bre16}. 

A major challenge in CR physics today is to explain the difference in the source spectra between H, He, and heavier elements, given that the diffusive shock acceleration theory predicts that the accelerated nuclei should have the same rigidity spectrum. The existence of some of these spectral differences is still a matter of debate, as any GCR source spectrum depends on the model of CR propagation in the ISM and the relative contribution of secondary CRs to the measured LIS. Thus, the AMS-02 collaboration measured that Ne, Mg, and Si LIS have a different rigidity dependence than He, C, and O, and argued that these two groups of elements constitute two different classes of primary CRs \citep{agu20}. But \citet{sch21} recently found that the AMS-02 data for Ne, Mg, and Si can be correctly reproduced using the same source spectrum as for C and O. Also, the particularity of the Fe spectrum measured with AMS-02 \citep{agu21,bos21} seems questionable \citep{sch21}. However, the data from Voyager~1 and AMS-02 leave no doubt that H and He have different source spectra (Section~\ref{sec:data}), and that both are also significantly different from those of the other elements \citep{tat18,evo19,sch21}. It should be interesting to study in a future work if these spectral differences could be related to the ionisation states of the elements in the preshock gas of SNR shocks propagating in the WIM and in SBs, and use this information to further constrain the origin of GCRs in the ISM.  

\section*{Acknowledgements}

VT acknowledges the International Space Science Institute (ISSI) for Teamwork 351 on ``\textit{The origin and composition of galactic cosmic rays}'', where this work was initiated. JCR and VT are grateful to the Russian Academy of Science for the invitation to Uzkoe of the International Working group on ``\textit{High energy processes in astrophysical objects: fundamental physics and new detector technologies}''. SG would like to thank Damiano Caprioli for fruitful discussions on the cosmic-ray injection process. SG and VT acknowledge support from Agence Nationale de la Recherche (grant ANR-17-CE31-0014). 
The work of  SR is partially supported by 
the {\sc Departments of Excellence} grant awarded by the Italian Ministry of Education,
University and Research ({\sc Miur}), the 
Research grant {\sl The Dark Universe: A Synergic Multimessenger Approach}, No. 2017X7X85K funded by the {\sc Miur} and by the 
Research grant {\sc TAsP} (Theoretical Astroparticle Physics) funded by Istituto
Nazionale di Fisica Nucleare. 

\section*{Data Availability}
The data underlying this article are available in the article.




\appendix

\section{Massive star winds}

\begin{figure}
\begin{center}
 \includegraphics[width=0.7\columnwidth]{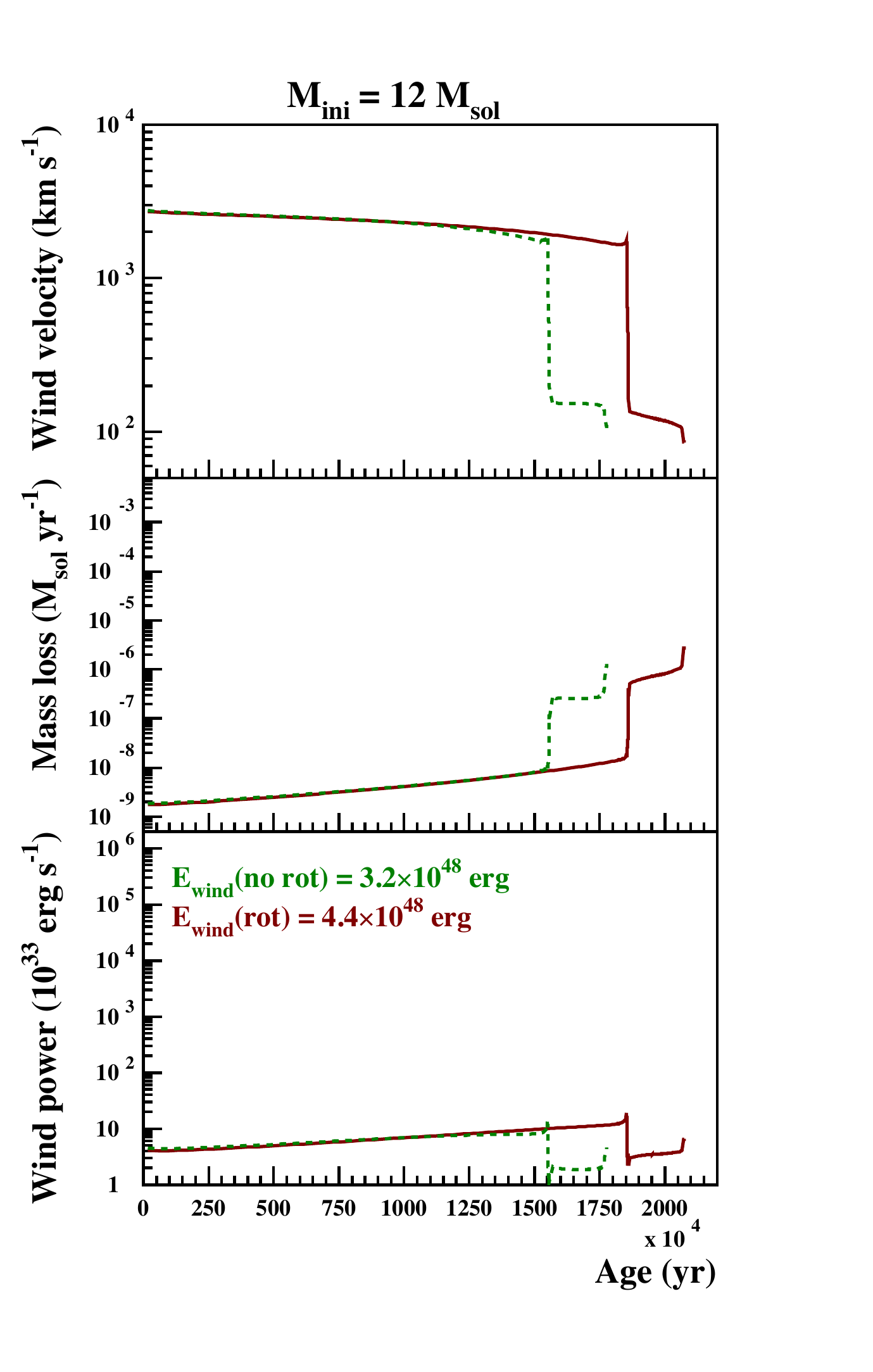}
 \caption{(\textit{Upper panel}) Wind terminal velocity, (\textit{middle panel}) star mass loss, and (\textit{lower panel}) wind mechanical power, as a function of stellar age, for two stars of initial mass $M_{\rm ini}=12~M_\odot$. Red solid curves: rotating star; green dashed curves: non-rotating star. The total kinetic energy of the winds integrated over the star lifetime is reported in the \textit{lower panel} for both the non-rotating and rotating stars.}
 \label{fig:wind12}
\end{center}
\end{figure}

\begin{figure}
\begin{center}
 \includegraphics[width=0.7\columnwidth]{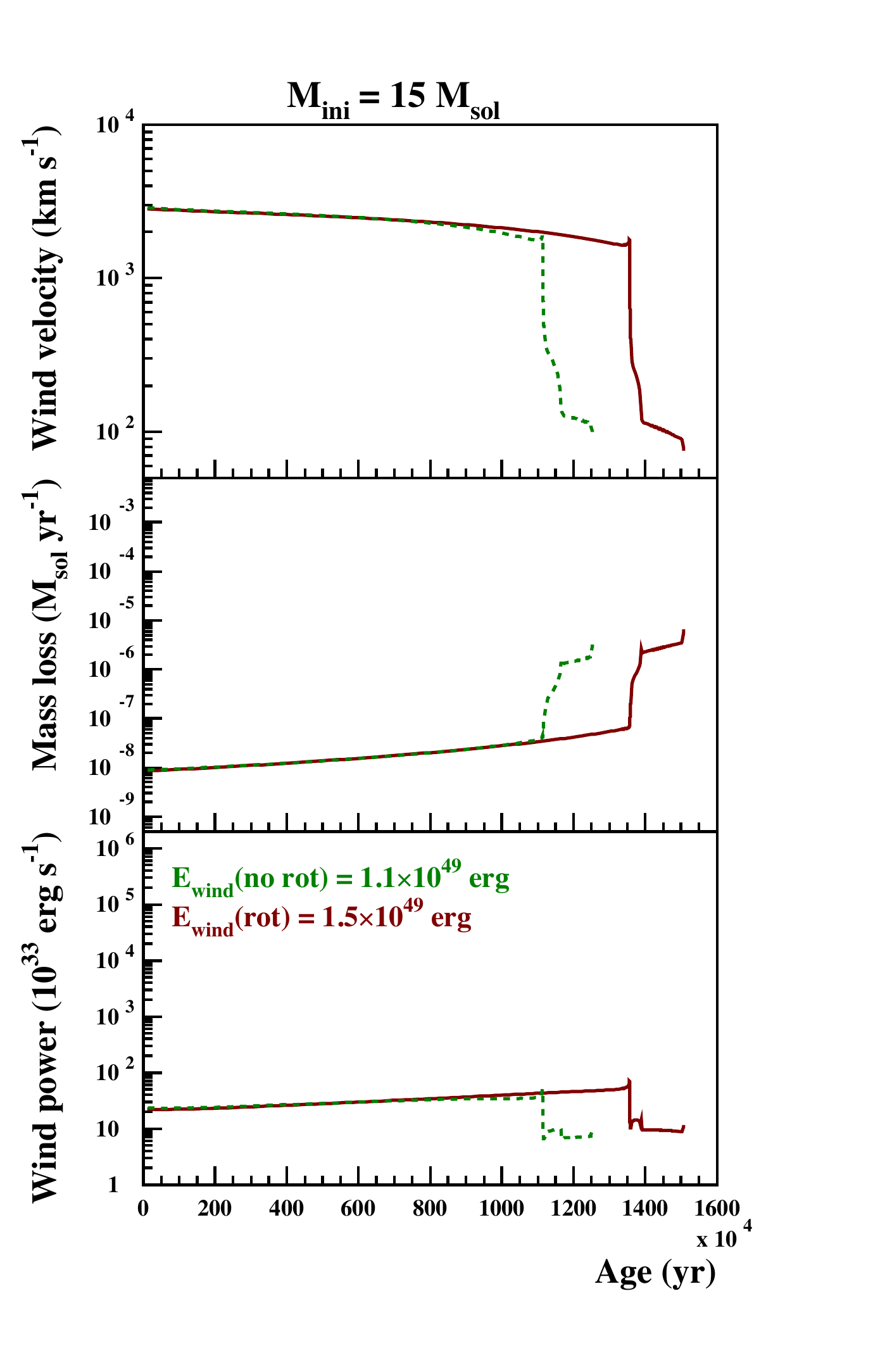}
 \caption{Same as Figure~\ref{fig:wind12} but for $M_{\rm ini}=15~M_\odot$.}
 \label{fig:wind15}
\end{center}
\end{figure}

\begin{figure}
\begin{center}
 \includegraphics[width=0.7\columnwidth]{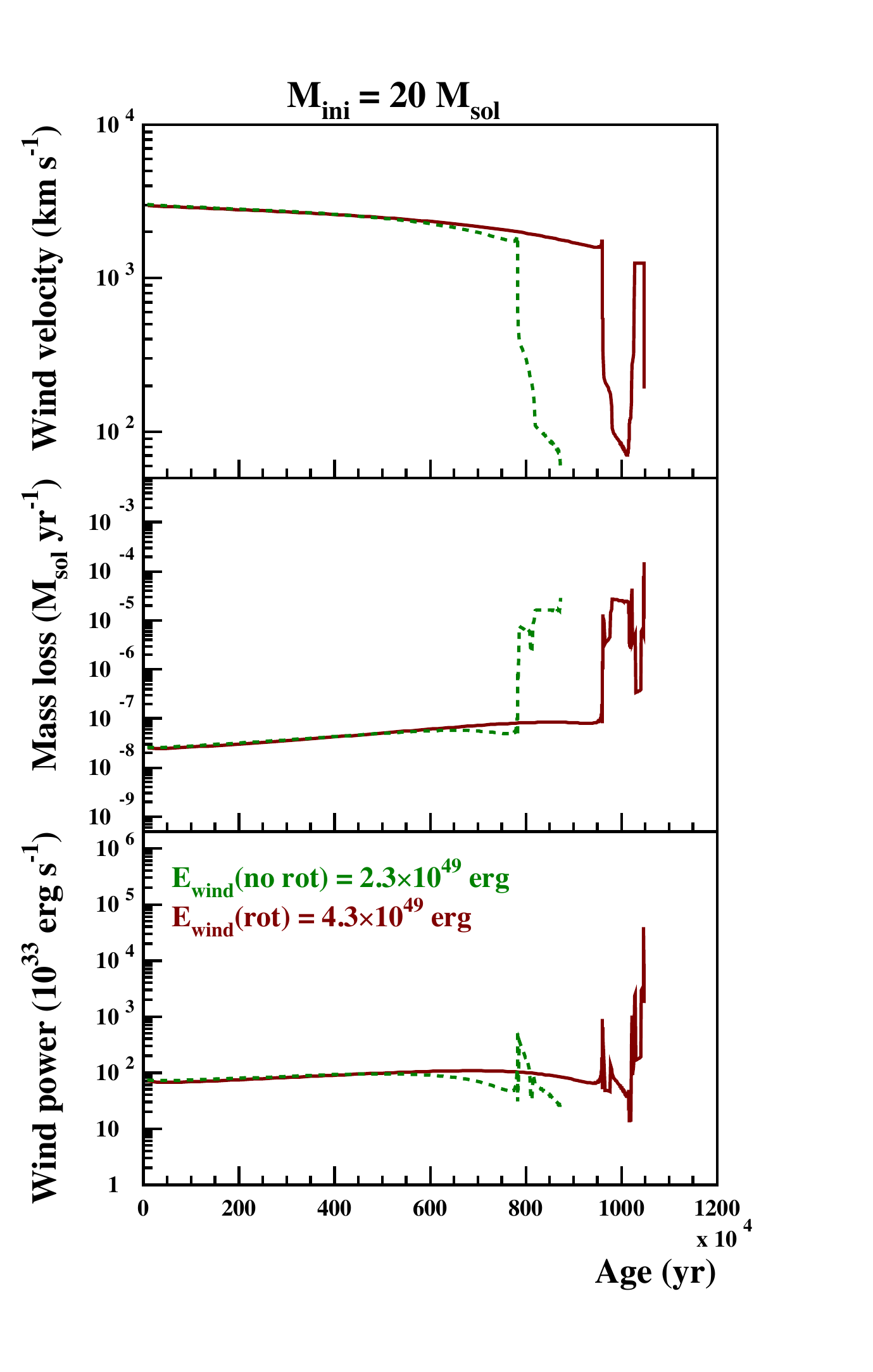}
 \caption{Same as Figure~\ref{fig:wind12} but for $M_{\rm ini}=20~M_\odot$.}
 \label{fig:wind20}
\end{center}
\end{figure}

\begin{figure}
\begin{center}
 \includegraphics[width=0.7\columnwidth]{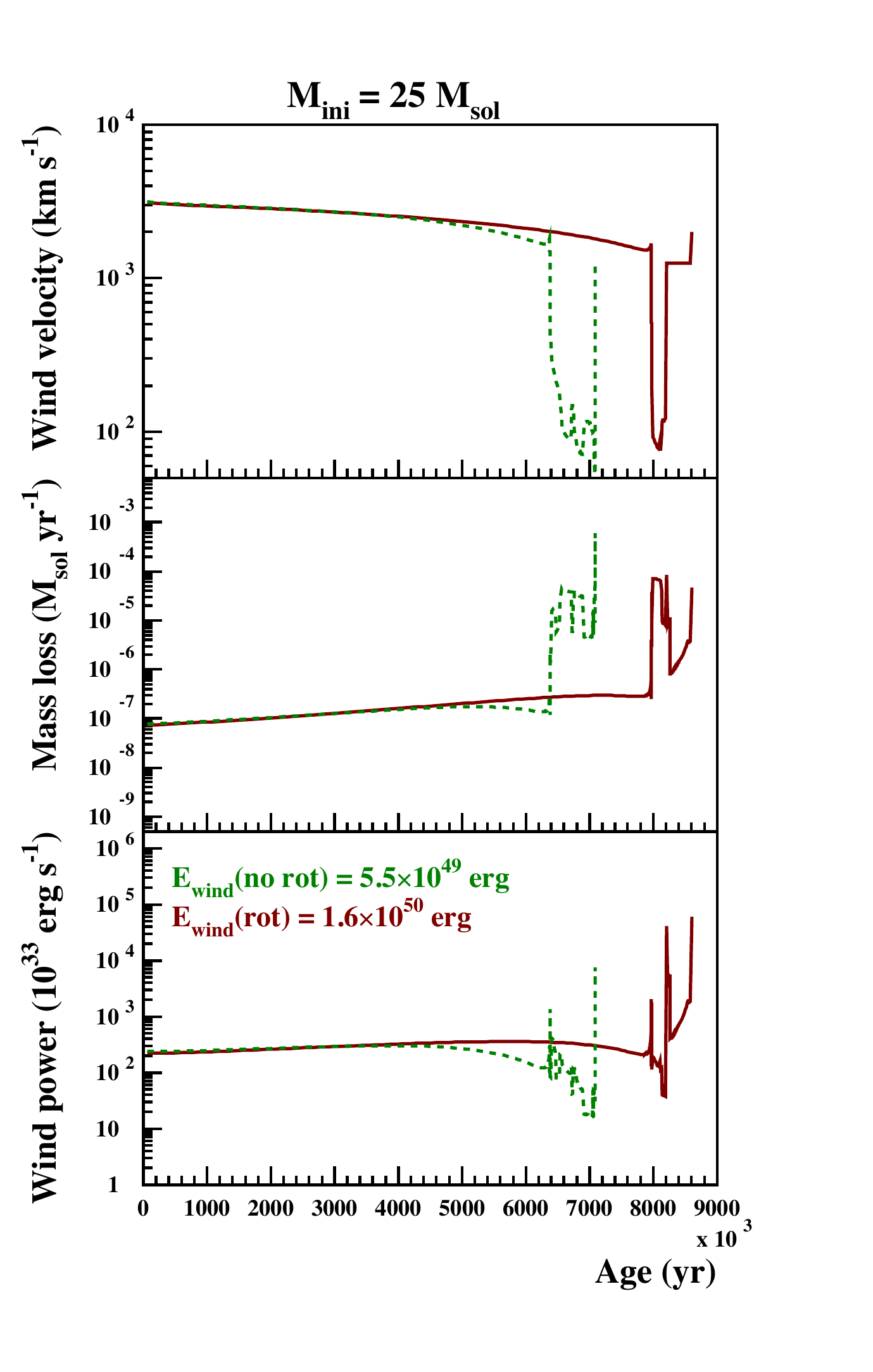}
 \caption{Same as Figure~\ref{fig:wind12} but for $M_{\rm ini}=25~M_\odot$.}
 \label{fig:wind25}
\end{center}
\end{figure}

\begin{figure}
\begin{center}
 \includegraphics[width=0.7\columnwidth]{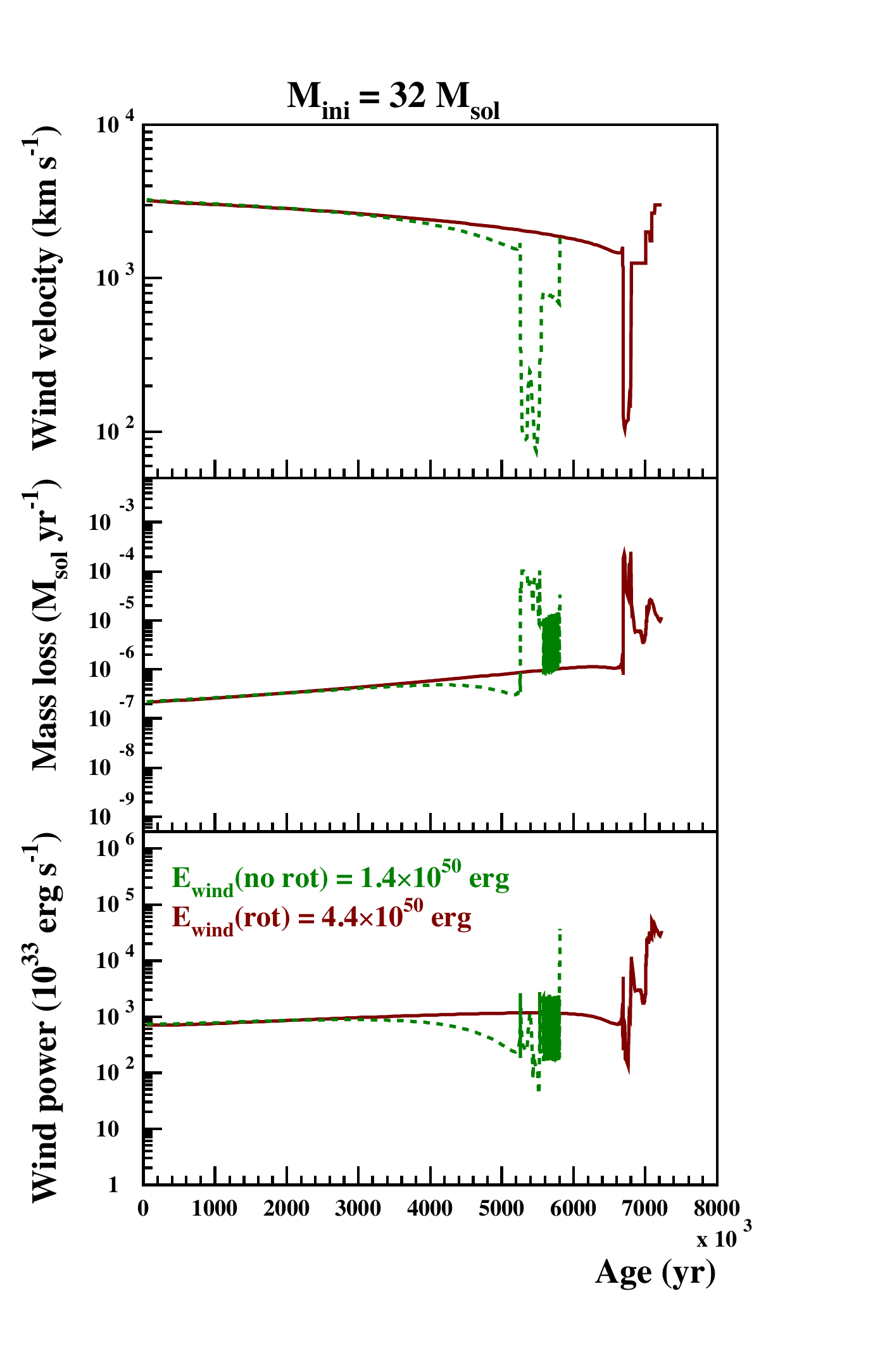}
 \caption{Same as Figure~\ref{fig:wind12} but for $M_{\rm ini}=32~M_\odot$.}
 \label{fig:wind32}
\end{center}
\end{figure}

\begin{figure}
\begin{center}
 \includegraphics[width=0.7\columnwidth]{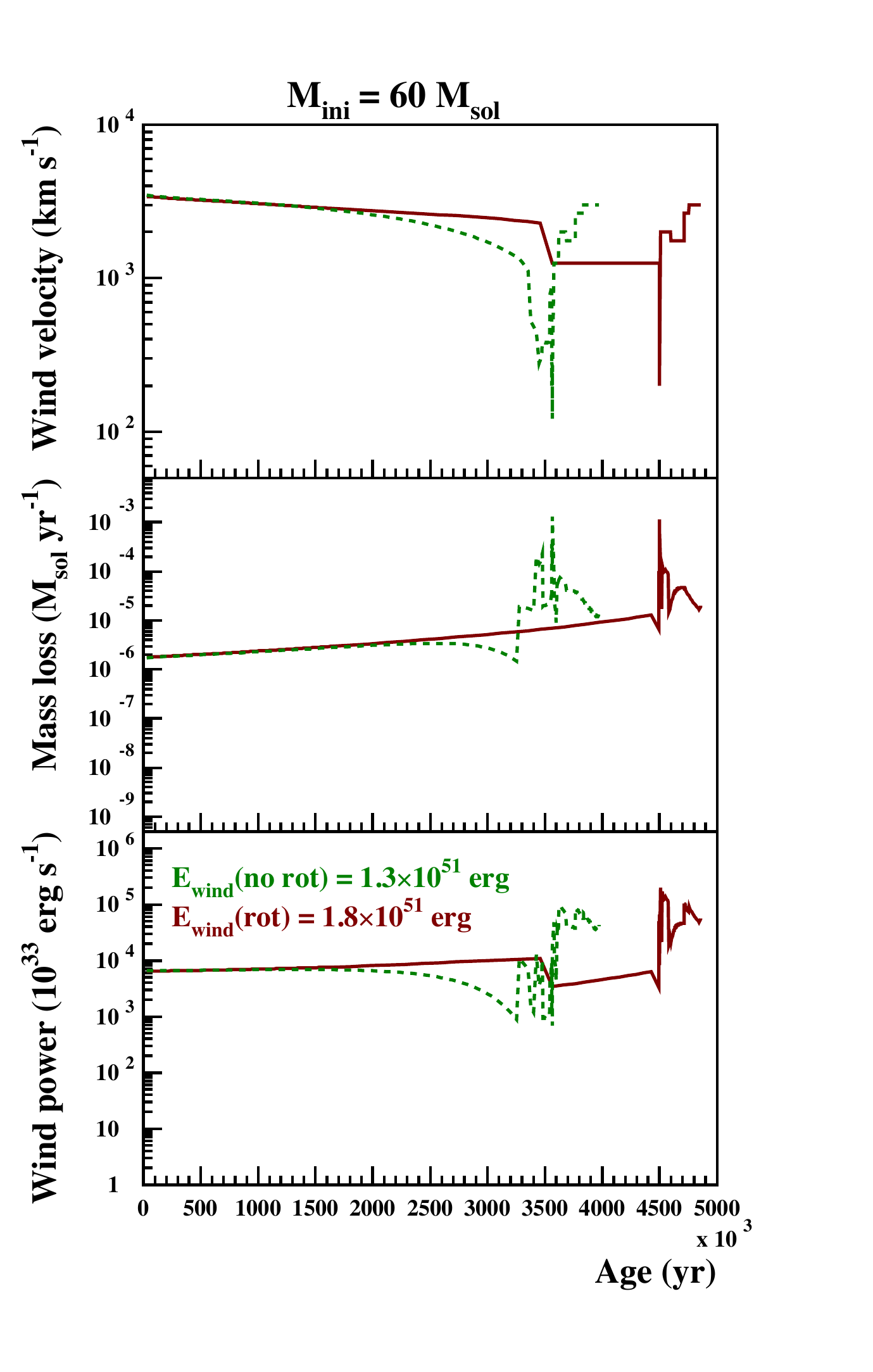}
 \caption{Same as Figure~\ref{fig:wind12} but for $M_{\rm ini}=60~M_\odot$.}
 \label{fig:wind60}
\end{center}
\end{figure}

\begin{figure}
\begin{center}
 \includegraphics[width=0.7\columnwidth]{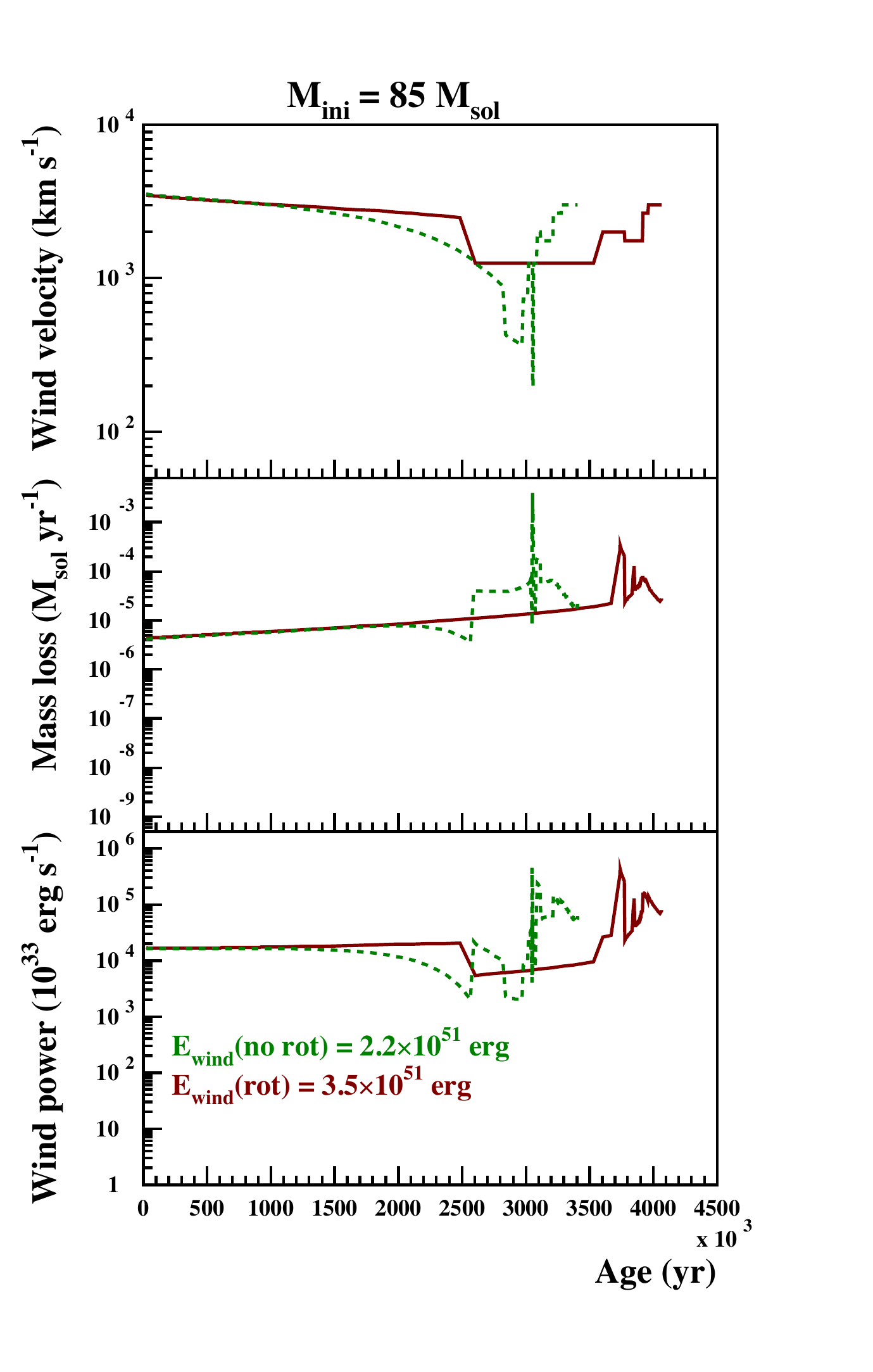}
 \caption{Same as Figure~\ref{fig:wind12} but for $M_{\rm ini}=85~M_\odot$.}
 \label{fig:wind85}
\end{center}
\end{figure}

\begin{figure}
\begin{center}
 \includegraphics[width=0.7\columnwidth]{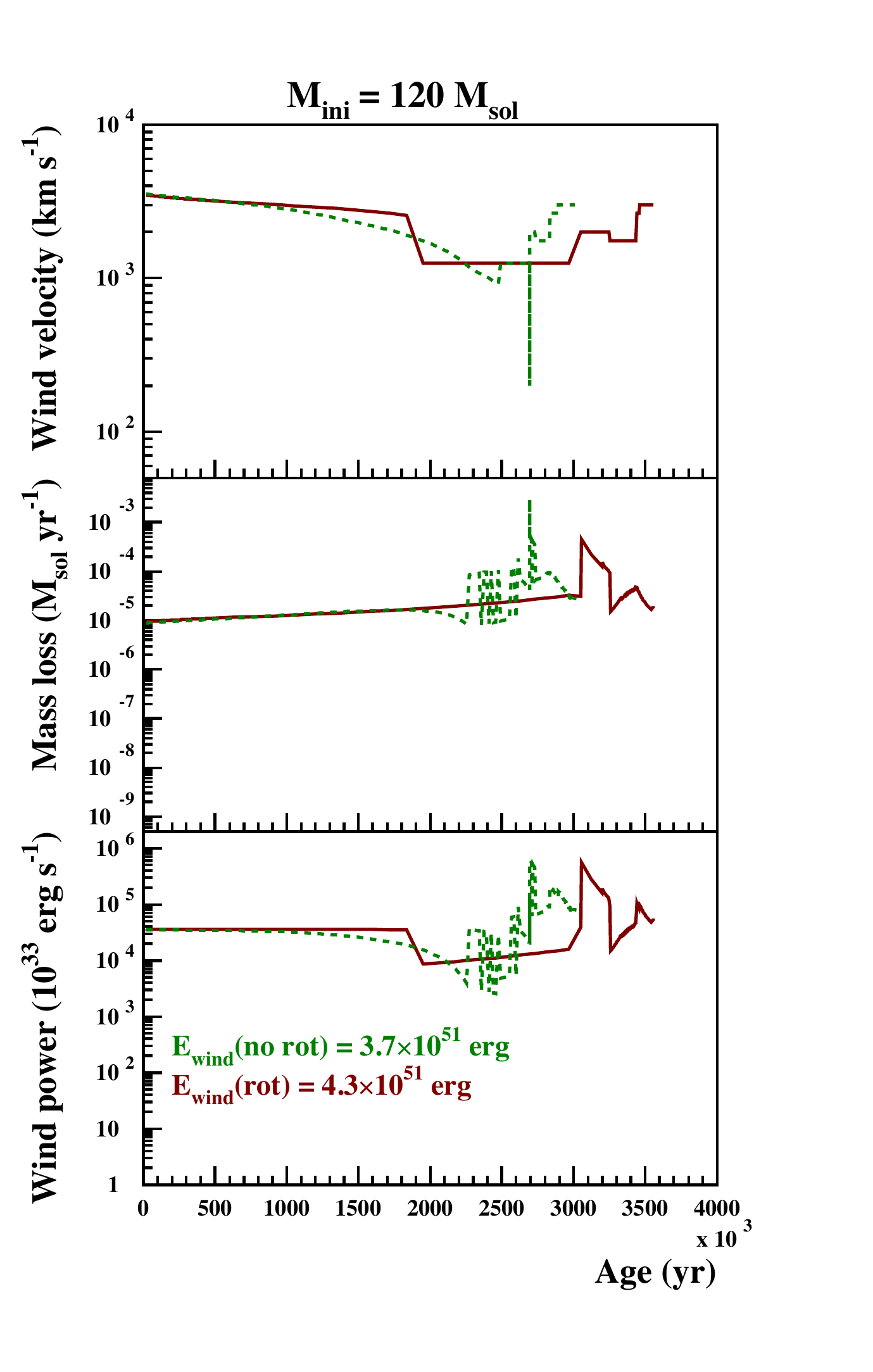}
 \caption{Same as Figure~\ref{fig:wind12} but for $M_{\rm ini}=120~M_\odot$.}
 \label{fig:wind120}
\end{center}
\end{figure}

Figures~\ref{fig:wind12} -- ~\ref{fig:wind120} show calculated properties of massive star winds from the stellar evolution models of \citet{eks12} and \citet{geo12} extracted from the Geneva Observatory database. Details of the calculations are given in Section~\ref{sec:22newind}.

\bsp	
\label{lastpage}
\end{document}